%
%
\input harvmac.tex
\hfuzz 15pt
\input amssym.def
\input amssym.tex
\input epsf\def\tfig#1{{
\xdef#1{Fig.\thinspace\the\figno}}Fig.\thinspace\the\figno
\global\advance\figno by1}


\input epsf

%





\def\p{\partial}

\def\a{\alpha}

\def\g{\gamma}

\def\e{\varepsilon}

\def\l{\lambda}

\def\s{\sigma}

\def\G{\Gamma}

\def\L{\lambda_{_{^L}} }



  \def\CA {{\cal A}}

  \def\CG {{\cal G}}

  \def\CL {{\cal L}}

  \def\CO {{\cal O}}

  \def\CT {{\cal T}}
  
  \def\CV {{\cal V}}
  \def\CW {{\cal W}}

\def\hG{\hat G}

\def\[{\left[}
\def\]{\right]}
\def\({\left(}
\def\){\right)}
\def\<{\left\langle\,}
\def\>{\,\right\rangle}
 \def\hf{ {\frac{1}{2}}}


\def\inv{^{-1}}

\def\hf{\textstyle{1\over 2}}

 \def\frac#1#2{ {{\textstyle{#1\over#2}}}}
\def\inv{^{\raise.15ex\hbox{${\scriptscriptstyle -}$}\kern-.05em 1}}

 \def\IN{ \Bbb N}
 \def\IP{\relax{\rm I\kern-.18em P}}

 \def\GL{ C^{\rm Liou}}
  \def\GM{ C^{\rm Matt }}

\def\rb{ \noindent $\bullet$\ \ }

  \def\zg{\gamma} 
\def\ze{\e}

\def\zu{\Upsilon}

\def\zum{\hat \Upsilon}

\def\la{\langle} \def\ra{\rangle}

\def\IR{{ \Bbb R} }
\def\IZ{{ \Bbb Z} }

\def\dC{C\kern-6.5pt I}

\def\CA{{\cal A}}              
              
\def\CG{{\cal G}}              
              \def\CL{{\cal L}}
              \def\CO{{\cal O}}
              
       \def\CT{{\cal T}}       
\def\CV{{\cal V}}       \def\CW{{\cal W}}

\def\CV{{\cal V}}

\def\bc{\bar{c}}

\def\gb{{\bf b}}
\def\gc{{\bf c}}
\def\bgc{\bar{\bf c}}

\def\muM{\mu_{_M}}

\def\LM{  \l_{_{M}} }
\def\ha{\hat a}


\chardef\tempcat=\the\catcode`\@ \catcode`\@=11
\def\cyracc{\def\u##1{\if \i##1\accent"24 i%
    \else \accent"24 ##1\fi }}
\newfam\cyrfam



\def\ijmp#1#2#3{{Int. J. Mod. Phys.} {\bf #1} (#2) #3}

\def\hepth#1{{hep-th/}#1}

\def\encadremath#1{\vbox{\hrule\hbox{\vrule\kern8pt\vbox{\kern8pt
 \hbox{$\displaystyle #1$}\kern8pt}
 \kern8pt\vrule}\hrule}}

\def\hfb{ \frac{b}{2}}

%
\lref\DO{ H. Dorn, H.J. Otto: Two and three point functions in
Liouville theory, {\it Nucl. Phys.} {\bf B 429} (1994) 375, hep-th/9403141.  }
\lref\ZZtp{A.B. Zamolodchikov and Al.B. Zamolodchikov, Structure
constants and conformal bootstrap in Liouville field theory,
{\it Nucl. Phys.} {\bf B 477} (1996) 577, hep-th/9506136.  }
\lref\Ta{J. Teschner, On the Liouville
three-point function.  {\it Phys.Lett.}, {\bf B 363} (1995) 65,
hep-th/9507109.  }
\lref\PTa{B.~Ponsot, J.~Teschner, ``Liouville bootstrap via harmonic
analysis on a noncompact quantum group'', hep-th/99111110;
``Clebsch-Gordan and Racah-Wigner coefficients for a continuous
series of representations of $U_q(sl(2,\IR))$'', {\it Comm.  Math.  Phys.}  {\bf 224}, 3 (2001), math-qa/0007097.}
    \lref\FZZb{V.~Fateev, A.~B.~Zamolodchikov and A.~B.~Zamolodchikov,
   ``Boundary Liouville field theory.  I: Boundary state and boundary  two-point function'', hep-th/0001012.  }
\lref\ZZPseudo{ A.~B. Zamolodchikov and A.~B. Zamolodchikov,
``Liouville field theory on a pseudosphere, \hepth{0101152}.  }
\lref\PTtwo{B.~Ponsot, J.~Teschner, ``Boundary Liouville field theory:
Boundary three point function'', Nucl.~Phys.~{\bf B 622} (2002) 309,
\hepth{0110244}.  }
\lref\hosomichi{K.~Hosomichi, "Bulk-Boundary Propagator in Liouville
Theory on a Disc", JHEP {\bf 0111} 044 (2001), \hepth{0108093}.  }
  \lref\GinM{ P. Ginsparg and G. Moore, ``Lectures on 2D gravity and
  2D string theory (TASI 1992)", hep-th/9304011.  }
\lref\DiFrancescoGinsparg{P.~Di Francesco, P.~Ginsparg and
J.~Zinn-Justin,``2-D Gravity and random matrices'', {\it Phys.\ Rept.}\ {\bf
254} (1995) 1, hep-th/9306153.  }
\lref\polchinski{J. Polchinski, ``What is string theory'', {\it
Lectures presented at the 1994 Les Houches Summer School ``Fluctuating
Geometries in Statistical Mechanics and Field Theory''},
\hepth{9411028}.}
\lref\KlebanovMQM{ I.~Klebanov, ``String theory in two-dimensions'',
hep-th/9108019.  }
\lref\JevickiQN{ A.~Jevicki, ``Developments in 2-d string theory'',
   hep-th/9309115.  }
  \lref\Witten{ E.~Witten, ``Ground ring of two-dimensional string
  theory'', {Nucl.\ Phys.}\ {\bf B 373}, 187 (1992), hep-th/9108004.  }
\lref\KlebPol{I.R.Klebanov and A.M. Polyakov, ``Interaction of discrete states in two-dimensional string theory'', {\it Mod. Phys. Lett.} {\bf 6} (1991) 3273, \hepth{9109032}.}
\lref\KMS{ D. Kutasov, E. Martinec, N. Seiberg, ``Ground rings and
their modules in 2-D gravity with $c\le1 $ matter", {\it Phys. Lett.}  {\bf
B 276} (1992) 437, \hepth{9111048}.  }
\lref\kachru{S. Kachru, ``Quantum rings and recursion relations in 2D
quantum gravity'', {\it Mod.  Phys.  Lett.}  {\bf A7} (1992) 1419,
hep-th/9201072.}
\lref\bershkut{ M. Bershadsky and D. Kutasov,
``Scattering of open and closed strings in (1+1)-dimensions",
{\it Nucl. Phys.} {\bf B 382} (1992) 213, \hepth{9204049}.  }
\lref\newhat{
  I.~R.~Klebanov, J.~Maldacena and N.~Seiberg,
  ``D-brane decay in two-dimensional string theory,''
  JHEP {\bf 0307}, 045 (2003)
\hepth{0305159}.
}
\lref\SeibergS{ N. Seiberg and D. Shih, ``Branes, rings and matrix
models on minimal (super)string theory", JHEP {\bf 0402} (2004) 021,
hep-th/0312170.  }
%
\lref\KostovCY{ I.~K.~Kostov, ``Boundary ground ring in 2D string
theory'', {\it Nucl.\ Phys.}\ {\bf B 689}, 3 (2004), hep-th/0312301.  }
\lref\BuKa{ D.~V.~Boulatov and V.~A.~Kazakov, ``The Ising model on
random planar lattice: The structure of phase transition and the exact
critical exponents,'' {\it Phys.\ Lett.}\ {\bf 186 B}, 379 (1987).  }
\lref\IOn{ I.~K.~Kostov, ``O(N) vector model on a planar random
lattice: Spectrum of anomalous dimensions'', {\it Mod.\ Phys.\ Lett.}\ {\bf
A 4}, 217 (1989).  }
\lref\KPb{ I.~K.~Kostov and V.~B.~Petkova, Non-rational 2d quantum
gravity: II. Target space CFT, hep-th/0609020.  }
\lref\KostovCG{ I.~K.~Kostov, ``Strings with discrete target space'',
{\it Nucl.\ Phys.}\ {\bf B 376}, 539 (1992), hep-th/9112059.  }
\lref\HiguchiPV{ S.~Higuchi and I.~K.~Kostov, ``Feynman rules for
string field theories with discrete target space'', {\it Phys.\ Lett.}\ {\bf
B 357}, 62 (1995), hep-th/9506022.  }
\lref\AlZ{Al.B. Zamolodchikov, The three-point function in the minimal
Liouville gravity, {Theor.\ Math.\ Phys.}\ {\bf 142}, 183 (2005); On the
three-point function in minimal Liouville gravity, \hepth{0505063}.  }
\lref\DF{Vl.S. Dotsenko and V.A. Fateev, ``Four point correlation
functions and the operator algebra in the two-dimensional conformal
invariant theories with the central charge $c < 1$'', {\it Nucl.  Phys.}
{\bf B 251} [FS13], 691 (1985).}
  \lref\DiK{ P. Di Francesco, D. Kutasov, ``World sheet and space time
  physics in two dimensional (super) string sheory", {\it Nucl.Phys.}  {\bf
  B 375} (1992) 119, hep-th/9109005.  }
\lref\KPlet{ I.~K.~Kostov and V.~B.~Petkova, Bulk correlation
functions in 2D quantum gravity, 
Theor. Math. Phys. {\bf 146} (1): 108 (2006), hep-th/0505078.
 }
  \lref\BZ{V. Pokrovsky, A. Belavin and Al.  Zamolodchikov, Moduli integrals, ground ring and four-point function in minimal Liouville gravity, in {\it Polyakov's string: Twenty five years after}, hep-th/0510214.}		
\lref\BM{ A.~Basu and E.~J.~Martinec,
  ``Boundary ground ring in minimal string theory,''
  {\it Phys.\ Rev.}\  {\bf D} {\bf 72}, 106007 (2005)
\hepth{0509142}.}
\lref\BZJ{ A.~A.~Belavin and A.~B.~Zamolodchikov, ``Moduli integrals
and ground ring in minimal Liouville gravity,'' JETP Lett.\ {\bf 82},
7 (2005) [Pisma Zh.\ Eksp.\ Teor.\ Fiz.\ {\bf 82}, 8 (2005)].  }
\lref\DavidHJ{ F.~David, ``Conformal field theories coupled to 2-D
gravity in the conformal gauge'', {\it Mod.\ Phys.\ Lett.}\ {\bf A 3}, 1651
(1988).  }
\lref\DistlerJT{ J.~Distler and H.~Kawai, ``Conformal field theory and
2-D quantum gravity or who's afraid of Joseph Liouville?'', {\it Nucl.\
Phys.}\ {\bf B 321}, 509 (1989).  }
\lref\Tes{J. Teschner, Remarks on Liouville theory
with boundary, hep-th/0009138.}
\lref\SeibergEB{ N.~Seiberg, ``Notes on quantum Liouville theory and
quantum gravity'', {\it Prog.\ Theor.\ Phys.\ Suppl.}\ {\bf 102}, 319
(1990).  }
  \lref\LianGK{ B.~H.~Lian and G.~J.~Zuckerman, ``New selection rules
  and physical states in 2-D gravity: Conformal gauge'', {\it Phys.\ Lett.}\
  {\bf B 254}, 417 (1991).  }
\lref\DFt{Vl.  Dotsenko and V. Fateev, Operator algebra of
two-dimensional conformal theories with central charge $c\le 1$, {\it Phys.
Lett.}  {\bf 154B}, (1085) 291.}
\lref\FGP{P. Furlan, A.Ch.  Ganchev and V.B. Petkova, ``Remarks on the
quantum group structure of the rational $c<1$ conformal theories'',
\ijmp{A6}{1991}{4859}.}
\lref\ST{A. Strominger and T. Takayanagi, ``Correlators in time like
bulk Liouville theory'', {\it Adv, Theor.  Math.  Phys.}  {\bf 7}, 369
(2003), hep-th/0303221.  }
\lref\Sch{V. Schomerus, ``Rolling tachyons
from Liouville theory'', JHEP {\bf 0311} (2003) 043, hep-th/0306026.  }
\lref\IMM{ C. Imbimbo, S. Mahpatra and S. Mukhi, Construction of
physical states of non-trivial ghost number in $c<1$ string theory,
{\it Nucl. Phys.} {\bf B 375} (1992) 399.  }
\lref\PZb{V.B. Petkova and J-B. Zuber, The many faces of Ocneanu
cells, {\it Nucl. Phys. } {\bf B 603} (2001) 449, hep-th/0101151.  }
\lref\AlZJ{ Al.  Zamolodchikov, Talk
delivered at the Workshop ``Liouville theory and matrix models'', RIKEN
(Wako, Saitama), June 2005.}
\lref\WitZw{ E.~Witten and B.~Zwiebach, ``Algebraic structures
and differential geometry in 2D string theory'', {\it Nucl. Phys.}
{\bf B  377} (1992) 55, hep-th/9201056.}

\lref\KlP{ I. R. Klebanov, A.  Pasquinucci, 
``Correlation functions from two-dimensional string Ward identities",
{\it Nucl.Phys.}  {\bf B 393} (1993) 261,  hep-th/9204052.}

\lref\Kl{ I. R. Klebanov, ``Ward Identities in Two-Dimensional String Theory", 
{\it Mod. Phys. Lett.} {\bf  A7} (1992) 723, hep-th/9201005.}

\lref\Verl{E.  Verlinde, ``The Master Equation of 2D String Theory", 
{\it Nucl. Phys.} {\bf B 381} (1992) 141, hep-th/920202.}
 

\overfullrule=0pt
\Title{\vbox{\baselineskip12pt
\hbox{}}}
{\vbox{\centerline
 {Non-Rational 2D Quantum Gravity: I. }
\centerline{}
 \centerline{  World Sheet CFT  }
\centerline{ }
 \vskip2pt
}}
  \centerline{
 I.K. Kostov$^{1}$ and V.B. Petkova$^2$
  }

 \vskip 0.5cm

\centerline{\it  $^1$ Service de Physique
Th{\'e}orique,
 CNRS -- URA 2306,}
 \vskip -2pt

\centerline{\it C.E.A. - Saclay,
  F-91191 Gif-Sur-Yvette, France
 }
   \bigskip

   \centerline{ \vbox{\baselineskip12pt\hbox
{\it  $^2$Institute for Nuclear Research and Nuclear Energy, }
\hbox{\ \  \it 72 Tsarigradsko Chauss\'ee,
1784 Sofia, Bulgaria }}
}


\vskip 1.5cm

\baselineskip=11pt
{{
\noindent
We address the problem of computing the tachyon correlation functions
in Liouville gravity with generic (non-rational) matter central charge $c<1$.  
We consider two variants of the theory.  The first is the  conventional one in 
which the effective matter interaction is given by the two matter screening charges.  
In the second variant  the interaction is  defined by the Liouville dressings of the 
non-trivial vertex operator of zero  dimension.  This particular deformation, referred
to as  ``diagonal'',  is motivated  by the comparison with the discrete approach, 
which is the subject of a subsequent paper.  In both  theories we determine the 
ground ring of ghost zero physical operators by computing its OPE action on the 
tachyons and derive recurrence relations for the  tachyon bulk correlation functions.  
We find 3- and 4-point solutions to these functional  equations for various matter 
spectra. In particular, we find a closed expression for the  4-point function of order 
operators in the diagonal theory.}}

\Date{}
\vfill
\eject

\baselineskip=14pt plus 1pt minus 1pt

 \newsec{Introduction}

\noindent 
The exact results in Liouville theory obtained in the last
decade \refs{\DO\ZZtp\Ta\PTa\FZZb\ZZPseudo\PTtwo-\hosomichi} allowed
to improve some old techniques developed in $c\le 1$ string theories
(reviewed in
\refs{\GinM\DiFrancescoGinsparg\polchinski\KlebanovMQM-\JevickiQN})
and find new links between the world-sheet and matrix model
descriptions.  In particular, the fundamental OPE identities, used in
\refs{\Ta,\FZZb,\ZZPseudo,\PTtwo,\hosomichi} to evaluate various
Liouville structure constants, are similar in nature to the ground
ring relations in string theories \refs{\Witten
\KlebPol\KMS\kachru-\bershkut}, \refs{\WitZw\Kl\Verl-\KlP}.  Recently the ground ring structure
was reconsidered in \refs{\newhat, \SeibergS, \KostovCY}, where it was
applied to study the solitons, or D-branes, in $c\le 1$ string
theories.

In this work we generalize and exploit this approach to derive and
solve finite difference equations for the tachyon correlators on the
sphere.  We recall that the ground ring is the ring with respect to
the operator product expansions (OPE), modulo $Q_{\rm BRST}$-exact
terms, of the physical operators of zero ghost number.
Physical operators of fixed ghost number, like the
tachyons, represent modules under the action of the ring
\refs{\Witten}.

So far the technique has been tested in the simplest model of 2D
string theory that can be considered as a marginal deformation by
Liouville interaction of a two-component gaussian field action with
background charges.  One can perturb in a similar way the matter
component of the gaussian field by the matter screening charges.  As a
result one obtains a gravitational analog of the Dotsenko-Fateev
Coulomb gas construction.  The correlation functions depend on two
coupling constants: the Liouville coupling (cosmological constant)
$\mu_{_L}$ and its matter counterpart $\mu_{_M}$, associated with the
matter screening charge.  Unlike most of the previous studies, which
deal with the minimal string theories, we shall consider non-rational
values of the matter central charge characterized by a real parameter
$b$,
 \eqn\ccharge{ c=1-6(\frac{1}{b}- b)^2 \,. 
 }

One of the motivations for this work was to compare the correlation
functions in the continuous (world sheet), and the discrete (target space)
approaches to the 2D quantum gravity.  This is an old problem and there are few
matrix model results on the correlation functions with a non-trivial
matter.  Such results are known only for the simplest examples of the
rational, minimal theories, as the Ising model \BuKa, recently
reconsidered in \AlZJ.
Moreover, there is no matrix model to
match the non-rational case, except the $O(n)$ matrix model \IOn, whose
poor field content is too restricted.  In a subsequent paper \KPb\ we construct
such 
a model, in which the
matter degrees of freedom are parametrized by a semi-infinite discrete
set, generalizing the $ADE$ string theories \KostovCG. In this matrix
chain model -- whose target space is the $A_\infty$ Dynkin graph,
one develops a finite diagram technique for the explicit computation
of the $n$-loop amplitudes; see \HiguchiPV\ for an early application
of this technique in the rational case.  Shrinking the loops, one
extracts the $n$-point local correlation functions.  However, this 
procedure is not unique on a fluctuating lattice. Moreover, it happens that
the most natural definition of the local fields leads to a different interpretation
of the matter screening than that in the conventional theory.
Namely, the charge
conservation condition
involves only multiples of the matter
background charge $e_0=1/b-b$,  and the Liouville dressings of the
``order parameter'' fields on the diagonal of the infinite Kac table close under fusion.
This has led us to introduce and study
another variant of the $c<1$ gravity in the continuum.  Instead of the
matter screening charges, $\!${\it i.e.$,\!$} the tachyons of matter
charge $-b$ and $1/b$, the interaction terms for the matter field are
now generated by the two Liouville dressings of the vertex of charge
$e_0=1/b-b$.  The effective action of this ``diagonal'' string theory
is no more a sum of Liouville and matter parts.  Nevertheless it is
possible to extend to this theory the ground ring technique and to
find solutions of the corresponding functional equations.  A class of
these solutions reproduces the 4-point tachyon correlators found in
the discrete approach.

The paper is organized as follows:

After some preliminaries collected in section 2, we determine in
section 3 the generic tachyon 3-point function as a product of the
$c>25$ Liouville \DO, \ZZtp\ and a  $c<1$ matter OPE structure constants.  The latter constant 
is given by an expression derived as in \Ta, which
in particular reproduces the Coulomb gas  OPE constant of \DF; see also \AlZ.

The ground ring is discussed in section 4.  Since its elements are
built of vertex operators, such that  both the matter and Liouville parts
are labelled by degenerate Virasoro representations, one can compute
their OPE with the tachyons using the free field Coulomb gas in the
presence of integer number of screening charges.  We determine the
action of the two generators of the ground ring, $a_-$ and $a_+\,,$  on a
tachyon of arbitrary momentum, taking into account both Liouville and
matter interactions.  This amounts in the computation of the general
3-point function of two tachyons and one ring generator, the details
are collected in Appendix A.1-2.  The result confirms the
expectation, see
\SeibergS\ for the rational case, that the
deformed ring is isomorphic to $sl(2)\times sl(2)$ type fusion ring.
In the ``diagonal matter'' string theory the deformed ground ring
element $a_+a_-$
generates a diagonal $sl(2)$ projection, preserving in particular the
order parameter tachyons.

We use this action  in section 5 to derive recurrence functional
relations for the bulk tachyon 4-point correlators. The equations are written for the correlators satisfying the ``chirality rule'' in the terminology of \GinM\ and they extend the ones
previously obtained for the case 
of gaussian matter  \refs{\kachru,
\bershkut, \KostovCY}.  The contact 3-point terms in these relations require
the computation of a set of  free field  4-point functions containing a ring generator and an integrated
tachyon, see  sections  A.3, A.4 of the Appendix.
  Appendix A.5 contains a computation of some chiral OPE
constants relevant for the boundary ground ring,  in particular the OPE coefficients
of the boundary ground ring generators and a tachyon of generic momentum.

In the remaining four sections we look for solutions of the equations
for the 4-point tachyon correlation functions.  In section 6 we
reproduce the correlators found in \DiK\ for the case of gaussian matter
field.  
The set of fixed chirality solutions are shown to serve as a local basis for 
another,   fully symmetric in the momenta correlator,  also described 
 in \DiK.
 We interpret this 
symmetry as locality requirement  
and discuss the relation between the two types of correlators and the respective equations they satisfy. 
In section 7 we solve the functional equations for a class of
correlators such that the total matter (or total Liouville) charge can be compensated by integer
number of screening charges.  
In section 8 we
find correlators in the diagonal theory in which one or all four
tachyons is degenerate, {\i.e.},  its momentum labels a degenerate
$c<1$ Virasoro representation on the diagonal of the infinite Kac table.
The case of degenerate fields in the conventional, non-diagonal
theory, is considered in section 9.  Here we find 4-point functions
with one matter (or Liouville) degenerate and three generic momenta.
A formula for the 4-point  correlators with four degenerate fields is
conjectured by analogy with the diagonal case. 

The 
results for the 4-point tachyon correlators as functions of the 
momenta $P_1,P_2,P_3,P_4\, $ are summarized by a ``partial wave
expansion'' formula sketched in Fig.1.

\epsfxsize=280pt
\vskip 20pt
\centerline{\epsfbox{ 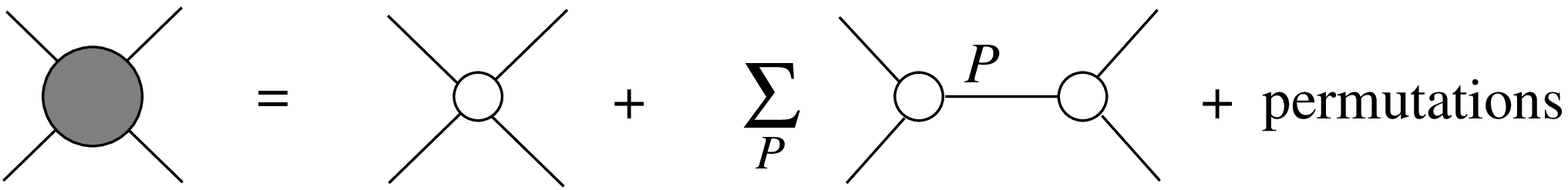   }}
\vskip 5pt

\centerline{\ninepoint Fig.1 : The general structure of the 4-point
function }
\vskip  10pt

\noindent The 4-point function is a sum of 1-particle
irreducible (1pi) piece and a sum of the contributions of the three
channels:
\eqn\summar{ \CG^{(\e)}_{P_1,P_2,P_3,P_4} 
\sim
 \hat \CG^{\rm 1pi}_{P_1,P_2,P_3,P_4} -
\e\, \sum_P\,\(N_{P_1,P_2,P}\, P\, N_{-P,P_3,P_4}^{'}\, + {\rm
permutations}\)\,, 
}
where $\e$ is a sign determined by the chiralities.
The  form of the 1pi term $\hat \CG^{\rm 1pi}$ and the interpretation of the 3-point ``fusion multiplicities'' in \summar\
depend on 
the considered spectrum of
momenta.  The case of
gaussian matter  in \DiK\  corresponds to 
 a single
contribution in each of the three channels in \summar.  
The formula \summar\  is symmetrised with the intermediate momentum $\e\, P $ replaced by $|P|$. This universal choice solves the locality requirement 
preserving the fusion rules, which in the conventional theory   typically match those
determined by  the underlying $c<1$ (or $c>25$) local correlators.  It is further supported
by  a recursive procedure extending the initial identities to equations  for the local correlators.
In the diagonal theory the term $\hat \CG^{\rm 1pi}$  is proportional to the
corresponding 4-point fusion multiplicity $N_{P_1,P_2,P_3,P_4}$, 
also expressed in terms of the 3-point vertices. 
The diagonal theory correlators are not a
special case of those   in the conventional string theory.

 We
conclude in section 10 with a summary of the results and a discussion
on the open problems of this investigation.

\medskip

This paper is a detailed and extended presentation of the results
announced in the short letter \KPlet\ and reported at conferences in
Dubna, Bonn, Varna and Santiago de Compostela.  In the mean time, two
papers appeared, which partially overlap with our results.
Ref. \BM\  deals with the boundary ground ring relations in the minimal $c<1$ gravity.
In ref.  \BZ\  a
different method for evaluating the bulk tachyon correlation functions
in 2D gravity is developed.  The class of 4-point correlators computed
in \BZ, namely those containing one matter degenerate field, are of
the type discussed in our section 9; see the text for a comparison.
We have been also informed by V. Fateev about an unpublished recent
work of him on the direct computation of some particular examples of
such correlators.


 \newsec{Preliminaries: effective action, observables}

 \subsec{\it Effective action}

\noindent
Consider Liouville gravity on 
the Riemann
sphere.  The effective action in the conformal gauge and locally flat
reference metric $\hat{g}_{ab}=\delta_{ab}\ $ is a perturbation of the
gaussian action
   \eqn\actg{\eqalign{ \CA ^{\rm free} & = {1\over 4 \pi}\int d^2 x \[
   (\p_a \phi)^2 + (\p_a \chi)^2+ (Q \phi + i e_0\chi ) \hat R\,\sqrt{\hat g}
    \]  +
{1\over \pi}\int d^2 x \[ {\bf b}\p_{\bar z}{\bf c} +{\bf \bar
b} \p_z {\bf\bar c}\] .
	}}
Here $\phi$  is the  Liouville field,   $\chi$ is the matter field,  
and  $\{{\rm\bf b, c}\}$  is a pair of reparametrization ghosts   
of scaling dimensions $\{2,  -1\}$.
 The reference scalar curvature
 $ \hat R $ is localized at the infinite point.  The Liouville and
 matter fields background charges $Q$ and $e_0$ are parametrized by a
 real constant $b $,
  \eqn\bgcharges{ Q= \frac{1}{b} +b, \qquad e_0= \frac{1}{b} -b \,.  }  
  We will consider the generic situation when $b^2$ is not a rational
  number.  With the choice \bgcharges\ the full central charge is
  \eqn\totch{ c_{\rm tot}\equiv c_M +c_L+ c_{\rm
  ghosts}=\big[13-6(b^2+\frac{1}{b^2})\big]+\big[13+6(b^2+\frac{1}{b^2})\big]
  -26 =0.  }
\medskip

\noindent
We consider a marginal deformation of the gaussian action \actg\ by
the Liouville interaction and its matter counterpart, which is one of
the two screening charges in the $c<1$ CFT,
  \eqn\deltaS{
    \CA  _{\rm int} 
   = \int
   d^2 x \(\mu_{_{L}}
e^{ 2b\phi }+ \muM  e^{-2ib\chi}  \)\,,
    }
  \eqn\deltaSd{ \tilde \CA_{\rm int}   = \int
  d^2 x \(\tilde\mu_{_{L}} e^{2\phi/b}+ \tilde \muM e^{2i \chi/b} \).  }
  In fact the interaction depends on the type of the correlators to be
  computed.  In some cases we shall take into account one or the two
  of the terms in the ``dual'' action, but it will be reduced to a
  source of integer number of screening charges.  We shall refer to
  the above theory, in which the Liouville and matter parts  of the action factorize,
  as the ``conventional'' $c<1$ string theory.  In sect.  4.1 we will
  also introduce another interaction which does not have this
  factorization property.

 \subsec{\it Vertex operators for the closed string tachyons}

\noindent
We shall consider the BRST invariant fields 
that correspond to the vertex operators $e^{2ie\chi}$ of the
matter CFT. In general, the matter vertex operators should be dressed
by Liouville vertex operators, $e^{2\alpha \phi}$,
\eqn\tachyons{
V_{e,\alpha } =
 \g(1-\a^2+e^2)\, 
\ e^{2i e\chi}\, e^{2\alpha  \phi },
}
so that the dressed operator has conformal weight $(1,1)$
\refs{\DavidHJ, \DistlerJT}:
\eqn\mashell{ 
\triangle_M(e)+\triangle_L(\a)=e(e-e_0) +\alpha (Q-\alpha
) = 1.  }
To have simpler expressions we normalized 
by the ``leg''  factors as in \DiK, where 
we used the standard notation $\g(x)=\G(x)/\G(1-x)$.

The condition \mashell\ is the on-mass-shell condition for the closed
string tachyons propagating in the euclidean $2D$ target space.  The
simplest examples satisfying it are the four operators in which only
one of the two vertex operators in \tachyons\ appears, $\!${\it
i.e.$,\!$} the sources of the Liouville and matter screening charges
in \deltaS, \deltaSd.  In general the solutions can be parametrized by
the tachyon target space momentum $P$ and the chirality $\e= \pm
1$\foot{For simplicity we shall assume, unless stated otherwise, that
the momenta are real, $P\in {\Bbb R}$.}.  The matter and Liouville
charges are expressed in terms of $P$ and $\e$ as
\eqn\alphapm{ 
e= \hf(e_0-P)\,, \quad  \a ={\hf}(Q-\e P)
= \e\, e + b^{\e}\,,  \qquad \e = \pm 1.}
We  denote by $V_{\a }^{\e}:=V_{\e \a-\e b^{\e}, \a }$ the vertex
operators \tachyons \ with $\a$ and $e$ related by \alphapm.
To compare with the microscopic theory it is  convenient
 to introduce also the alternative notation $\CV^{(\e)}_P$,
\eqn\newnor{
\CV_P^{(\pm)} \equiv V^\pm _\a=\ 
 \g(\pm b^{\pm 1} \, P)\
  e^{i(e_0-P)\chi  +(Q\mp  P)\phi}\,, 
}
 where    we used the relation
 \eqn\ealfP{
1-\a^2+e^2=b^{\e}(Q-2\a)= \e b^ \e P\, .
}
\noindent
We shall not restrict in general to the
``physical''  Seiberg  bound $\e P=Q-2\a>0$ \SeibergEB.\foot{We recall that in the rational minimal
gravity \LianGK\ the operators corresponding to the solutions of
\mashell\ with $e$ and $e_0-e$ are identified, and one is left with
the two Liouville dressings, distinguished by the sign of $Q/2-\a$.
We shall not assume in general such identification.}

The BRST invariant operators can be represented in two pictures:
either as (1,1)-forms integrated over the world-sheet, or as $Q_{\rm
BRST}$-closed 0-forms:
\eqn\twoforms{\eqalign{\CT_P^{(\pm)} \equiv T_\a^{\pm} =
\int \frac{d^2 x}{\pi}\,   V_\a^{\pm}
\qquad & {\rm or}  \qquad
\CW _P^{(\pm)}\equiv W _\a^{\pm}= {\rm \bf c \bar c}\   V_\a^{\pm}.
}}
In the $n$-point tachyon correlators, $n-3$ vertex operators are
integrated over the worldsheet, and three are placed, as usual, at
arbitrary points, say $0,1 $ and $\infty$, and the ghost zero mode
contribution $\la {\bf c} _{-1} {\bf c}_0 {\bf c}_1\ra$ is normalized
to $1$.  The correlation function should not depend on the choice of
the three operators.

 \subsec{\it Normalization of the couplings and duality
 transformations}

 \noindent
  It is also convenient to redefine the couplings in the
 effective action \deltaS\ according to the normalizations \newnor \
 of the vertex operators,
 \eqn\twopertops{\eqalign{
 \mu_{_L} \int e^{2b\phi} = \L\, T^+_b=  \L\,  \CT^{(+)}_{e_0},
  \ \  \ \  \mu_{_M} \int e^{-2i b\chi} =\LM\,    T_ 0^+=\LM\,
\CT^{(+)}_{Q}
\cr
\tilde\mu_{_L} \int e^{{2\phi \over b}} = \tilde \L\,   T_{1/b}^-
 =\tilde \L\,  \CT^{(-)}_{e_0}, \ \  \ \
 \tilde\mu_{_M} \int e^{{2 i\chi \over b}} = \tilde \LM\, T_0^-
 = \tilde \LM\,
\CT^{(-)}_{-Q} ,
}}
 where the new coupling constants are related to the old ones by
\eqn\renor{\eqalign{
{\L}  \ \ =\ \ \pi   \g(b^2)\, \mu_{_L} ,& \ \ \
\tilde \L\ \ =  \pi \g(\frac{1}{b^{2}})\, \tilde\mu_{_L },\cr
  \LM = \pi \g(-b^2)\, \muM,& \ \ \
\tilde \LM = \pi\g(- \frac{1}{b^{2}})\,  \tilde\muM.
}
}
All correlation functions in Liouville theory are invariant (for fixed
charges) w.r.t. the substitution \FZZb
 \eqn\DUALTY{  \eqalign{b&\to
  1/b,\cr  \L   & \to
\tilde\L= \L^{1/b^2}\,.
}
}
As we will see in the next section, for a consistent description of
the matter correlatation functions one should introduce the dual
matter coupling constant so that the functions in the $c<1$ theory
obey the symmetry
\eqn\dmc{\eqalign{
b &\to -1/b\,, \cr
 \LM & \to \tilde\LM
=
(\LM )^{-{1/b^2}}\,.
}}
The duality transformation \DUALTY\ (or \dmc, or their composition)
relates the tachyon correlators to those of a conjugated theory,
obtained by flipping the sign of $e_0$ (or $Q$, or both) respectively;
thus effectively we can restrict the real parameter $b$ to the region
$(0,1)$.  On the other hand, the composition of \DUALTY\ with $\chi\to
-\chi$ and $\LM \to \tilde \LM$ preserves the free action \actg\ and
interchanges the two interaction actions \deltaS, \deltaSd.  The same
effect yields the matter duality transformation \dmc, accompanied with 
$ \phi \to -\phi\,$ and $\L \to \tilde \L$.  In parameter space these duality
transformations  are 
formulated  as 
\eqn\dualtL{\eqalign{
&  \{\ b,   \L,\LM, \ P_i, \ \e_i\}\ 
 \to\  \{ \frac{1}{b}, \tilde \L, \tilde \LM, -P_i,  -\e_i\}\,, 
}}
\eqn\dualtm{\eqalign{
&  \{\ b, \L,\LM,\  P_i, \ \e_i\}\  \to\  \{-\frac{1}{b}, \ 
\tilde \L, \tilde \LM,  P_i, -\e_i\} .
}}

\newsec{The tachyon 3-point function as a product of Liouville and
matter 3-point functions}

\noindent 
In this simplest case the correlation function factorizes to
a product of the matter and Liouville three-point OPE constants
 \eqn\Gfactor{
 G_3^{\e_1\e_2\e_3}(\a_1,\a_2, \a_3)= \< W_{\a_1}^{\e_1} 
 W_{\a_2}^{\e_2} W_{\a_3}^{\e_3} \>=
 {
 \GL(\alpha _1,\alpha _2,\alpha _3)\GM(e_1,e_2,e_3)
  \over  
  \prod _{j=1}^3 
   \g( \alpha _j^2-e_j^2)}\,.
}
The case of the 3-point
function is unique in the sense that for $n>3$ the factorization holds
only before the integration over the $n-3$ moduli while for $n<3$
there is a residual conformal symmetry which  does not 
allow the direct
evaluation. 

 \subsec{\it The case $\sum e_i = e_0$   \  
 }

\noindent 
First we assume that $\sum e_i = e_0$, in which case the
matter 3-point OPE constant is equal to one.  For the Liouville
3-point OPE constant we take the DOZZ formula \refs{\DO, \ZZtp}
\eqn\Lc{
\GL(\alpha_1\,,\,\alpha_2\,,\,\alpha_3)=\( \L^{1/b} \,
b^{2e_0}\)^{Q-\alpha _1-\alpha _2-\alpha _3}\,
{\zu(b)\, \zu(2\a _1)\,\zu(2\a _2)\,\zu(2\a _3)\,\over
\zu(\a _{123}-Q)\,\zu(\a _{23}^1)\,\zu(\a _{13}^2)\,\zu(\a _{12}^3)}
}
with notation $\a _{12}^3=\a _1+\a _2-\a _3\,, \a _{123}=\a _1+\a
_2+\a _3$, etc.  Here $\a_i$ and $e_i$ are solutions \alphapm\ of the
 mass-shell condition \mashell.  Imposing the constraint
\eqn\constr{
\sum_i e_i=e_0\quad   \Leftrightarrow \quad 
\sum_i \e_i (\a_i -b^{\e_i})=e_0
}
 and using the basic property of the function $\zu=\zu_b=\zu_{1/b}$,
\eqn\recups{
{\zu(x+b^{\ze})/ \zu(x)}=b^{\ze(1-2b^{\ze}x)}\, \zg(x\, b^{\ze})\,,
\qquad \ze=\pm 1}
one checks that the constant \Lc\ reduces to
\eqn\Lcs{\eqalign{
\GL(\alpha_1\,,\,\alpha_2\,,\,\alpha_3)
&=
{
\L ^{  {1\over b} (Q-\a_1-\a_2-\a_3 )}
\over \prod_{i=1}^3   b^{\ze_i}\, \zg[b^{\ze_i}\,(Q-2\a _i)]}\,, \qquad {\rm for}\ 
\ \sum_i \e_i (\a_i -b^{\e_i})=e_0\,.
}
}
 Thus one finds for the 3-point
function \Gfactor, using \ealfP,
\eqn\Tripto{
G^{\e_1\e_2\e_3}_3(\a_1,\a_2,\a_3)={
\L^{  {1\over 
 b}
 (Q-\a_1-\a_2-\a_3 )
}\over 
 b^{\ze_1 +\ze_2+\ze_3}}.
}
\noindent
\rb {\it Comment:}

{\ninepoint \noindent To compare with the expressions in \DiK\ and the
perturbative Coulomb gas computation let us choose, say, the action
\deltaS. The constraint \constr\ combined with
$\sum_{i=1}^3\a_i-Q=-sb$ for a positive integer $s$ implies
$2\a_3=1/b- s b$ or $2\a_3=-(s-1)b$ for the choice of chiralities
$^{(++-)}$ or $^{(--+)}$ respectively.  (For the other two choices
$^{(\pm\pm\pm)}$ the above two conditions are inconsistent.)  The
Coulomb gas computation of the {\it unnormalized} by the leg factors
3-point functions gives for generic values of $\a_1,\a_2$ a finite
expression for the case $^{(++-)}$ and zero for the case
$^{(--+)}$.\foot{As a consequence, the derivatives with respect to
$\mu_L$, i.e, the (unnormalized) $n$-point tachyon correlators
$G_n^{--++ \cdots +}(\a_1,\a_2,\a_3,b, \cdots,b) $ are all vanishing.}
On the other hand in both cases
the {\it normalized} by leg factors correlators are rendered finite by
the overall zero mode divergence factor \DiK\ and this corresponds to
our expression \Tripto.  For the dual interaction \deltaSd\ the roles
of $^{(++-)}$ and $^{(--+)}$ are interchanged.}

 \medskip

\subsec{\it A general  formula for the matter 3-point function}

\noindent 
Now consider the case of arbitrary matter charges
$e_1\,,e_2\,,e_3$, when the matter OPE constant is no more equal to
one.  The general $c<1$ matter 3-point OPE constant satisfies the
identities
\eqna\fr
$$\eqalignno{
{\GM (e_1-b\,,\,e_2\,,\,e_3)\over \GM (e_1\,,\,e_2\,,\,e_3)} &={
1   \over b^4\LM}\, {\gamma(b(2e_1-b))\,\gamma(b\,2e_1)\,
\gamma(b(e_{23}^1+b))\over \gamma(b (e_{123}-e_0))\,
\gamma(b(e_{12}^3))\,\gamma(b(e_{13}^2))}\,\cr 
& {}  & \fr {} \cr 
{\GM (e_1+{1\over b}\,,\,e_2\,,\,e_3)\over \,\GM
(e_1\,,\,e_2\,,\,e_3)} &= {b^4\over \tilde \LM}
\, {\gamma(-{1\over
b}(2e_1+{1\over b}))\,\gamma(-{1\over b}2e_1)\, \gamma(-{1\over
b}(e_{23}^1-{1\over b}))\over \gamma(-{1\over b}
(e_{123}-e_0))\,\gamma(-{1\over b}(e_{12}^3))\, \gamma(-{1\over
b}(e_{13}^2))} .}$$
The change of variables $b\to -1/b\,, \LM \to \tilde\LM $ interchanges
the two relations \fr{}.  These functional relations come from the
locality requirement on the 4-point matter functions with one of the
two simplest degenerate fields $e={b\over 2}\,, e=-{1\over 2b}$
inserted.  Their derivation is analogous to the one for the Liouville
case in \Ta, where the DOZZ formula \Lc\ was reproduced as the unique
solution of the $c_L > 25$ functional relations for positive,
irrational $b^2$.  Identifying the dual coupling constant as in \dmc,
the solution of \fr{}\ is expressed in terms of the Liouville constant
$\GL$ in \Lc, with $\a_i= \e_i e_i+b^{\e_i}, i=1,2,3$, \KPlet{}\foot{The derivation of the matter
constant $C^{\rm Matt}(e_1,e_2,e_3)$ with a different choice of the
normalization, has been carried out independently by Al.
Zamolodchikov, published in  \AlZ. }
\eqn\conj{ \eqalign{
\GM(e_1\,,\,e_2\,,\,e_3)=&
 { \LM^{-{1\over b}\(e_0-\sum_ie_i\)}\over
\prod_{i=1}^{3} b^{\e_i}\, \gamma(b^{\e_i}(Q-2\alpha_i))}\
{ \L^{{1\over b}\(Q-\sum_i\a_i\)} 
\over \GL(\alpha_1\,,\,\alpha_2\,,\,\alpha_3)}
}\,. 
}
The relation holds for any choice of the three signs $\e_i$.  The
overall constant is fixed by
\eqn\mnor{
\GM(e_1\,,\,e_2\,,\,e_3)=1\ \ {\rm  for} \  \ \sum_i e_i=e_0\,,
} 
which is checked using \Lcs. 

 The formula \conj{}\ 
  is  obtained alternatively as
 analytic continuation of the particular (thermal) Dotsenko-Fateev constant
 computed with one of the matter screening charges, in full analogy
 with the derivation of \Lc\ in \refs{\DO, \ZZtp}.  For $\sum_i e_i-
 e_0=mb-{n\over b}\,, $ $ n,m$ non-negative integers, the expression
 \conj\ for the matter structure constant is finite for generic $b^2$
 and reproduces the 3-point Dotsenko-Fateev constant in (B.10) of
 \DF{}, times the powers $(-\muM)^m(-\tilde \muM)^n$.
 In other words,  in the Coulomb gas range of the
 three parameters $e_i$ \conj{}\ can be looked as a compact representation of the DF  constant. 
Introducing the function 
\eqn\mups{
\zum_b(x):={1\over \zu_b(x+b)}={1\over \zu_b(-x+{1\over b})}=\zum_b(e_0-x)
=\zum_{1\over b}(-x)
}
we can rewrite \conj{}\ in a form analogous to the DOZZ formula \Lc\ \foot{This expression and the function \mups\ 
have been earlier considered, see e.g.  \Sch, without discussion of the relation to the 
$c<1$  DF constant. }
\eqn\mDOZZ{
\GM(e_1\,,\, e_2\,,\,e_3)=\( \LM^{1/b} \,
b^{2Q}\)^{e_1+e_2+e_3-e_0}\,
{\zum(0)\, \zum(2e_1)\,\zum(2e_2)\,\zum(2e_3)\,\over
\zum(e_{123}-e_0)\,\zum(e_{23}^1)\,\zum(e_{13}^2)\,\zum(e_{12}^3)}\,.
}
The functional relations  \recups\ are replaced by
\eqn\recupsm{
{\zum(x-b)/ \zum(x)} = \g(b x)\, b^{1-2bx}\,, \ 
\ {\zum(x+\frac{1}{b})/ \zum(x)} = \g(-\frac{1}{ b} x)\,
b^{-1-2x/b}\,.
}
The logarithm of the function $\zum(x)$ admits an integral
representation as the one for the logarithm of $\zu$, with $Q$
replaced by $e_0$ (so that it is invariant under the change $b \to
-1/b$),
\eqn\mdozz{
\log \zum_b(x)=-\int_0^\infty {dt\over t} \( \big({e_0\over
2}-x\big)^2 \, e^{-t} 
 - {\sinh^2({e_0\over 2}-x) {t\over 2}\over
\sinh b {t\over 2}\, \sinh {t\over 2b}} \)=\log \zum_{-\frac{1}{b}}(x)\,,
}
 which converges (for $b>0$) in the strip $ -b < $ Re $ x < {1\over
 b}$.

\medskip

\noindent
\rb  {\it Examples:}
 \eqn\ex{\eqalign{
&\GM (\frac{b}{2} ,e,e_0- e + \frac{b}{2})
=-\muM \,\pi\,{\zg((2e-e_0)b)\over \zg(b^2)\, \zg(2eb)}\,, \cr
&\GM (-\frac{1}{ 2b},e,e_0-e-\frac{1}{ 2b})
=-\tilde\muM\,\pi\, {\zg((e_0-2e){1\over b})\over \zg({1\over
b^2})\, \zg(-2e{1\over b})} \,.
}}
\medskip

\rb The Liouville three-point OPE constant satisfies the reflection
 property \ZZtp
 \eqn\reflL{ \GL(\alpha_1\,,\,\alpha_2\,,\,\alpha_3) = {1\over
 b^2}\, \L ^{Q-2\alpha_1\over b}\, {\gamma(\frac{1}{b}(2\alpha_1-Q))
 \over \gamma(b(Q-2\alpha_1)) }\,
 \GL(Q-\alpha_1\,,\,\alpha_2\,,\,\alpha_3) \,.
 }
 This identity has been used to write the r.h.s. of \conj{}\ in
 various equivalent ways corresponding to the different choices
 \alphapm\ of the relation between $e_i$ and $\a_i$.  Analogously,
  \eqn\reflm{ \GM (e_1\,,\,e_2\,,\,e_3)=b^2\,
 \LM ^{2e_1-e_0\over b}\,{ \gamma(b (2e_1-e_0)) \over \gamma( {1\over
 b} (2e_1-e_0)) }\ \GM (e_0-e_1\,,\,e_2\,,\,e_3).  }
 In particular 
 for $e_1=e_2=e, e_3=0$ \reflm\ implies
  \eqn\tp{ \GM
 (e\,,e\,,0)=b^2\, \LM ^{2e-e_0\over b}\,{ \gamma(b (2e-e_0)) \over
 \gamma( {1\over b} (2e-e_0)) }\,.  }
 The last formula reproduces 
  the 2-point constant
 found in \ST. As  pointed out in \ST, the choice $e_3=0$ does
 not force $e_1=e_2$, or $e_1=e_0-e_2$; see also \Sch\ for a different
 solution in the case $c=1$ and a different interpretation of the
 identity operator which avoids this problem, not essential for our
 purposes.\foot{ We recall that the DF constants \DF, \DFt\  
 for rational $b^2$ 
 are nonvanishing 
 for some  values in the minimal spectrum, but beyond the 
restrictions of the fusion rules.
In that sense these constants
by themselves do not determine the fusion. 
However,  due to certain identities satisfied by the constants  and the fusing matrix elements,
 there occur cancellations in the block expansion of the local 4-point functions, so that each channel is consistent  with the fusion rules; 
see e.g. \FGP. }
 The construction of a consistent non-rational matter theory for arbitrary momenta is beyond the
 scope of this paper.

\subsec{\it The   tachyon 3-point function for generic momenta}
  \noindent

\noindent 
Having evaluated the matter three-point function \conj, we
insert it together with \Lc\ in \Gfactor\ and obtain a simple
expression for the tachyon 3-point function. If we factorize the dependence on the coupling 
constants $\L,\LM $ and the powers of $b$, denoting  the remaining momentum-dependent
factor by $N_{P_1,P_2,P_3} $,  we get
\eqn\ttp{\eqalign{
\CG ^{\ze_1\ze_2\ze_3 }(P_1\,,\,P_2\,,\,P_3)\, &= {\L
^{{1\over 2b} ( \sum_i \ze_iP_i -Q)}\, \LM ^{{1\over 2b} (e_0-
\sum_iP_i )} \over 
\, b^{\ze_1+\ze_2+\ze_3} } \,
N_{P_1,P_2,P_3}\cr
&=b^{-\sum_i\ze_i} \, 
  \L ^{{1\over b}(Q-\sum_i \a_i)}\,
  \LM ^{-{1\over b}(e_0-\sum_i e_i)}
\,, \cr
}}
or $N_{P_1,P_2,P_3} =1$. This    
    reduces to \Tripto\ for $\LM=1$.  
   The simple 3-point function \ttp{}\
   satisfies reflection properties inherited from those of its
   Liouville \reflL\ and matter \reflm\ parts.  That is, in the
   expectation value \Gfactor\ given by \ttp\ the following identities hold:
\eqn\drefl{
 \CW_P^{(+ )}=  b^{-2}\, \L^{P/b}\ \CW_P^{(-)}=b^{-2}\,  \LM^{-P/b}\
\CW_{-P}^{(-)}=\(\L/\LM\)^{P/b}\ \CW_{-P}^{(+)} .
}
 Notice however that the 
 relations \drefl\ do not
necessarily hold within the 4-point functions, as will be  discussed below.

\medskip

\rb Given the expression \ttp{}\ for the 3-point constant, the 2-point
tachyon correlators are conventionally defined \DO\ as integrals of
3-point ones over some of the interaction constants.  For example, the
2-point tachyon function for $e_1+e_2=e_0$ is determined from
$G_3^{++-}(b,\a,\a) =-\partial_{\L} G_2^{+-}(\a,\a)\,,$
$\a=e_1+b=Q/2-P/2$, { i.e.}
\eqn\twopP{\eqalign{
 G_2^{+-}(\a,\a)&=\CG_2^{+-}(P, -P)
  =- {\L^{P/ b}\over  
 P}
 =-{\L^{{1\over b}(Q-2\a)}\over   (Q- 2\a)}\cr
 &
=(Q-2\a)\,  b^2 \L^{\frac{1}{b}(Q-2\a)}\, \langle 0| \gc(z)\p_z \bar \gc(\bar z)\p_{\bar z} W_{\a}^+(z,\bar z)  W_{Q-\a}^+(z', \bar z')|0\rangle\,.
}}
 The composition of matter times Liouville reflections reproduces up to a sign the 
 correlator $G_2^{+-}(e,e)$ determined analogously from $G_3^{++-}(0, \a, Q-\a)=-\partial_{\LM} G_2^{+-}(\a,Q-\a)$.

The same convention leads to the partition function $Z$ and its dual
$\tilde Z$ defined as
 \eqn\partfa{\eqalign{
-\partial_{\L}^3 Z(\L,\LM, b) 
=<W^+_{b}W^+_bW^+_b>\  &\Rightarrow \  Z(\L,\LM, b) =
- {b\over 
 Q e_0}\, \L^{Q\over b}\LM^{-e_0\over b}\,, \cr
 -\partial_{\tilde \L}^3 \tilde Z(\L,\LM, b) 
=<W^-_{\frac{1}{b}}W^-_{\frac{1}{b}}W^-_{\frac{1}{b}}>\
& \Rightarrow \tilde   Z(\L,\LM, b) 
=  {1\over  b\,Q e_0} \tilde \L^{b\,Q}\tilde \LM^{b e_0 }= {1\over  b\,Q e_0}\L^{Q\over b}\LM^{-{e_0\over b}}\,, \cr
Z( \tilde \L,  \tilde \LM, \frac{1}{b})= \tilde  Z(\L,\LM, b) &= -\frac{1}{b^2}\, Z(\L,\LM, b).
}}
 \noindent We can then introduce   normalized  functions 
 \eqn\normt{\eqalign{
{G_3^{\e_1\e_2\e_3}(\a_1,\a_2,\a_3;  \L, \LM, b)\over Z(\L,\LM,b)}&=- b^{-\sum_i \e_i -1}\ e_0\,Q\,
  \L ^{-\frac{\sum_i \a_i}{ b}}\,
  \LM ^{\frac{\sum_i e_i}{ b}}\,, \cr
{ G_3^{\e_1\e_2\e_3}(\a_1,\a_2,\a_3;  \L, \LM, b)\over \tilde Z( \L, \LM, b)}&= b^{-\sum_i \e_i +1}\,e_0\,Q\, 
 \L ^{-\frac{\sum_i \a_i}{ b}}\,
 \LM ^{\frac{\sum_i e_i} {b}}.
}}
These correlators are interchanged by the duality
transformations  \dualtL, \dualtm, 
which  become 
 equivalent  since \normt\  is  expressed only through the 
 variables \foot{The first duality relation is satisfied  by the 
 constants \ttp{}, 
but the second holds up to a  sign, {\it i.e.} 
$\CG^{\{\e_i\}}(\{P_i\}; b,\L,\LM)=\CG^{\{-\e_i\}}(\{-P_i\};\frac{1}{b}, \tilde \L, \tilde \LM)=-\CG^{\{-\e_i\}}(\{P_i\};-\frac{1}{b}, \tilde \L, \tilde \LM)$.  Nevertheless to simplify notation we shall  work with the unnormalized 
correlators or with other normalizations. Another possible though less intuitive definition 
is to relate the partition function to the correlator $\langle W_b^-W_b^+W_b^+\rangle$ ; according to \drefl\ this removes a factor $b^{-2}\LM^{-e_0/b}$ from $Z$.
}
   \eqn\duality{ \{b^2, \L,\LM, b P_i, \e_i \} \to \{\frac{1}{b^2}, \tilde  \L,\tilde \LM, - \frac{P_i}{b}, -\e_i \} \,.
}
Furthermore each of the simple constants \normt\  is invariant
under analytic continuation $b\to \pm i b$, transforming the charges and the coupling constants as
\eqn\dualmL{\{\ b, \L, \LM, e_i, \a_i\}
\ \to\ \{  \pm i b,    \LM,  \L,   \mp i \a_i,
\mp i  e_i\}
}
or, in terms of target space momenta, $ P_i \to \mp i \e_iP_i, \ \e_i
\to \e_i$.  This   matter-Liouville duality transformation reflects the
invariance of the actions \actg, \deltaS\ and \deltaSd\ under the
respective interchange of the matter and Liouville fields $ \{\chi,
\phi\}\to \{\pm \phi, \mp \chi\}$.

\medskip

\rb The final simple expression for the 3-point functions \ttp{}\
satisfies the identities
\eqn\recr{\eqalign{
&\LM \, G_3^{-++}(\a_1, \a_2-\hfb , \a_3)=
G_3^{-++}(\a_1-\hfb , \a_2, \a_3)
=\L \, G_3^{-++}(\a_1,  \a_2+ \hfb , \a_3)\,,
}
}
and hence,
\eqn\recrr{
 G_3^{-++}(\a_1, \a_2-b , \a_3)={\L\over \LM}\,
G_3^{-++}(\a_1, \a_2 , \a_3)\,. 
 }
  The last identity is also a direct consequence of the matter functional
  relations \fr{}\ and the corresponding Liouville ones \Ta, and can
  be used itself as a relation determining the 3-point tachyon
  correlator.

However the  r.h.s. of the matter \fr{}\  or the  corresponding Liouville functional
  relations may become singular,  so both 
 the functional relation \recrr\ and the simple solution \ttp{}
are valid for generic momenta.  For momenta such that some of
the factors in  \Gfactor\ becomes singular, there is a
$0\times \infty $ indeterminacy. 
This ambiguity leads us to
reconsider the problem of determining the tachyon 3-point function and not  rely on factorization. Then the 
arbitrary  ``multiplicity" factors $N_{P_1,P_2,P_3}$ in the first line in \ttp{}\  must satisfy a
pair of difference equations, to be derived below as part of the set
of functional identities for the $n$-point tachyon correlators.  These
equations are weaker than \recr, \recrr:
\eqn\homrel{\eqalign{
N_{P_1+ b^{\e},P_2,P_3}+N_{P_1- b^{\e},P_2,P_3}&=
N_{P_1,P_2+ b^{\e} ,P_3}+N_{P_1,P_2- b^{\e} ,P_3}\qquad \ \e=\pm 1\, .
}}
 The expression \ttp{}, $\!${\it i.e. $\!$} $ N_{P_1,P_2,P_3} =1$,  is
 only the simplest of their solutions.
We shall deal with basically two deviations from this generic
solution.  One is the case when the factor $N_{P_1,P_2,P_3}$ has the
meaning of a fusion multiplicity and can take values 1 or $0$.  In the
second, this factor will be rather a distribution.
Thus in the simplest example of a gaussian matter
$N_{P_1,P_2,P_3}=\delta(P_1+ P_2+P_3- e_0)$ replaces the normalization condition 
\mnor.
The duality relations for the  3-point correlators are preserved if $N_{P_1,P_2,P_3}(b)=N_{P_1,P_2,P_3}(-\frac{1}{b})=N_{-P_1,-P_2,-P_3}(\frac{1}{b})$.

 \newsec{The  ground ring }

 \def\gc{{\bf c} }

\noindent 
The ground ring operators are 
BRST  invariant
fields obtained by applying raising operators of level $rs-1$ to the
product of two degenerate matter and Liouville fields with Kac labels
$r,s$.  The resulting operators have conformal weights $(0,0)$, see
\IMM\ for an explicit construction of some of the corresponding
states.  The ring is generated by the lowest two operators 
$a_{\pm}=a_{\pm}(z)\,\bar a_{\pm}(\bar z)\,$ \Witten, 
\eqn\bulkGR{ \eqalign{ 
a_-(z)= &: \Big(\gb(z)\gc(z) - \frac{1}{b}
\p_z\big[\phi(z)+i \chi(z)\big]\Big) e^{ -b(\phi(z) -i\chi(z))}:\cr 
a_+(z)= &: \Big( \gb(z)\gc(z) -b\, 
\p_z\big[\phi(z)-i \chi(z)\big]\Big)  e^{- {1\over b}(\phi(z)+i\chi(z))}:\,   
}}
The derivatives
$\p_z a_\pm$ and $\p_{\bar z} a_\pm$ are $Q_{\rm BRST}$-exact, and
therefore any amplitude that involves $ a_\pm$ and other BRST
invariant operators does not depend on the position of $ a_\pm$.  
This property 
allows to write recurrence equations for the correlation
functions from the OPE of $a_\pm$ and the tachyons $W_{\a}^{\e}$
\refs{\KMS, \kachru, \bershkut, \KostovCY}, which will be generalized below.

\subsec{ The action of the  ring generators on the tachyons }

\noindent
The recurrence  equations were initially derived for the free fields with no
interaction, or at most accounting for the perturbative first order
contribution of the Liouville interaction.  The momenta were therefore
assumed to satisfy the charge conservation, or ``neutrality''
condition
 \eqn\neut{
\hf \sum_i (e_0-P_i) \equiv \sum_i e_i =e_0.
}

More generally, treating the Liouville and matter screening charges in
\deltaS\ and \deltaSd\ as perturbations amounts to modifying the
original ring generators as
\eqn\newapm{\eqalign{ a_-
&\to a_- \(1 - \L\, T^+_b+...\) ={\hat a}_-\,,\cr 
a_+ &\to a_+\(1 - \tilde \L\,
T^-_{{1/b}}+...\)=\hat{a}_+\, .
}}
Summarizing, 
 for generic momenta, $\!${\it i.e.$,\!$} taking any complex 
values  excluding the lattice 
\eqn\Latt{
\CL  :=\IZ b +\IZ \frac{1}{b},
}
one finds that 
  the action 
  of the ring generators on the tachyons $W_{\a}^{\e}= {\rm
  \bf c \bar c}\ V_\a^{\e}$ of given chirality contains two terms, up
  to $Q_{\rm BRST}$ commutators:
     \eqn\mpd{\eqalign{
\ha_-
   \,  W^{+}_{\a}
&=   - \L\,
W^{+}_{\a +  {b\over 2} } - \LM  W^{+}_{\a -  {b\over 2} }\cr
\ha_-
\, W_{\a}^{-} & = - \L\LM\,  W_{\a +  {b\over 2 } }^{-}  -   W_{\a-  {b\over 2 } }^{-} \,
}}
  \eqn\pmd{\eqalign{
\ha_+
\,
  W^{-}_{\a}& =    - \tilde \L\
W^{-}_{\a+  {1\over 2 b} } -  \tilde\LM
 W^{-}_{\a-  {1\over 2 b} } \cr
\ha_+
\, W_{\a}^{+}& =   - \tilde\L\tilde\LM\,  W_{\a + {1\over 2 b} } ^{+}   -   W_{\a -  {1\over 2 b} } ^{+}
\,,
}}
or, in the alternative notation,
      \eqn\mpdp{\eqalign{
\ha_-
   \,  \CW^{(+)}_{P}
&=   - \L\,
\CW^{(+)}_{P -  b } - \LM  W^{(+)}_{P+ b }\cr
\ha_-
\, \CW_{P}^{(-)} & =-
   \CW_{P-  b}^{(-)}
   - \L\LM\,  \CW_{P + b }^{(-)}   \,
}}
  \eqn\pmdp{\eqalign{
\ha_+
\,
  \CW^{(-)}_{P}& =    - \tilde \L\
\CW^{(-)}_{P+  {1\over  b} } -  \tilde\LM
 \CW^{(-)}_{P -  {1\over  b} } \cr
\ha_+
\, \CW_{P}^{(+)}& = -
   \CW_{P +  {1\over  b} } ^{(+)}
   - \tilde\L\tilde\LM\,  \CW_{P - {1\over  b} } ^{(+)}  \,.
}}
  The  duality transformations  interchange the two ring generators and 
  the two pairs of module relations.

 The OPE coefficients in \mpd, \pmd\ are found either by direct
 evaluation of the 3-point function of the ring generator and two
 tachyons in the presence of a number of screening charges, or by
 exploiting the factorization to the related known $c<1$ and $c>25$ Coulomb gas
 3-point constants, see Appendix A for more
 details.  
   The  coefficients,
 say in \mpd, $C^{(\e, \e)}_{-\frac{b}{2}\, \a}{}^{\a'}$, are expressed as products of the corresponding
 matter and Liouville constants 
\eqn\fcons{\eqalign{
C^{(\e, \e)}_{-\frac{b}{2}\, \a}{}^{\a - \eta \frac{b}{2}}&=-\LM^{1+\e \eta \over 2}\L^{1-\eta\over 2}\,
= \ \ 
 {\g(b^{\e}(Q -2\a))\over
\g(b^{\e}(Q-2\a+ \eta b))}\,  {(Q-2\a)^2\over b^2} \cr
 \times  &C^{\rm Matt}(\frac{b}{2}, e ,
e_0 -e +\e \eta \, \frac{b}{2})\ \hat \GL(-\frac{b}{2}, \a ,
Q-\a+\eta\,\frac{b}{2})\,, \qquad 
\eta=\pm 1\,.
}}
Similar formula holds for \pmd.  
 The matter OPE constants in \fcons\ are either $1$, as in \mnor, or
given by the first example in \ex{}.  The constants $\hat{C}^{\rm
Liou}(\a_1,\a_2, \a_3)$ with $\sum_i\a_i-Q=(\eta-1)\,\frac{b}{2}$ are  the analogous $c>25$ 
Coulomb gas
expressions, which are alternatively obtained as residue  of the  
 Liouville   constant $\GL$ in \Lc.

\medskip

\rb {\it The case of momenta in the lattice $\CL$  }
\medskip

\noindent 
For some momenta $P\in \CL$ on the lattice   \Latt, the free field 3-point
function determining the OPE coefficients is nonvanishing  for more than two values, leading to 
additional terms in \mpd\ and \pmd. This  typically requires an integer power of one of the
screening charges in the dual interaction action, while the
 generic OPE \mpd\ and \pmd\ correspond to deformations
with the respective actions \deltaS, \deltaSd; 
see Appendix A.2 for details.  The
additional OPE terms correspond to reflections with respect to the
matter or Liouville, or both, charges of the terms in \mpd\ and \pmd.
In the first two cases the chirality is inverted.

  Let us restrict the consideration to momenta labelled by degenerate
  matter representations
\eqn\degen{\eqalign{
& \CL_{M}^{\pm} :=
 \Big\{P\equiv e_0-2e=\pm ({n\over b}-mb)\Big\}_{ m,n\in \IN}\, \subset \CL.
 }}
For one of the signs in \degen, $P=P_{m,n}={n/b}-mb$, the r.h.s. of
\mpd\ and \pmd\ contains two more terms, \foot{This happens as well for
the border lines $m=0$, $ n=0$ outside of \degen.} while for the other
sign, $P=mb-n/b$, the generic formulae \mpd, \pmd \ hold.  For
example,
\eqn\fmpd{\eqalign{
&W^+_{\a\pm {b\over 2}}\ \   \rightarrow\ \  
 W^+_{\a\pm {b\over 2}}
+
{\LM^{{1\over b}(2e\pm b-e_0)}\over b^2}\, \ \ 
W^-_{\a\pm {b\over
2}}
  \qquad\   2\a-Q= \frac{ (1\pm 1) b}{2} +m b - \frac{n}{b}\,, \cr
}}
\eqn\fmpdd{\eqalign{
&W^-_{\a\pm {b\over 2}} \rightarrow\  W^-_{\a\pm {b\over 2}}
+
b^2\, \LM^{{1\over b}(2e\mp b-e_0)}\,W^+_{\a\pm {b\over
2}} 
\,,   \qquad  2\a-Q=\frac{(1\mp 1)b}{2} - \frac{m b}{2}+ \frac{n}{2b}.
}}
 In \fmpd\ and \fmpdd\ appear the combinations invariant under matter
 reflection,
\eqn\rinv{\eqalign{
{\tilde W_{\a}}^+: &=b \LM^{{1\over 2b}(e_0-2e)}\, W^+_{\a}
+ \frac{1}{b}\,
\LM^{{1\over 2b}(2e-e_0)}\,W^-_{\a}\cr
&=b \LM^{{1\over 2b}(e_0-2e)} \, \g[b(Q-2\a)]\Big(e^{2i e\chi}+
\GM(e,e,0)\,
e^{2i (e_0-e)\chi}\Big) e^{2\a\phi}
}}
or, in terms of momenta, 
$$
{\tilde\CW}_{P_{m,n}}^{(\pm)}
=b^{\pm1}\, \LM^{{1\over 2b}P_{m,n}}\, \CW^{(\pm)}_{P_{m,n}}
+ \frac{1}{b^{\pm 1}}\,
\LM^{-{1\over 2b}P_{m,n}}\,\CW^{(\mp)}_{-P_{m,n}}\,, \qquad P_{m,n}=n/b-mb\,.
$$
The relative constant for the unnormalized vertex operators in this
linear combination is given by the 2-point function \tp.

A similar argument can be carried out for the momenta labelled by
degenerate Liouville representations
\eqn\degenL{
  \CL_{L}^{\pm} :=
\Big\{\e P\equiv Q-2\a=\pm (mb+{n \over b})\Big\}_{ m,n\in \IN}\, \subset \CL \,.
 }
In this case Liouville reflected terms appear in the OPEs which
correspond to the plus sign in \degenL.

In fact the appearance of the Liouville or matter 
 reflected  points is universal  and is a consequence of the properties \reflL\ and \reflm\ of the 3-point functions \Lc\ and \mDOZZ; the only peculiarity of the degenerate cases discussed here is that both  OPE coefficients are given 
 by  a Coulomb gas 3-point correlator, while in general the reflection images correspond to functions satisfying
 a relation obtained by a reflection from the Coulomb gas charge conservation condition.
Taking this into account in particular removes the above 
 asymmetry of the OPEs for the tachyons of momenta $P_{m,n}$ and
 $-P_{m,n}$, related by a matter charge reflection.
To make sense  of these relations one should  be able to identify in the correlators
the matter reflected 
tachyons  in \rinv\ (or the Liouville reflected ones in the case of degenerate Liouville case \degenL).
At this stage we shall merely assume that the action of the ring
generators is given again by the generic formulae \mpd\ and \pmd. 
This is analogous to what is done 
 in field theory, 
where only one of the two
charges of the same dimension 
 is included 
 in the block decomposition of the 4-point functions. 

\subsec{The ground  ring  at non-rational $b^2$}

 Assuming that the tachyons at the border lines of the degenerate set
\degen\ vanish (at least in the averages), one gets a semi-infinite
set, in one to one correspondence with the irreps of $sl(2)\times
sl(2)$.  The modules of given chirality are generated from the
corresponding tachyon of momentum $P=e_0$ serving as an identity.  After
absorbing the constant $\L$ in the normalization of the vertex
operators, the relations \mpd\ and \pmd\ are equivalent to the multiplication rule of
the characters of $sl(2)$ irreps
of dimensions respectively
$m$ and $n$,  with the character of the fundamental representation of dimension $2$.
It  allows to represent any  character as a polynomial of the
fundamental one -   the above rule is the  functional identity defining the Chebyshev polynomials $U_{m-1}$ of second kind.  Analogously
  \mpd\ and \pmd\ imply (setting $\LM=1$)
     \eqn\Cheb{
\CW_{P_{mn}} ^{(\e)}
=  \L^{{1\over 2b}  \e (P_{mn} -e_0)} \
U_{m-1}(\frac{1}{2}\CO_{21}))\, U_{n-1}(\hf
\CO_{12}  )\
\CW_{e_0}^{(\e)},
}
with
$ \CO_{21}= - \L^{-{1\over 2}} \hat a_-,
\
  \CO_{12}= - \tilde \L^{-{1\over 2}}\hat a_+
$.
  The polynomial acting on the tachyon $ \CW_{e_0}^{(\e)}$ represents
  the ground ring element $\CO_{mn}$.  The formula \Cheb\ derived from
  \mpd\ and \pmd\ confirms the Ansatz in \SeibergS\ used in the
  context of the minimal string theory, see also \BZ. The equalities
  \mpd, \pmd, and hence \Cheb, all hold true up to $Q$-exact terms,
  which in general disappear only in the 3-point tachyon functions.

\subsec{Further OPE channels}

\noindent
The two-term relations
\mpd\ and \pmd\ describe the OPEs of the ring
generators $a_{\pm}$  perturbed by the screening charges in the action.
 In presence of one or more tachyons, given by an integrated
vertex operators, as happens in any n-point function with $n>3$, the OPE 
will contain  more terms. Indeed now any  integrated tachyon  
serves as a ``screening charge''.  
Instead 
 of computing explicitly in an operator form the Q-exact  terms, appearing in the product of free exponential fields,  and then moving  them to the
right or left, one can 
compute all possible OPE relations which send a  
 tachyon $W_\a^{\e}$ to a 
 tachyon $W_{\a'}^{\e'}$
 \eqn\tatota{
  a_{\pm}  W ^{\e}_{\a}  T^{\e_1}_{\a_1}\dots  T^{\e_t}_{\a_t} \ \to \ W ^{\e'}_{\a'} \,.
 }
 Thus the  effect of the  skipped Q - exact terms (if any) is already accounted for  in  the added new OPE channels, which
in turn are valid again up to such terms.

The mass-shell condition implies a relation on the possible combinations of chiralities and momenta
in \tatota,  see Appendix A.3
 for a summary of the consequences of these constraints.  The coefficients of these OPEs are computed from  free field correlators, see Appendix A.4, 
  here we summarize these results.

 The simplest
example consistent with the  mass-shell condition is given by the OPE
relations \refs{\bershkut, \kachru}
\eqn\aminus{
 a_-  W ^{+}_{\a}  T^{+}_{\a_1}
 =  W  ^{+}_{\a+\a_1 -{b\over 2}} \,, \qquad 
    a_+  W ^{-}_{\a}  T^{-}_{\a_1}
=   W^{-}_{\a+\a_1-{1\over 2b}}\,.
}
The relations \aminus\  have been already used for particular values of $\a_1$ and the chiralities
in the derivation
of the linear terms in the first lines of \mpd, \pmd.
They
are  generalized for generic values of $\a$ to  a whole series, with $k=0,1,2,\dots, $ 
\foot{The first nontrivial example  $k=1$ of these  OPE coefficients has
been computed by P. Furlan.  }
\eqn\aminusda{
 a_-  W ^{+}_{\a}  T^{+}_{\a_1}\, {(T^{-}_{0}T^{-}_{1/b})^k\over k!^2}
 = W  ^{+}_{\a+\a_1  -{b\over 2}+ {k \over  b}} \,,\ \ \ 
 }

\eqn\aplusda{
    a_+  W ^{-}_{\a}  T^{-}_{\a_1}\,{(T^{+}_{0}T^{+}_{b})^k\over (k!)^2}
	=  W^{-}_{\a+\a_1-{1\over 2b} +k b}\,.  \ \ \ 
}
Taken for different $k$ the relations demonstrate the effect of the 
 $Q$-exact terms, 
\eqn\qterms{\eqalign{
a_+  W ^{-}_{\a}  T^{-}_{\a_1}\,& (T^{+}_{0}T^{+}_{b})^k= (0+...) T^{-}_{\a_1}\,(T^{+}_{0}T^{+}_{b})^k=(W  ^{-}_{\a+\a_1  -{1\over 2b}}+...)\,(T^{+}_{0}T^{+}_{b})^k
\cr
&=(p!^2 {k \choose p}^2 W  ^{-}_{\a+\a_1  -{1\over 2b}+ p  b}+...)\,(T^{+}_{0}T^{+}_{b})^{k-p}= k!^2\, W  ^{-}_{\a+\a_1  -{1\over 2b}+ k b} + ...
}}
 For the product of  interacting
 fields we obtain combining with \mpd, \pmd, 
 \eqn\opeonet{\eqalign{
 \hat a_-\, W^+_{\a}T^+_{\a_1}&=  \sum_{\a'}\,  C^{(++)}_{-\frac{b}{2} \,\a}{}^{\a'}\ W^+_{\a'}\, T^+_{\a_1}+
 \sum_{\a'}\, C^{(+++)}_{-\frac{b}{2} \,\a\, \a_1}{}^{\a'}\ W^+_{\a'}\cr & =
 (  - \L\,
W^{+}_{\a +  {b\over 2} } - \LM  W^{+}_{\a -  {b\over 2} })\, T^+_{\a_1}+  \sum_{k=0}
  (\tilde \L\, \tilde \LM)^k\,
   \, \CW^+_{P+P_1-\frac{2 k+1}{b}}\cr
   \hat a_+ \, W_{\a}^-T_{\a_1}^- &= \sum_{\a'}\, C^{(--)}_{-\frac{1}{2b} \,\a}{}^{\a'}\ W^-_{\a'}\,  T^-_{\a_1}+
  \sum_{\a'}\, C^{(---)}_{-\frac{1}{2b} \,\a\, \a_1}{}^{\a'}\ W^-_{\a'}\cr & =(- \tilde \L\
\CW^{(-)}_{P+  {1\over  b} } -  \tilde\LM
 \CW^{(-)}_{P -  {1\over  b} } ) \, T^+_{\a_1}+
 \sum_{k=0} 
   (\L\,  \LM)^k
     \, \CW^-_{P+P_1+(2 k +1)b}\,.
}}
The 4-point OPE coefficients in \opeonet\ are expressed in terms of products of matter and Liouville Coulomb gas 3-point constants, see formula (A.30) below. 
The relations  
hold for values for which each of those constants is well defined. In particular \aminusda\ 
extends to degenerate values of  $\a$ (with shifted compared with \degen, \degenL\ notation),
\eqn\degm{
P=\frac{n+1}{b}\pm (m+1) b\,, \ m,n\in \IZ_{\ge 0}\,,
}
for any $k \le n$. Similarly \aplusda\  extends  for  $k\le m$ if  
\eqn\degmd{
P=\mp \frac{n+1}{b}-  (m+1) b\,, \ m,n\in \IZ_{\ge 0}\,.
}
  For these values
 the  infinite sums  in \opeonet\ 
 truncate to the first $n+1$ (respectively $m+1$) terms, 
  see Appendix A.4. We can interpret the set $\{\CW  ^{+}_{P_1-\frac{1}{b}+P- {2k \over  b}}\,, k=0,1,2,  \dots \}$ as the states of a $sl(2)$ Verma module of h.w.
 $b P=n+1 $.  The state $k=n+1$ of weight $ -bP$ 
 corresponds to  the singular vector 
and it is  set to zero for the $n+1$- dimensional irrep.
\medskip
\rb 
 The kinematical mass-shell constraints
 imply that for generic momenta $P_\a, P_{\a_1}\not \in \CL$ and $P_\a
 +P_{\a_1}\not \in \CL$ the identities \opeonet\ exhaust all OPEs in presence
 of one integrated tachyon. 
 Furthermore  under the same type of restrictions
 there is no  contribution of two or more such integrated  tachyons to the OPE of the interacting fields. Therefore for a product of 
 $p$ tachyons $T_{\a_i}^\e$ there are $p$ terms in the OPE of the type in \opeonet.

 However for values in $\CL$
there are more solutions already for the case of one integrated tachyon.
In particular its  momentum   $P_1$  can be given  any $c<1$ or $c>25$ degenerate value.
The OPE coefficients in this case are  computed 
 for generic values of $\a$ and  any integers  $n, m  \ge 0$:
 \eqn\opeonetde{\eqalign{
 \hat a_-\, W^-_{\a}\,T^+_{\a_1}- &\sum_{\a'}  C^{(--)}_{-\frac{b}{2} \,\a\, \a_1}{}^{\a'}\ W^-_{\a'}\, T^+_{\a_1}=
 \sum_{\a'} C^{(-+-)}_{-\frac{b}{2} \,\a\, \a_1}{}^{\a'}\ W^-_{\a'}\cr & =
  \sum_{s=0}^n
 \tilde \LM^{n-s }\,\tilde \L^{s}\, 
  W^-_{\a+  \frac{(m+1) b}{2}+\frac{2s- n}{2b} } 
   \cases{  \LM^{m+1}\, 
    & if $  P_1= {n+1\over b}- (m+1) b  $\cr
    \L^{m+1}\,      & if $  P_1= {n+1\over b} + (m+1) b  $}
 \cr
 \hat a_+\, W^+_{\a}\, T^-_{\a_1}-  &\sum_{\a'}  C^{(++)}_{-\frac{1}{2b} \,\a}{}^{\a'}\ W^+_{\a'}\, T^-_{\a_1}=
 \sum_{\a'} C^{(+-+)}_{-\frac{1}{2b} \,\a\, \a_1}{}^{\a'}\ W^+_{\a'}\cr & =
  \sum_{k=0}^m
 \LM^{k }\, \L^{m-k}\, 
 W^-_{\a+  \frac{(m-2k) b}{2}+\frac{ n+1}{2b} } 
    \cases{ \tilde  \LM^{n+1}\,   
  & if $  P_1= {n+1\over b}- (m+1) b  $\cr
  \tilde   \L^{n+1}\,    & if $  P_1= -{n+1\over b} - (m+1) b  $}
 }}

 More generally, one can have a product of any number of arbitrary tachyons
 with partial sums of momenta in  $\CL$, depending on the chiralities.  The simplest computable examples with two integrated tachyons are given by 
\eqn\di{\eqalign{
&a_{-}     \, W_{\a}^{-}
 \,T_{\a_1}^{+}\, T _{\a_2}^{+}
=-
 W_{\a+{b\over 2}}^{-}\,, \qquad \  \a_1 +\a_2=b\,,\cr
&a_{+}  \, W_{\a}^{+}
 \,T_{\a_1}^{-}\,
   T_{\a_2}^{-}
=-
W_{\a+{1\over 2b}}^{+}\,, \qquad  \a_1+\a_2= \frac{1} {b}\,,
}}
or $P_1+P_2=2/b, 2 b$ respectively.  These identities  were used for $\a_1=0$
in the derivation of the last terms in
\mpd\ and \pmd; for this value  they reduce to \opeonetde. There are also cases in which the chirality of the tachyon in the OPE is  inverted. 
  In our consideration below
 we shall restrict to combinations of momenta which allow at most the basic series in \opeonet, \opeonetde.


 \subsec{Diagonal  ground ring}


\noindent
As we mentioned in the Introduction, it is possible to construct a discrete model
of non-rational 2D quantum gravity in which the order operators, {\it i.e.}  the
degenerate fields labelled by the diagonal ($m=n$) of the infinite Kac table,
have a simple realization as observables.

It happens that in this theory the 4-point function of order fields contains only
order fields in the intermediate channels. Therefore in  the corresponding CFT
the order field tachyons must form  a closed algebra under OPE. This is not possible
in the matter CFT on a  rigid surface, where the OPE of the diagonal fields
generates the whole spectrum of degenerate fields. The question arises, is it
possible, after switching on the Liouville field,  that the order fields form a closed
algebra? 
We shall argue that such a theory exists.

First we notice that the ground ring element $\CO_{2,2}$ obtained by
combining \mpd\ and \pmd, has four term OPE with the tachyons of given
chirality.  They involve shifts of the momenta with $\pm Q$ and $\pm
e_0$.  To preserve the diagonal $m=n$ of \degen\
we need rather a
projection to the two terms with shifts by $\pm e_0$.  Indeed, such
a projection exists but it requires a different deformation of the
free field ring elements.  This new theory is defined by an
interaction which contains the two Liouville screening charges, as
well as the two possible Liouville dressings of the non-trivial vertex
operator with zero dimension: $T_{1/b}^{+}=\CT_{-e_0}^{(+)}$ and
$T_{b}^-= \CT_{-e_0}^{(-)}\,$: 
\eqn\deltadiag{\eqalign{
 \CA^{\rm int}& =   \int\Big(\mu_{_L} e^{2b\phi}
  +  {\tilde \mu}_{_L}   e^{2/b\phi}
  - \frac{\pi^2  }{e_0^2} \, 
   \mu_{_M}{\tilde \mu}_{_M}\, e^{2i e_0 \chi}
  \(b^2 \mu_L e^{2 b\phi}+ b^{-2}  \tilde \mu_L  e^{2/b\phi}\).
 \cr
&= \L  T^+_{b} +\L'
 \, T_{1/b}^+ 
 + \tilde \L T^-_{1/b} +
 \tilde \L' \,  T^-_{b}
  \,, \quad\qquad \L'=\tilde \L \LM \tilde \LM\,.
}}
The matter charges in the correlators computed with this action can be
screened only by multiples of $e_0=\frac{1}{b}-b$, whence the name
``diagonal'', by which we refer to it.\foot{In the same way one can consider
an interaction theory described by the two matter screening charges
and their Liouville reflected counterparts $T_Q^{\pm}$.
\eqn\sinL{
 \CA^{\rm dg}_{\rm int} 
=  \int\Big(\mu_{_M} e^{-2i b\chi}
 +  \tilde \mu_{_M}   e^{{2i\over b}\chi}
  -{\pi^2\, \mu_{_L} {\tilde \mu_{_L}}  \over Q^2}\, e^{2 Q \phi}\big(b^2 
\mu_{_M}
    e^{-2i b\chi} +{\tilde
\mu_{_M} \over b^2}  e^{{2i\over b}\chi}  \big)\Big)
}}
 The duality transformations
\DUALTY, or \dmc, with a simultaneous change of sign of one of the two
fields, as discussed above, exchange the Liouville screening charges
 as well as the two new terms, so that the action is invariant.  On the other
hand, the matter-Liouville transformations $\{b, \L, \LM, \phi, \chi
\}\ \to \ \{\pm i b, \LM, \L, \mp\chi, \pm \phi \}$ map it to the
action \sinL.

The deformation \deltadiag\ leads to an operator $a_-a_+ \to A$ with
the following OPEs 
\eqn\compra{\eqalign{
&A\,W_{\a}^+ =\L\, W_{\a-\frac{e_0}{2}}^+ +{\tilde \L}\LM \, \tilde \LM\, 
W_{\a+\frac{e_0}{2}}^+\,,\cr
& A \,W_{\a}^- =\tilde \L\,
W_{\a+\frac{e_0}{2}}^- +
 \L \LM \, \tilde \LM\, \,W_{\a-\frac{e_0}{2}}^-\,.
}}
These relations are obtained combining the free field formulae used in
the derivation of \mpd, \pmd\ and \aminus; see Appendix A.1.
Comparing with the composition of \mpd\ and \pmd, the product
$T^-_{1/b} \,T^-_{0}\, T^+_0$ of two matter and one Liouville
screening charges, which leads to the shift $\a \to \a+e_0/2 $, is now
traded for the tachyon $T^+_{1/b}=\CT^{(+)}_{-e_0}$.  This explains
the expression
 $\lambda'=
 \tilde \L \tilde \LM \LM $
   for the coupling constant in \deltadiag, 
   \compra.

Note that  we now need all the four terms in the interaction 
 action \deltadiag\ in order to determine the OPEs of the ring
 generator with tachyons of both chiralities, in contrast with any of
 the relations \mpd, or \pmd.  Let us stress that at this stage we
 will consider the diagonal action \deltadiag\ as a formal tool, which
 provides us in a systematic way with certain rules.  In particular,
 we will not discuss its possible semiclassical limits.

It appears that in this theory the  mass-shell condition applied to the
potential OPE channels is much more restrictive.  Thus there are no
additional terms in the OPE as far as we consider either generic
momenta, or the set of diagonal  momenta $P=k  e_0$.
Similarly for the momenta of interest
(generic, or diagonal degenerate) there are no more OPE terms in the
presence of integrated tachyons besides \aminus.  
The operator $a_+a_-$, perturbed by the diagonal
action \deltadiag,   generates  a  $sl(2)$ type ring,  
as does each of the operators $a_{\mp}$ 
perturbed by the actions \deltaS\ and \deltaSd.  Applying \compra\ to
the set of order parameter fields 
we get a formula analogous to \Cheb, representing the diagonal
ring elements as  Chebyshev polynomial of 
the generator 
$A$.

 \newsec{Functional  relations for the closed string
tachyon amplitudes }

\subsec{3-point solutions of the ring identities}

\noindent
In this section we shall apply the free field computed OPEs of the ring generators  
assuming that they
hold in a general tachyon correlator.

The general 4-point function with one of the ring generators and
arbitrary three tachyons $W_{\a_i}$ can be computed in two ways
exploiting the operator product expansions \mpd, \pmd.  
This leads to
the finite difference identities \homrel\ for the tachyon 3-point
correlators.  Similarly in the diagonal theory \compra\ implies the
relation
\eqn\homreldi{\eqalign{
& N_{P_1+ e_0,P_2,P_3} +N_{P_1- e_0,P_2,P_3}=N_{P_1,P_2+e_0,P_3}
+N_{P_1,P_2-e_0,P_3}\,.\cr
}}
  The simplest solution of  
 \homreldi, as that for the identities \homrel, is
 $N_{P_1,P_2,P_3}=1$.  To fix  here the overall normalization constant we
 assume that the correlators with zero overall matter charge are the
 same in both theories, $\!${\it i.e.$,\!$} they are given by the
 normalized with the leg factors Liouville 3-point constant \Lcs.

Non-trivial solutions exist whenever the momenta take values
corresponding to degenerate Virasoro representations.  Let us first
consider the correlators in the diagonal theory for tachyons with
momenta $P=\IZ e_0$.  We require that the 3-point function vanishes
whenever one of the momenta is zero, that is, outside of the set of
degenerate values.  We choose for definiteness the sign in \degen,
taken for $m=n$, to coincide with the chirality $\e_i$.  Then the
diagonal ring relation \homreldi\ for the 3-point function
\eqn\pardiag{
N_{P_1, P_2, P_3} = 
N_{m_1,m_2,m_3}\,, \qquad P_i = \e_i m_i e_0\leftrightarrow
\a_i=\frac{Q}{2}-m_i \frac{e_0}{2}\, ,
}
turns into the standard recurrence relation for the tensor-product
decomposition multiplicities of the irreps of $sl(2)$ of finite
dimensions $m_k$:
\eqn\sltwo{\eqalign{
&N_{m_1,m_2,m_3}= \cases{ 1 & if $\matrix{
&|m_1-m_2|+1\le m_3\le  m_1+m_2-1 \cr  &
{\rm and}\  \ m_1+m_2+m_3=  {\rm odd}
}$\ ;\cr 0 & otherwise .}  \cr
}}
Any of the two sides in \homreldi\ is  equal to   the 4-point multiplicity
$N_{m_1,m_2,m_3, 2 }$,
where,
\eqn\fourpmd{\eqalign{
&N_{P_1, P_2, P_3, P_4}=\sum_{m=1}  N_{m_1,m_2,m}\
N_{m,m_3,m_4}=N_{m_1,m_2,m_3,m_4}\cr
 &=\hf \(\min(m_1+m_2,m_3+m_4)-\max(|m_1-m_2|, |m_3-m_4|)\)\,.
}}
Similarly for general degenerate momenta \degen\
 the identities \homrel\  are solved by the product
\eqn\sltwopr{
N_{P_1, P_2, P_3}=N_{m_1,m_2,m_3} N_{n_1,n_2,n_3}\,, \qquad
P_s=
\pm( n_s/b -  m_sb )\,,
}
assuming the vanishing of the tachyons on the border
lines $m=0$, or $n=0$ of \degen{}; see also \SeibergS\ for the
rational case.  
The solution is symmetric with respect to matter
charge reflections $P_s\to -P_s$ and thus can be identified as a
correlator of the invariant combinations \rinv.  These $sl(2)\times
sl(2)$ decomposition multiplicities are the fusion multiplicities in
the quasi-rational matter theory at generic values of $b^2$, described
by the infinite set of fields of momenta $P_{m,n}$.  The same
solution of \homrel\ is found if the tachyon momenta 
take the Liouville degenerate values $\e_s P_s= \pm(n_s/b+m_s b)$ 
as in \degenL.

 The fusion multiplicities \sltwopr\ coincide, when restricted to the
 diagonal $m_s=n_s$, with those obtained in the diagonal theory,
 \pardiag\ and \sltwo.  However the 4-point fusion multiplicities
\eqn\fourpmul{
N_{P_1,P_2,P_3, P_4}=\sum_{m,n=1} N_{P_1,P_2, P_{m,n}}
 \, N_{P_{m,n},P_3,P_4} =
N_{m_1,m_2,m_3,m_4}\,N_{n_1,n_2,n_3,n_4} 
}
taken for such values differ from their  counterparts \fourpmd\
in the diagonal theory.

For  other 3-point solutions see also Appendix B. 

\subsec{Recurrence relations} 

\noindent 
We next apply the OPE relations inserting a 
ring
generator in the 4-point function
 \eqn\fptf{
G^{\e_1\e_2\e_3\e_4}_4
(\a_1,\a_2,\a_3,\a_4)=
 \<   W^{\e_1}_{\a_1}(0)\
 \  W ^{\e_2}_{\a_2}(1) \ \
   \! T ^{\e_3}_{\a_3}\
  \ W^{\e_4}_{\a_4}(\infty)\>\,.
  } 
  The relation one gets
   for  $a_-$
   approaching the first or the second tachyons, reads
  \eqn\wardid{\eqalign{
  \sum_\a C_{-\frac{b}{2} \,\a_1}^{(\e)}{}^{\a}&\ G_4^{(\e)}(\a, \a_2,\a_3, \a_4) +\sum_\a C_{-\frac{b}{2} \,\a_1 \a_3}^{(\e)}{}^{\a}\  G_3^{(\e)}(\a, \a_2, \a_4) \cr
  &= \sum_\a C_{-\frac{b}{2} \, \a_2}^{(\e)}{}^{\a}\  G_4^{(\e)}(\a_1, \a,\a_3, \a_4) +\sum_\a C_{-\frac{b}{2} \, \a_2 \a_3}^{(\e)}{}^{\a}\  G_3^{(\e)}(\a_1, \a, \a_4)
}}
where we have omitted for simplicity the explicit dependence of  the tachyon
correlators $G^{(\e)}$ and the OPE coefficients $C^{(\e)}$ on each of the  chiralities.
The OPE relations \mpd\  determine the first terms in both sides of \wardid. 
On the other hand  \opeonet, and for special values of the momenta, \opeonetde,
give the  explicit expressions for the 4-point  OPE coefficients, which determine the inhomogeneous, 3-point  ''contact'',  terms  $\sim  G_3^{(\e)}$  in \wardid.  For generic momenta the ring relation \wardid\ generalizes straightforwardly to a correlator with an arbitrary number  $p-3$ of integrated tachyons $T_{\a_i}^{+}$ with  summations  over  $(p-1)$-point contact terms. 
If however some partial sums of momenta ``degenerate'', i.e., lie on the lattice $\CL$,  there are other possible solutions of the mass-shell conditions,  as explained in Appendix A.3, and hence  
potentially new
$m$-point contact terms, $3\le m\le p-1$. The OPE coefficients would require the computation of higher
$p-m+3$-point 
free field functions
matrix elements, 
generalizing the $p=4=m+1$ 
case of  Appendix A.4.

 The inhomogeneous associativity identities \wardid\ can be interpreted as  string analogs 
 of the  duality equations for the local 4-point correlators of the $c<1$ or $c>25$ Virasoro theory. Given the OPE  coefficients $C^{(\e)}$ and a choice of the 3-point 
  terms, the set of  these  relations  
 determines the 4-point tachyon correlators. What has also to be added to this set of recursive  difference equations is a choice of some boundary conditions, {\it i.e.} particular  known   values of the   4-point tachyon correlators.

 \medskip
 \rb 
 We shall  now specialize the contact terms in \wardid\ for two basic classes of 
4-point  tachyon correlators. For the 3-point tachyon correlators in \wardid\ we shall take the generic 
solution \ttp{}. Let us choose  $\e_2=\e_3 
 =1=-\e_1$.   Then the  first  series in \opeonet\ contributes 
to the r.h.s. of \wardid\  (with $W_{\a}^+T_{\a_1}^+$ now denoted  $W_{\a_2}^+T_{\a_3}^+$).  As we have mentioned above, 
 if the tachyon $W_{\a_2}^+$
is labelled by the  degenerate momentum  \degm\
of matter or Liouville type,  
 the r.h.s. of \opeonet\  terminates.  Hence  only the first $n+1$ contact terms have to be taken into account  in the r.h.s. of \wardid. 
 Using 
  the relation \recr\  satisfied by the generic solution \ttp{}\ 
  one  obtains $(n+1)$ times  a power  of the coupling constants $\L,\LM$.  For generic values 
$P_1\,,  P_3
  \not \in \CL$, and $
P_2+P_3 \not \in \CL$
these are the only contact terms in \wardid\ and we obtain  the  equation
\eqn\contgena{\eqalign{
&G_4^{-+++}(\a_1,\a_2,\a_3, \a_4)+ \L\LM\,
G_4^{-+++}(\a_1+b,\a_2,\a_3, \a_4)\cr 
&-\L\,G_4^{-+++}(\a_1+\frac{b}{2},\a_2+\frac{b}{2},\a_3, \a_4)-\LM\,
G_4^{-+++}(\a_1+\frac{b}{2},\a_2-\frac{b}{2},\a_3, \a_4)\cr 
&=\sum_{k=0}^n 
\,(\tilde \L \tilde \LM)^k\,
  G_3^{-++}(\a_1+\frac{b}{2},
\a_2+\a_3-\frac{b}{2}+\frac{k}{b},\a_4)\cr
&=
- (n+1)\,  G_3^{-++}(\a_1+\frac{b}{2},\a_2 +\a_3 -\frac{b}{2},   \a_4) \,.
}}
In the last line we have used  the relation \recr\  satisfied by the generic solution \ttp{}; in general one should keep the r.h.s. of the first equality. 

 There is another class
of correlators in   which 
the OPE relations  \opeonet\  
produce finite number of contact terms in \wardid. These are the  correlators with all generic momenta, but restricted by an overall  
 charge conservation condition.  It can be  a relation involving the two matter charges
\eqn\neutgg{
\sum_{i=1}^4 e_i -e_0 = mb-\frac{n}{b} \  \leftrightarrow \ 
\sum_{i=1}^4 P_i = 
2e_0 -2mb +\frac{2n}{b}\,, \ \ m,n \in \IZ_{\ge 0} \,,
}
or it can be   $c>25$ charge conservation
condition 
\eqn\neutggL{ \sum_i
\a_i -Q=-m b - {n/ b} \  \leftrightarrow \ 
\sum_{i=1}^4 \e_i P_i = 
2Q + 2mb +\frac{2n}{b}\,, \ \
m,n\in \IZ_{\ge 0}\,. }
For a fixed $n$ in \neutgg\  (or \neutggL) $k$ of the charges
$T^-_0$ (or $T^-_{1/b}$), $ k=0,1,\dots, n,$  can be assigned to the OPE  in 
\opeonet. The general  identity  \wardid\ 
takes again the form of  \contgena.

\medskip
 The equation \contgena\ is recursive with $m$, e.g., in the case of the  matter type charge conservation condition \neutgg, the  $\LM$-independent and $\LM$-dependent  terms in the l.h.s. correspond to $m$ and $m-1$
respectively.  In the case of degenerate $P_2$ \degm,  terms with the three values $m, m\pm 1 $ appear. 

If on the other hand the degenerate field appears as an integrated tachyon, {\it i.e.}  in our notation $P_3$ is degenerate
as in the r.h.s. of  \opeonetde, then (changing the notation 
 $n \to n_3$  in  \opeonetde{}),  this OPE relation  leads to new $(n_3+1)$ contact terms.
 This implies, using once again   the simple solution \ttp{}, that the coefficient  in the r.h.s. of \contgena\  is modified to $(n-n_3)$.   
For $n_3=n$  
the contact term in the r.h.s. disappears and the relation \contgena\ becomes homogeneous.

In particular in the class of correlators with $n=0=n_3$ 
the r.h.s of \contgena\ simplifies to one or zero contact terms, which we write 
as  
 \eqn\rrminus{
\eqalign{
&G_4^{-+++}(\a_1-\frac{b}{2}  ,\a_2,\a_3,\a_4)+
\L\,\LM\,G_4^{-+++}(\a_1+ \frac{b}{2}  ,\a_2,\a_3,\a_4)
\cr
& -\L
G_4^{-+++}(\a_1, \a_2+\frac{b}{2} ,\a_3,\a_4)-
\LM\,G_4^{-+++}(\a_1, \a_2-\frac{b}{2} ,\a_3,\a_4)\cr
&=-
G_3^{-++}(\a_1,\a_2+\a_3-\frac{b}{2} ,\a_4)\cr &
 +\sum_{m_3=0} (\LM^{m_3+1}  \,\delta_{\a_3,b+\frac{m_3 b}{2}}\,  +\L^{m_3+1}  \,\delta_{\a_3, -\frac{m_3 b}{2}})\,
 G_3^{-++}(\a_1+\frac{(m_3+1) b}{ 2} ,\a_2, \a_4)\,.
 }}
The relation dual to \rrminus\ reads  
\eqn\rrplus{
\eqalign{
&G_4^{+---}(\a_1-\frac{1}{2b} ,\a_2, \a_3,\a_4)+
\tilde\L\,\tilde\LM\,G_4^{+---}(\a_2+\frac{1}{2b},\a_2, \a_3,\a_4)\cr
&-\tilde\L\,
G_4^{+---}(\a_1,\a_2+\frac{1}{ 2b}, \a_3,\a_4)-
\tilde\LM
 \,G_4^{+---}(\a_1, \a_2-\frac{1}{ 2b}, \a_3,\a_4)\cr
&= -G_3^{+--}(\a_1, \a_2+\a_3-\frac{1}{
2b},\a_4)\cr &
+\sum_{n_3=0}(\tilde\LM^{n_3+1} \,\delta_{\a_3,{1\over b}+ \frac{n_3 }{2b }}+ \tilde\L^{n_3+1}  \,\delta_{\a_3,- \frac{n_3 }{2b} })\,
G_3^{+--}(\a_1+\frac{(n_3+1)}{ 2b}, \a_2,\a_4)\,.
}}
 In these simplified equations only 
one of the matter and one of the Liouville charges is effectively contributing, namely the pairs in the action 
\deltaS, or \deltaSd\ respectively.
 The equations \rrminus, \rrplus\ become homogeneous if,  in particular,  $T_{\a_3}^\e$
coincides with one of the four screening charges.  
 For example, if
$\a_3=0$ ($e_3 = -b$), the
simplest solution of the homogeneous relation \rrminus\ is
\eqn\tpsol{\eqalign{
 G_4^{(-+++)}(\a_1,\a_2,0,\a_4)&=-{1\over 
  b^2} \big(\sum_{i} e_i-e_0+b\big)
 \,\L ^{{1\over b}(Q-\sum_{i} \alpha_i)}\,
   \LM^{-{1\over b}(e_0 -\sum_{i} e_i)}\cr
  &=  - {\p\over \p  \LM  }  G_3(\a_1,\a_2,\a_4)\,,
}}
while  for $ a_3=b$
($e_3=0$)  it is 
\eqn\simpleex{\eqalign{
G^{(-+++)}(\a_1,\a_2,b,\a_4)
&=
{1\over 
 b^{2}}\,(\sum_{i} \alpha_i-Q-b)\,
   \,\L ^{{1\over b}(Q-\sum_{i} \alpha_i )}\,
{  \LM}^{-{1\over b}(e_0-\sum_{i} e_i)}  \cr
  &= -{\p\over \p  \L  }   G(\a_1,\a_2,\a_4)\,.
    }}

  Setting $\LM=0$ in \rrplus\  (or $\tilde \LM=0$ in \rrminus) reduces  furthermore \rrminus, \rrplus,
  so that
   they apply to correlators with momenta satisfying  trivial matter condition  $m=0=n$  
    in
   \neutgg.    Note that besides the main contact terms, each of these   equations with two terms in the l.h.s.
   still contains  
 an     ''accidental'' contact  term,  e.g., $m_3=0$ in \rrminus,
    missed in the old considerations. 
 Taking into account such  terms in the 
 (4+m)-point generalisations of  the reduced two term  equations, derived with only the Liouville interaction included,  leads to an alternative derivation of  the four term  identities \rrminus, \rrplus.

  In a similar way, one derives generically homogeneous relations fusing
the ring generators with two  tachyons of  the same chirality.
E.g., for $\e_s=1\,, s=2,3,4$ 
  \eqn\rrminhom{
\eqalign{
&\L G_4(\a_1,\a_2 ,\a_3+\frac{b}{2},\a_4)+
\LM\,\,G_4(\a_1,\a_2,\a_3-\frac{b}{2},\a_4)
\cr
& =\L
G_4(\a_1, \a_2+\frac{b}{2} ,\a_3,\a_4)+
\LM \,G_4(\a_1, \a_2-\frac{b}{2} ,\a_3,\a_4),
 }}
 cancelling the difference of the two contact terms. We stress that this  and the above discussed
 cancellations occur when the simplest constant 3-point solution \ttp{}\ is used; in general we should keep the full linear combinations of 3-point contact terms. 

\medskip

\rb The functional equations take a more compact form after rescaling
 \eqn\Gfourf{ G_n^{(\e)}\equiv  b^{^{-\sum_i\e_i }}\,
\L^{ {1\over b}(Q-\a)}\, \LM^{{1\over b} (e-e_0)}\ 
{\hat  G}_n^{(\e)}, \qquad \a=\sum_{i=1}^n \a_i\, \quad
e = \sum_{i=1}^n e_i.  }
The normalized correlators $\hG_n$ do not depend on the constants $\L,
\LM$ as is standardly checked by shifting $\phi \to \phi-\frac{\log
\L}{2b}\,, \chi \to \chi+\frac{\log \LM}{2b i}\,.$ The rescaling by
the power of $b$ in \Gfourf\ is equivalent to a change of the leg
factor normalization
\eqn\renormal{
V_{\a}^{\e} \to \hat{V}_{\a}^{\e}= b^{\e } \, 
V_{\a}^{\e}\,
}
which  removes the chirality-dependent power of $b$ in the 3-point function \ttp{}. 
This normalisation  does not change the OPE ring identities \mpd, \pmd, but changes
the coefficients in front of the contact terms.
With a slight abuse of
notation, in what follows we shall write $G_n$  for the corresponding correlators
with just the powers of $b$ removed, $\!${\it i.e.$,\!$} the ones
differing from $\hat G_n$ only by the powers of $\L\, $ and $\LM$.
For further reference we write the ring relations \rrminus\ and
\rrplus\ also in terms of the target space momenta.  For
the rescaled functions
$$ \hat \CG_4^{(\pm)}(P_1 | P_2,P_3,P_4)\equiv 
  \hat  G^{ \mp \pm \pm \pm}(\a_1,\a_2,\a_3,\a_4),
 \qquad \e_iP_i=Q-2\a_i,$$
 the equations take the form
\eqn\difec{
\eqalign{
&2(\cosh b \p_{ P_1}-\cosh b \p_{P_2})\
\hat \CG_4^{(+)}(P_1|P_2,P_3,P_4)
=- b (n+1)N_{P _1,P _2+P _3- 
b^{-1},P _4}\,,\cr
&2(\cosh \frac{1}{b} \p_{ P_1}-\cosh  \frac{1}{b} \p_{P_2})\
\hat \CG_4^{(-)}(P_1|P_2,P_3,P_4)
=-  \frac{1}{b} (m+1)N_{P _1,P _2+P _3 +
b,P _4}\, 
 }}

\medskip

\rb In the diagonal  theory  the ring relations read 
\eqn\rrmindiag{
\eqalign{
&\tilde \L G_4^{-,+,\e,\e}(\a_1+\frac{e_0}{2}  ,\a_2,\a_3,\a_4)+
\L \LM^{-\frac{e_0}{b}}\,
G_4^{-,+,\e,\e}(\a_1- \frac{e_0}{2}  ,\a_2,\a_3,\a_4)
\cr & -\L G_4^{-,+,\e,\e}(\a_1, \a_2-\frac{e_0}{2} ,\a_3,\a_4)-
{\tilde \L}\LM^{-\frac{e_0}{b}}\,
G_4^{-,+,\e,\e}(\a_1, \a_2+\frac{e_0}{2} ,\a_3,\a_4)\cr
&=\cases{- G_3^{-,+,\e}(\a_1,\a_2+\a_3-\frac{Q}{2}
,\a_4)\,, & if $\e=+$\cr
\ \ G_3^{-,+,\e}(\a_1+\a_3-\frac{Q}{2},\a_2 ,\a_4)\,, & if $\e=-$}\, .
}}
Here  we shall   normalize the 
correlators (taking also into account  the rescaling \renormal),
as
\eqn\diagresc{
G_4^{(+)}=
-\frac{b}{e_0}\,\L^{{1\over b}(Q-\a)}\, \LM^{{1\over b} (e-e_0)}\,
\hat G_4^{(+)}, \  \ 
G_4^{(-)}=
\frac{1}{be_0}\, \L^{{1\over b}(Q-\a)}\,\LM^{{1\over b} (e-e_0)}\,
\hat G_4^{(-)} .
}
 With  this normalization 
  \rrmindiag\   becomes   
\eqn\rrplupdia{
\eqalign{2\(\cosh e_0 \p_{P_1}-\cosh {e_0 \p_{P_2}}\)
   \hat  \CG_4^{(\e)}(P_1 |P_2,P_3,P_4) =\e e_0\, 
   \hat  \CG_3^{(\e)}(P _1|P _2+P _3 ,P _4)\,.
}}
 The solutions of the  mass-shell condition,
 restricted to diagonal momenta $P_3\in \IZ e_0\subset \CL$
 allow
 only $P_3=0$ as a possible momentum leading to an "accidental"
 contact term.  This is the tachyon $T_{Q/2}$ of no definite
 chirality.

In the same way one derives 
relations  
  with
 $\alpha_3$ exchanged with $\a_2$ or $\a_1$ respectively.  
The  derivation can be repeated also with the third field taken at
 infinity.  The collection of these identities for generic momenta imply a set of symmetry
 relations for the 
 contact terms in the r.h.s., e.g. for $\e=1$,  
 \eqn\symta{
\CG_3(P_1,P_2+P_3,P_4)=
\CG_3(P_1,P_2+P_4,P_3)=
\CG_3(P_1,P_4+P_3,P_2).
}


\newsec{Solutions of the ring relations in the absense of matter
(or Liouville) screening charges}  


\noindent
In this section we will describe solutions of the ring generated
functional equations in the simplest case of only Liouville or only
matter perturbation. 

\subsec{\it Solutions with matter charge conservation}

 \noindent 
 In the case of gaussian matter field (formally $\LM=0=\tilde \LM)$
 the neutrality condition \neut\ holds.  The l.h.s. of the functional relations
\rrminus, \rrplus\  reduces to a difference of two terms. For generic momenta the equations extend \KostovCY\
straighforwardly to $p$-point correlators satisfying the  ``chirality rule'', {\it i.e.},
 one of the tachyons has the opposite chirality $-\e$  to the chirality $\e$ of the 
 other $p-1$ ones; these are  in fact the only correlators comparable with the
 microscopic approach.  If we restrict  to the resonant correlators, satisfying  also the Liouville type conservation condition $\sum_{s=1}^p\a_s=Q$, the equations simplify  with only the $\L$ - independent term surviving in the l.h.s. The r.h.s. is recursively reduced to a 3-point function and the solution is a constant independent of the momenta. In  general
 the functions are symmetric with respect to  $p-1$ of the charges and since they have to reproduce as a special case the resonant amplitudes 
 they depend only on the sum $\a=\sum_{s=1}^p\a_s$ of the
 Liouville charges.
 Taking into account the new normalization in \renormal\  the equations
 read, 
\eqn\oldring{\eqalign{ 
\hG_{p}^{(\e)}(\a)-\, \L\, 
\hG_p^{(\e)}(\a+b^\e)=-(p-3)b^\e\, \hG_{p-1}^{(\e)}(\a)\,, \qquad \a=\sum_{s=1}^p\a_s\,.
}}
Starting with the 4-point case, the solution involves an arbitrary
solution of the homogeneous equation, $\!${\it i.e.$,\!$} a periodic
in $b$ or $1/b$ function.  We shall use as a boundary condition the
known expressions \simpleex\ and its dual,  in which one of the
Liouville charges is $b$ or $1/ b$.  This leaves us with a linear
function of $\a$ 
\eqn\derivL{\eqalign{
G^{(\e)}
(\a_1,\a_2,\a_3, \a_4) 
&=
\L^{{1\over b}(Q- \a)} (\a-Q-b^\e)= \L^{{1\over 2b}(-2Q +\sum_i\e_i P_i)}(b^{-\e}-\hf \sum_i\e_i P_i)\,. 
}}
The two choices of boundary
conditions
 are correlated with the two effective actions,   
  \deltaS\ and \deltaSd.
The $p\ge 4$ 
relation \oldring\ is solved by
\eqn\nsoll{\eqalign{
G_p^{(+)}(\a)&=
\(-b\,\p_{\L}\)^{p-3}
\L^{{1\over b}(Q - \a)+ p-3   }\,,\cr
G_p^{(-)}(\a)&=\(-\frac{1}{b}\,\p_{\tilde \L}\)^{p-3}
{\tilde \L}^{b(Q - \a)+ p-3   }\,,
}}
recovering formula $(2.53)$ of \DiK. The solutions of the two equations are
interchanged by the duality transformations \dualtL\ and \dualtm, up
to a sign in the second case; it disappears for the properly
normalized correlators as in \normt.

 The formulae \derivL\ are valid also for the 4-point functions with $
 \sum_i\e_i=\pm 4$, which are constants, because of the matter charge
 conservation condition.  These constants and the solutions with $
 \sum_i\e_i= \pm 2$ in \derivL\ are related with an inhomogeneous
 analog of the Liouville reflection relation, in contrast with what we
 had for the 3-point functions in \drefl.  For example,
\eqn\Lrefl{\eqalign{
&G^{++++}(\a_1,\a_2,\a_3, \a_4)=
\cr
&=\L^{{1\over b}(Q-2\a_1)}
G^{-+++}(Q-\a_1,\a_2, \a_3, \a_4) +
(Q-2\a_1)
\,\L^{{1\over b}(Q-\sum_{i=1}^4 \a_i)}\,.
}}
The second term in the last equality compensates the contact term in
the ring relation \rrminus\ for the 4-point function of type
$^{(-+++)}$ in the r.h.s. of \Lrefl, so that the l.h.s. satisfies a
homogeneous equation without contact terms, as it should.  So far we
have excluded from the discussion the correlators with two equal
chiralities, $\!${\it i.e.$,\!$} of the type $^{(++--)}$.  The reason
is that the contact terms depend on the choice for the integrated
tachyon and one obtains an inconsistent set of relations.  These
correlators have been neglected in the earlier considerations, {\it
e.g.} \DiK\ and \GinM, basically because of the vanishing of the
unnormalized perturbative expressions, as discussed above in the
comment after \Tripto.   Furthermore the correlator of type $ ^{(+---)}$
is also trivial constant when determined by  the action \deltaS, and similarly
for the correlator of type $^{(-+++)}$ computed with \deltaSd, since 
 the corresponding functional equations are homogeneous.
\medskip
\rb
 The chirality rule satisfying 
 solutions of the two type equations 
 are related by  pairs of inhomogeneous Liouville reflections
  \eqn\reflpro{\eqalign{
 \hat \CG^{-+++}(P_1,P_2,P_3,P_4)&= P_1+b = 
  -\hat \CG^{+---}(P_1, P_2,P_3,P_4) +Q
 \cr
 & =\hat \CG^{+-++}(P_1,P_2,P_3,P_4) + (P_1-P_2)\cr
  &=  \hat \CG^{---+}(P_1, P_2,P_3,P_4) + (P_1+P_4-e_0)\cr
 & =  \hat \CG^{+---}(P_1, P_2,P_3,P_4)  +2 P_1 - e_0\,. 
 }}
  The matter reflections do not make sense since they  violate the  charge conservation condition
 \neut.
 Using 
 this condition $\sum_i P_i=2e_0$, we can 
rewrite \derivL\  as 
\eqn\fopbb{
\CG^{(\e=-\e_1)}(P_1|P_2,P_3,P_4) =
\L^{{1\over 2b}(-2Q+ \sum_{i=1}^4 \e_i P_i)} 
\big(Q -\e_1\, \sum_{s\ne 1} (e_0-P_1-P_s)\big)\,
}
i.e., in the form of  \summar, with   $N_{P_1,P_2,P_3}= 1\,, $ if $\sum_i P_i= e_0$ and $N_{P_1,P_2,P_3}= 0$ otherwise. 

 If we restrict the momenta to the range $\e_s (P_s-\frac{e_0}{2}) >0\,, \ s\ne 1$
  (physical 
 for  $\e\, e_0>0$ and  implying  $\e_1 (P_1-\frac{e_0}{2})>0$ as well), 
 the correlator
\fopbb\ reproduces the three channel
 expansion formula of
\DiK,
\eqn\fourpDK{\eqalign{
\CG(&P_1,P_2,P_3, P_4) =
 \L^{ {1\over 2b}   (\sum_{i=1}^4 |P_i|-2Q)}\,
 \hat \CG(P_1,P_2,P_3, P_4)\,, \cr
&\hat \CG(P_1,P_2,P_3, P_4)=\hf \,
\(Q- |P_1+P_2 -e_0|
-|P_1+P_3 -e_0|-|P_1+P_4 -e_0|\)\,.
}}
   This formula holds irrespectively of which of the four momenta is
chosen with opposite sign  since,  unlike
\fopbb{},  it is symmetric in them, but at the price that it is not analytic.  Vice versa, 
 for any choice of the
signs of the combinations $$\{P_{st}:=P_s +P_t-e_0\}_{s,t=1,...,4}$$ compatible with the
conservation condition $\sum_i P_i=2e_0$, the formula \fourpDK\ recovers 
 one of the correlators
with $\sum_i \e_i =\pm 2$.  In other words \fourpDK\  is a symmetrization over the chiralities,
  \eqn\symch{
 \hat \CG(P_1,P_2,P_3,P_4)=\sum_{\e_t} \prod_{s\ne t} \theta(\e_t (P_t+P_s- e_0))\, \hat \CG^{(-\e_t)}(P_1,P_2,P_3,P_4)
 }
where $\theta(x)=\cases{1& if  $x\ge 0$\cr 0 & if  $x<0$ }$. Here  $\sum_{\e_t} =\sum_t\sum_{\e_t=\pm}$ and  $\e_t$  in $\hat \CG^{(-\e_t)}$ indicates as 
before the chirality   opposite to the remaining three.\foot{
{\ninepoint \noindent 
 We could restrict to the subdomain in which the signs of $P_i-e_0/2$, instead of the chiralities,  satisfy the chirality rule.
 Then
 replacing  the step function factor in the symmetrization formula \symch\   with 
 $\prod_t \theta(\e_t (P_t-\frac{e_0}{2}))$ makes the correspondence of the two types of correlators ``local'', i.e., depending only on the individual momenta.  However the subdomain  is not preserved by the shifts,
 combinations 
  with
 two positive and two negative signs of  $\{P_t-\frac{e_0}{2}\}$ may appear.}}
 
 The permutation symmetry with respect to the   four  matter charges $e_1,e_2, e_3, e_4$
is an analog of the locality of the 4-point euclidean correlation
functions, so we shall refer to formulae of this type as ``local''
or ``physical''.  The other symmetric combination with a plus relative sign corresponds
to  generically  unphysical momenta $\e_i (P_i -e_0)<0\,. $ 
 The local correlators do not depend on the chiralities and so they are invariant under Liouville reflections. 
Formula \fourpDK\   is reproduced  in the discrete model
framework.  
 For one of the momenta coinciding with  $e_0$ 
\fourpDK\  reduces to a derivative
of the ``physical'' 3-point function 
\eqn\bphys{ 
\CG(P_1,P_2,P_3, e_0) 
=  -b\p_{\L} \CG_3(P_1,P_2,P_3) = -b\p_{\L}  \L^{\frac{1}{2b}(-Q+\sum_s |P_s|)}\,
}
if  $ e_0>0$ and analogously a derivative  with respect to $\tilde \L$ for  $ e_0<0$. On the level of 3-point functions the physical 
tachyons are identified by the fixed chirality fields $\CW^{+}_P$ for $P>0$ or $\CW^{-}_P$ for $P<0$.  
The functional identities rewritten for the correlators of these physical representatives contain in general
$P$-dependent  powers of $\L$, coming from a  Liouville reflection as in \drefl, whenever unphysical 
value $\e P<0$ is reached. As it is clear from \symch\  this local representation of the physical
fields is not possible on the level of the 4-point function. 
Note that there are other  symmetric
combinations  locally reproduced by the eight solutions of  the equations. They are   
obtained by replacing 
$ \sum_i \e_i P_i $ with $\sum_i |P_i| $ in \derivL: one gets two combinations which are  interchanged
under  the  transformation \dualtL. 
What distinguishes the  correlator  \fourpDK\ is that it preserves the simple fusion rule of the underlying local matter theory in each of the $s,t,u$ channels so that the notion of  ``locality" of the 4-point  tachyon ``correlation numbers" matches that of  standard locality. Furthermore with the chosen normalisation \renormal\ this correlator 
is self-dual with respect to the Liouville type transformation \dualtL;   in 
 the initial normalization  the  two
analogs of \fourpDK\  differ  only by an overall power of $b^2$. 
If  we further normalize with the partition functions \partfa\ 
 we  can define   two  correlators, depending only  on $\{b^2,  bP_i\}$, 
 which are exchanged by \duality\   
\eqn\fourpm{\eqalign{
&{}^{(n)}\CG_4(P_1,P_2, P_3,P_4;\L,\LM, b)
=- {e_0 Q\over 2 b^2 }
\L^{ {1\over 2b}  
 (\sum_{i=1}^4 |P_i|-4Q)}\,\LM^{e_0\over b}\,\( \frac{ Q}{2b}- \sum_{s\ne  1} | \frac{1}{2b}P_{1s}|\)\cr
&={}^{(n)}{\tilde \CG}_4(-P_1,-P_2,-P_3,-P_4;\tilde \L,\tilde \LM, \frac{1}{b})={}^{(n)}{\tilde \CG}_4(P_1,P_2,P_3,P_4;\tilde \L,\tilde \LM, -\frac{1}{b}) \,.
}}

\rb    The difference identities for the correlators with
definite chiralities like \fopbb\
do not preserve
the physical regions, neither the region determined by the set of inequalities 
above. Accordingly \fourpDK\ does not satisfy   globally the  equations which apply by definition only to the
partially symmetric, fixed chirality correlators. 
One can compute directly the shift relations for the local correlators  from the explicit expression \fourpDK, or derive them from the initial identities.

 It is instructive to  compare the two types of equations.
If the shift crosses the boundary of the momenta region  in which a given fixed chirality correlaror represents the
local one, the shifted correlator  can be  replaced  via pairs of Liouville reflections \reflpro\ 
by the proper local representative 
in the new region. In this replacement there appear  linear in the momenta terms which 
can be   moved to the r.h.s. and interpreted as a 
 modification of the  contact terms.
 By the same mechanism  any of the homogeneous relations
 acquires a non-trivial r.h.s., if the shift crosses the boundary of the corresponding region. 
  Since the  coefficients in the linear relation 
  \symch\    project to different regions of momenta, this effectively implies that the (4-point) OPE coefficients  in the analog of \wardid\  for the local correlators will be no more constants but will depend themselves on the momenta.   The rule which is extracted from the 
  explicit expression \fourpDK\  is that whenever the boundary is crossed, a Liouville  reflected tachyon in the OPE appears,  ``dressed''  with the inverse propagator \twopP.    Namely
    (taking as usual $b>0$)  we compute from \fourpDK\
  \eqn\moeqs{\eqalign{
& - \hat \CG_4(P_1,P_2,P_3,P_4) + \hat \CG_4(P_1+ b,P_2- b,P_3,P_4)\cr
 &=\sum_{s=3,4}\((\theta(P_{2s})
  - \theta(-P_{2s}))
    \, \frac{b}{2}
 + \theta(-P_{2s}+ b)\theta(P_{2s})\,(P_{2s}- b)\)\cr
  }}
 We shall take  $P_{2s} >0\,, s= 3,4\,, \ P_{34}>0$  so that  the first correlator in the l.h.s. of \moeqs\  is  identified  with $\CG^{-+++}$. Then the 
 r.h.s. of \moeqs\ reduces to
  \eqn\wardidloc{
  b + \sum_{s=3,4} \theta(-(P_{2s}- b) )\, (P_{2s}- b)= b -  \sum_{s=3,4} \theta(-P_{2s}+ b) \, \hat \CG_2(P_{23}-b,-P_{23}+b)^{-1}
\,.
}
  If  both  shifted momenta change sign, $P_{2s}-b<0\,, s= 3,4$ the shifted correlator in the l.h.s. is identified with  $\CG^{+-++}$. 
 If only $P_{23}-b<0$, while  $P_{24}-b>0$ (i.e., $P_{13}+b<0$), this shifted correlator  is of type
$\CG^{---+}$. 
Irrespectively of the signs 
of $P_{2s}-b$ the first  term in  \wardidloc\ corresponds to the standard constant  contact term.
For negative  $P_{23}-b$ the physical tachyon in the 3-point  function $\hat \CG_3(P_1+b,P_{23}-b,P_4)$  has to be identified with the 
Liouville reflection of the  tachyon $W_{\a_2+\a_3-b/2}^+=\CW_{P_{23}-b}^+$. However the new contact term cannot be identified simply with the product of this 3-point correlator and the 
 Liouville  reflected OPE constant
$C_{-\frac{b}{2} \, \a_2 \a_3}^{++-}{}^{Q- \a_2-\a_3+\frac{b}{2}} = b^{-2} \L^{P_{23}-b\over b}$
(see (A.32) below); rather it is related to the   derivative of this constant with respect to $\L$.

We stress that   \moeqs\ is just an alternative rewriting 
of  the initial shift relation  as a relation for the local correlators;
otherwise the new terms  $P_{23}+P_{24}-2b=(P_2-b)- (P_1+b)$  (or $P_{23}-b=(P_1+b) +P_4-e_0$) in \wardidloc\  are precisely 
the inhomogeneous terms of the Liouville reflections in \reflpro, needed to represent the  shifted correlator in the l.h.s. by a function of type 
$\CG^{+-++}$ (or of type $\CG^{---+}$) respectively.

There are several  remarks in order concerning the identity  \moeqs, \wardidloc:

 i)  The appearance of  $P_{24}$-dependent terms, besides  $P_{23}$,   serves as a symmetrization  as in  the simple relation $\symta$ (in which the OPE coefficients are set to $1$).  
  These terms correspond to  correlators in which the fourth tachyon is represented
 by an integral $T_{\a_4}$; the shift equation for 
 the  symmetrised  correlator 
  does not distinguish the two situations. 
Or, alternatively, the relation \moeqs\ 
represents the   
 ``splitting"  of the local 5-point function with a ring generator  into various  products of  $3$- times $4$-point,  times the inverse of  a 2-point,  correlators.   
Effectively   the shift  equation rewritten for the symmetrized correlator  manifests 
the short distance expansion around all the three points $0,1,\infty$.\foot{{\ninepoint \noindent This is analogous to 
  the general discussion in \WitZw\ 
 (see also \refs{\Kl\Verl-\KlP}),  where functional relations for the tachyon correlators in the $c=1$ theory without interactions are derived starting from Ward identities of   non-scalar currents.   The resonant  amplitudes described in these  works are too simple to actually make a distinction between the two types of equations but an extension of the method might be appropriate for the problem under consideration. }}

ii)   On the other hand, 
once extracted from \wardidloc,   
these modified contact terms can be    used to extend the r.h.s. of    the  general $4$-term ring relation \contgena\  for $n=0$.
 Namely for $P_{st}>0\,, s,t=2,3,4$ (ensured by the physical values $\e_sP_s>e_0/2$)  we have 
\eqn\recu{\eqalign{
&\hat G(P_1+b,P_2-b,P_3,P_4) - \hat G^{-+++}(P_1,P_2,P_3,P_4) \cr
&+\hat G^{-+++}(P_1+b,P_2+b,P_3,P_4)- \hat G^{-+++}(P_1+2b,P_2,P_3,P_4) \cr
&=b +\sum_{s=3,4} 
\theta(-P_{2s}+b) (P_{2s}-b) \,,  \ \  \qquad P_{2s}>0\,, s=3,4\,, P_{34}>0\,.
}}
 This   extended relation can be taken as a definition of the local correlator  in  a range of momenta,  larger than  the physical range  in which it is represented by the fixed chirality 
 correlators. 
Combining \recu\ with  the initial fixed chirality  relation \contgena\ 
we obtain alternatively, relabeling  the momenta,
\eqn\recua{\eqalign{
\hat G(P_1,P_2,P_3,P_4) 
&= \hat G^{-+++}(P_1,P_2,P_3,P_4)  
+\sum_{s=3,4} \theta(-P_{2s} ) \, P_{2s} \,, \cr
  &\ \  {\rm for } \  P_{2s}+b>0\,, s=3,4\,, \, P_{34}>0\,.
}}
The momenta of the  three  fixed chirality correlators in \recu\ 
are in  the  range,  in which these correlators coincide with the local correlators so that \recu\ 
can be also interpreted as a shift relation for local 
correlators.  Similarly one can derive  shift relations  and their duals in other regions of momenta,  ``neighbouring"  the physical range. In particular, the dual  of  \recu,  extending the second equation in   \difec\ for the case $m=0$,  
implies 
\eqn\recuaa{\eqalign{
\hat G(P_1,P_2,P_3,P_4) 
&= \hat G^{+---}(P_1,P_2,P_3,P_4)  
- \sum_{s=3,4} \theta(P_{2s} ) \, P_{2s} \,, \cr
  &\ \  {\rm for } \  P_{2s}-\frac{1}{b}< 0\,, s=3,4\,, \, P_{34}<0\,.
}}
We shall exploit all  these relations   in section 7 below.
\medskip
\subsec{\it  Distribution type  solution  of the two term ring  relations.}

  We can also interpret the solution of the ring relation in distribution sense,
accounting for  the charge conservation condition \neut\ by a $\delta$-function.
The correlators are
expressed in terms of the $p$-point ``multiplicities" for gaussian
matter \eqn\detp{\eqalign{ & N_{P_1,P_2, \dots ,P_p} =
N_p(P_1+P_2+\dots +P_p), }}
\eqn\detpa{
N_p(P)
=\delta(P-(p-2)e_0)\,,   \qquad p\ge 3\,.
}
We interpret the 3-point multiplicity $ N_{P_1,P_2,P_3}$ as the factor
modifying the generic 3-point constant \ttp{},  
i.e.
as the matter part of the 3-point correlator instead of \mnor.  It
satisfies the second relation in \recr\ and its dual, which are the
3-point ring relations in the absence of matter screening charges.
Now \fopbb\ is replaced by the integral representation
\eqn\fopbP{\eqalign{
&\hat \CG^{(\e)}(P_1| P_2,P_3,P_4) = 
 \hf \Big[ N_{P_1,P_2,P_3,P_4} \, Q
- \e  \< P\>_{P_1; P_2,P_3,P_4}
\Big]\,,
}}
where  
\eqn\chanfor{
  \< P\>_{P_1; P_2,P_3,P_4}:=    \int_{-\infty}^{\infty} dP
  \(N_{P_1,P_2, P} \, P \, N_{-P,P_3,P_4}
 + \Big\{\matrix{{\rm  permutations}\cr
          P_2\to P_3, P_4}\Big\} \)\,.
}

\noindent
The analog of the local  4-point function \fourpDK\  
is obtained  replacing $\e P \to |P|=|Q-2\a|$ in the three    channels in \chanfor. 
 \medskip
\rb    The ring relation \contgena\  remains a relation with two terms in the l.h.s. for the correlators  satisfying \neutgg\ with fixed $m=0$ and nontrivial $n\ne 0$. 
 A second  boundary condition 
is provided by the correlator with $\a_1=0, e_1=1/b$, $\!${\it i.e.$,\!$}
the negative chirality  field is given by the
dual matter charge $T^-_0$
\eqn\tbcond{\eqalign{
G_4^{(+)}(0,\a_2,\a_3,\a_4)_{\big| e_1=1/b}&= (-\frac{1}{b}\p_{\tilde \LM})G_3(\a_2,\a_3,\a_4)=-{1\over b}(n+1)\,\L^{{1\over
b}(Q-\a)}\,{\tilde
\LM}^{n}\,
}}
and similarly 
\eqn\tbcondd{
G_4^{(-)}(0,\a_2,\a_3,\a_4)_{\big| e_1=1/b}= (-b \p_{ \LM})G_3(\a_2,\a_3,\a_4)= -(m+1) b\,\L^{{1\over
b}(Q-\a)}\,\LM^{m}\,.
}
The  solutions  (to be  discussed in more detail below in section 7.  generalize \derivL\
 \eqn\twobcondf{\eqalign{
&G_4^{(+)}(\a_1,\a_2,\a_3,\a_4)= \L^{{1\over b}(Q- \a)} {\tilde \LM}^{n}\,  (n+1)(\a-Q-b+\frac{n}{b})\,
\cr
&G_4^{(-)}(\a_1,\a_2,\a_3,\a_4)= \L^{{1\over b}(Q- \a)} \LM^{m}\,  (m+1)(\a-Q-\frac{1}{b}+m b)\,.
}}

\subsec{\it  Solution  with   Liouville   charge conservation}


\noindent 
One can obtain similarly the solutions of the ring relations
in the absence of Liouville screening charges, so that $\sum_{s=1}^p
\a_s-Q=0$, or $\sum_{s=1}^p \e_s P_s =(p-2)Q$.  Such a constraint
again goes beyond the normalization assumptions which led us to
\ttp{}\ since \Lc\ is singular, i.e., it rather corresponds to  the Coulomb gas constant
 $
\hat{C}^{\rm
Liou}$ obtained as a residuum of \Lc. 
 Nevertheless the 
 final expression in \ttp{}\ satisfies the
($\L=0$) equations for the tachyon 3-point functions and can be taken
as the solution in this case.  Then the analog of \nsoll\ is given by
a derivative with respect to the matter constant $\LM$
\eqn\msoll{\eqalign{
 G^{  (+)}_p(\a_1,\a_2,\cdots,
\a_p)&= \(- b\,{\p _{\LM}}\)^{p-3} \LM^{-{1\over
b}(e_0-\sum_{i=1}^p e_i)+p-3}\,\cr  
}}
and a similar formula for the opposite chiralities.  
The normalized with  the partition functions $Z_M(\L, \LM,b)=-Z(\L, \LM,b)$  and $\tilde Z_M(\L, \LM,b)=-\tilde Z(\L, \LM,b)$  solution \msoll\  
and  its dual   are related to  the normalized correlators  \nsoll, 
by the matter-Liouville
duality  \dualmL, now  equivalent to $\{ b^2, \L, b\a, b e\} \to  \{- b^2, \LM, be, b\a\}$ . The analog of the formula \fopbb\ 
reads \eqn\mDiK{\eqalign{
 \hat G^{(\e)}_4(\a_1,\a_2,\a_3,\a_4)=
\e\big(e_0-e -\e b^{\e})=-\e\(\frac{e_0}{2} -\e\,
\sum_{s=2}^4(\a_1+\a_s-\frac{Q}{2})\)\,.
}} 
One can
introduce also  different  analogs of the ``local'' correlator \fourpDK, 
now symmetric with respect to the four Liouville momenta $\a_1 ,
\a_2,\a_3,\a_4$. 
This correlator can be used to define another "local" extension of the general ring relations, analogously to \recu. This case is however more speculative since  
we lack a selection rule of the type of Seiberg inequality and moreover
 we have no independent 
information on the generic  $c<1$  n-point correlators.  

 \newsec{ The 4-point function   for  fixed  number of  screening charges } 
 \subsec{The fixed chirality solutions}
 \noindent
In this section we analyse the difference equations \difec\  in the
case when the total sum of momenta is restricted by integer numbers of
matter screening charges as in \neutgg.  Reducing 
recursively with $m$ to the two term identities for $m=0$ 
discussed above 
one obtains
\eqn\rrgen{\eqalign{
 G_4^{(\e)}(\a; & m,n)-\L^{b^{\e-1}}\, G_4^{(\e)}(\a+b^{\e};m,n)
 \cr
&=\sum_{l=0}^m\sum_{k=0}^n \LM^l\, (\tilde \L\tilde \LM)^kG_3^{(\e)}(\a_1+{l-1\over 2 } b, \a_2+\a_3-{l-1\over 2 } b +{k\over b},\a_4;m- l,k)
\cr &
=-b^{\e}
 \, (m+1)(n+1)\,
\L^{{1\over b}(Q-\a)}\LM^{m}\tilde \LM^{n}\, ,
}}
 where $\a=\sum_{s=1}^4\a_s$ is as before  the total Liouville charge. The recursive equations for the 3-point functions are similarly reduced to 
 the second  identity in \recr\  and its dual. In the last line of \rrgen\  we have 
 inserted again the simplest  solution \ttp{}\ of these identities.  
 
 The equation \rrgen\  admits a
 solution generalizing  \twobcondf,  and which can be cast into a ``three
 channel'' expansion form, consistent with a $sl(2)\times sl(2)$ type
 decomposition rule:
 \eqn\twobcond{\eqalign{
&\hat G^{(\e)}(\a;m,n)=
 (m+1)(n+1)(\a-Q-b^{\e}+\frac{n}{b}\,\frac{1+\e}{2}+mb\,\frac{1-\e}{2})=\cr
&
 \hf(m+1)(n+1)(Q+mb +\frac{n}{b}) +{\e\over 2} \,\sum_{s\ne 1}
 \sum_{k=0}^{m}\,\sum_{l=0}^{n}\,
\big(P_1+P_s-e_0 + 2k b -\frac{2l}{b}\big)\, .
}}
Let us take for
definiteness $\e=+1$. The solution \twobcond\  reduces to \derivL \ for 
 $m=0=n$,  and to \twobcondf\ for $m=0$. 
For $n=0$ the correlator is compared with another solution of
\rrminus, namely the solution \msoll, \mDiK\ with Liouville charge
conservation $\sum_i\a_i-Q=\a-Q=0$, $\!${\it i.e.$,\!$} it is
consistent with a  third boundary condition given by the matter charge
$T^+_0$,\foot{The two
simultaneous restrictions on the matter and Liouville charges lead to
a value $P_1\in \CL$, which implies a  new accidental contact
term in \mpd, see 
section  A.3. The solution here is 
consistent with \mpd\ becoming a homogeneous relation.}

\eqn\sbcond{\eqalign{
&G^{(+)}(\a;m,0)_{\big|\a=Q}=-b(m+1)\LM^m
=(e_0- e-b)\LM^{{1\over
b}(e-e_0)}\,.
\cr
}}
The boundary conditions described 
do not fix uniquely the arbitrariness in  the solution 
 of the homogeneous difference equation -  a term   of the type
  $m n (m+1)(n+1) P(m,n)$ with an arbitrary polynomial $P(m,n)$ is still  allowed.
 As we shall see below,  \twobcond\  is the solution
 ``smoothly''  related to another class of solutions of \rrgen, the ones with one degenerate field,
   which are constructed recursively starting from a boundary value.

Let us summarize. We have imposed three boundary conditions corresponding to  derivatives of  
3-point correlators
with respect to $\L,\LM,\tilde \LM$ - for $\e=1$ (or  $\tilde \L,\LM,\tilde \LM$ - for $\e=-1$). These are the   coupling constants  in  a  three-term interaction
which includes 
one of the Liouville 
and  both matter  screening charges. 
The ``doubled'' matter interaction contributes perturbatively, i.e, with integer powers of the screening charges.

The duality transformation \dualtL\ exchanges the two solutions
\twobcond\ $\e=\pm1$.
Pairs of inhomogeneous 
 Liouville reflections  interchange  
 the solutions of different  chiralities,   generalising the relations \reflpro.   

\medskip 
\rb We  note that there is 
a special case involving a non-integer
number of screening charges.  
For $n=0, \e=1$ or for
$m=0, \e=-1$, the fixed chirality solutions \twobcond\ of the
functional equations \rrminus\ and \rrplus\ can be written in a form
which allows to extend them to arbitrary (non-integer) values of $m$
or $n$ respectively,
\eqn\gensol{\eqalign{
G^{(+)}_4&=\L^{{1\over b}(Q- \a)} \LM^{{1\over b}(e-e_0)}\,
 {1\over b}(e-e_0+b)(\a-Q-b)=
 -b\,\p_{\L}\,\p_{\LM}\L^{{1\over b}(Q- \a)+1} \LM^{{1\over b}(e-e_0)+1} 
\,\cr
G^{(-)}_4&= {\tilde \L}^{b(Q- \a)} {\tilde \LM}^{ b(e_0-e)} \,
 b(e_0-e+\frac{1}{b})(\a-Q-\frac{1}{b})=
 -\frac{1}{b}\,\p_{\tilde \L}\,\p_{\tilde \LM}
 {\tilde \L}^{ b(Q- \a)+1} {\tilde \LM}^{b(e_0-e)+1} 
\,.
}}

\subsec{Correlators satisfying the locality requirement}

Let us now look for  a  ``local'' 4-point function, symmetric in the
four matter charges, which reproduces for certain range of their
values the fixed chirality correlators \twobcond.  Now the symmetry constraint has many solutions.
We observe
that  the representation in the last line of \twobcond\  takes the form of \summar\
with  fusion multiplicities determined by the charge conservation condition, {\it i.e.}
$N(P_1,P_2,P_3)=1$ if $P_i$ satisfy $\sum_i P_i=2e_0-2kb +2\frac{l}{b}$ with some integers $k,l$
between $0$ and $m$ or $n$, and $N(P_1,P_2,P_3)=0$ otherwise. These are the fusion rules 
of the underlying local matter correlators of this type and it is  natural to solve the symmetry requirement, i.e., to determine the 
 correlator of ``local" tachyons  so that to 
preserve these fusion rules.
Of the two possible
such combinations we choose the one which reduces to \fourpDK\ for $m=0=n$, 
\eqn\chmodges{\eqalign{
&{\hat  \CG}
(P_1, P_2, P_3,P_4; m,n)=\cr
&=\hf (m+1)(n+1)(Q+mb+\frac{n}{b}) - \hf \sum_{s=2}^4
\sum_{k=0}^{m}\,\sum_{l=0}^{n}\, 
|P_1+P_s-e_0 + 2k b -\frac{2l}{b}| \cr
&= \hf (m+1)(n+1)(Q+mb+\frac{n}{b})
\cr
&\hskip 3cm - \sum_{s=2}^4
\sum^m_{_{{r=-m \atop {\rm mod}\ 2}}}\,
\sum^n_{_{{t=-n \atop  {\rm  mod} \ 2}}}
|e_1+e_s-\frac{e_0}{2}- \frac{mb}{2}+ \frac{n}{2b} - \frac{r
b}{2} -\frac{t}{2b}|\,.
}}
The symmetry under permutations of the momenta is ensured by the
charge conservation condition \neutgg.  
 The duality properties of the properly normalised correlators \chmodges\ are analogous to those 
in \fourpm ; to ensure that the transformation \dualtL\  and \dualtm\ become identical, we should include a power of $b$ under the modulus in \chmodges.
The shift equations satisfied by the  local correlators are derived from the explicit expression \chmodges.

We shall now give another argument in support of the formula \chmodges.

Clearly unlike the simplest example \fourpDK\  discussed in section 4, we now lack a complete ``atlas'' of fixed chirality solutions 
to match locally \chmodges\ in all regions of the momenta. Consider the case $n=0$. In the physical region   $P_{st} =P_s+P_t-e_0>0$, $s,t=2,3,4$ (or equivalently $P_{1 i}+2m b <0\,,  i =1,2,3 $)
 the local correlator is represented by 
 the solution   
 $G^{-+++}(P_1,P_2,P_3,P_4;m,0)$ in \twobcond. On the other hand we can use the extended identities as \recua\ to find a representation of the local correlator in the  "vicinity" of 
 any region described by the eight fixed chirality correlators. The identities \recu,  \recua\ imply
 that in the extended  range of momenta the shift  relation \rrgen\ is replaced by a relation for the local correlators, namely
 \eqn\recub{\eqalign{
-  \hat G(P_1,P_2,P_3,P_4;m,0) &+ \hat G(P_1+b,P_2-b,P_3,P_4;m,0) 
\cr = 
&2b +\sum_{s=3,4} \theta(-P_{2s} +b) \, (P_{2s}-b) \,, \qquad \ 
  {\rm for } \  P_{st}>0\,, s,t=2,3,4\,.    
 }}
The  
proposed correlators \chmodges\ 
do indeed satisfy \recua\ and the shift  equation \recub, as well as all other similar identities.
 In fact  these equations 
 determine 
completely  the local correlators for $m=1\,, n=0$   (or  the correlators for $m=0\,, n=1$), 
 taking also into account the inhomogeneous Liouville reflection
relations generalizing \reflpro. The solution is identical to the one prescribed by 
\chmodges.  Then  in the next  step we can use this solution, as we did in the case $m=0=n$,  in order to extend further the general ring relations as identities for the local correlators, generalising \recu. This in particular determines the 
local correlator in the  "next to the nearest" range, extending \recua,
\eqn\recuare{\eqalign{
\hat G&(P_1,P_2,P_3,P_4; m,0) 
= \hat G^{-+++}(P_1,P_2,P_3,P_4;m,0) \cr 
&+\sum_{s=3,4} \(\theta(-P_{2s} ) \, P_{2s} + \theta(-P_{2s} -2b) \, (P_{2s}+2b)\)\,, 
 \quad  {\rm for } \  P_{2s}+3b>0\,, s=3,4\,, \, P_{34}>0\,.
}}
In principle one  can reproduce  in this way recursively
the correlators \chmodges\  for the two thermal cases $n=0$, or $m=0$. Furthermore  we can 
combine the two types of shift relations.  Thus
 starting again from the local correlator $\hat G(P_1,P_2,P_3,P_4;1,0)$, which is represented by $\hat G^{+---}(P_1,P_2,P_3,P_4;1,0)$ in the range $\{P_{st}+2b<0\,, s=2,3,4\}$,
 we can compute the r.h.s. 
of the second  identity  in \difec\ and use the new contact terms  to extend this identity   for arbitrary $n$ and the fixed $m=1$,
\eqn\recudu{\eqalign{
&\hat G(P_1-\frac{1}{b},P_2+\frac{1}{b},P_3,P_4;1,n) - \hat G^{+---}(P_1,P_2,P_3,P_4;1,n) \cr
&+\hat G^{+---}(P_1-\frac{1}{b},P_2-\frac{1}{b},P_3,P_4;1,n-1)- \hat G^{+---}(P_1-\frac{2}{b},P_2,P_3,P_4;1,n-1) \cr
&=\frac{2}{b} -\sum_{s=3,4} 
(\theta(P_{2s}+\frac{1}{b}) (P_{2s}+\frac{1}{b}) +\theta(P_{2s}+2b+\frac{1}{b}) (P_{2s}+ 2b+\frac{1}{b})) \,, 
\  P_{st}+2b<0\,, s=2,3,4\,;  \cr
&{\rm or}, \cr
&\hat G(P_1,P_2,P_3,P_4; 1,n) 
= \hat G^{+---}(P_1,P_2,P_3,P_4;1,n) \cr 
&+\sum_{s=3,4} \(\theta(P_{2s}) \, P_{2s} + \theta(P_{2s} +2b) \, (P_{2s}+2b)\)\,, 
     \  {\rm for }  
  \  P_{2s}+2b-\frac{1}{b}<0\,, s=3,4\,,  P_{34}+2b<0\,,  \cr
}}
etc., confirming \chmodges.

One  finds also   symmetric with respect to the Liouville labels $\a_i$  tachyon correlators  - they preserve the fusion rules of the Coulomb gas  $c>25$ theory.

\subsec{ Distribution type solutions}

Furthermore  a distribution type solution 
  generalizing \fopbP\ is obtained by multiplying \twobcond\
 with $ \delta(P-2e_0 +2mb -2n/b)\,$ and summing over nonnegative
 $m,n$.  The $n$-point ``multiplicities'' are again distributions, 
 depending only on the total momentum $P$, but instead of \detpa\ they
 are given by semi-infinite  double sums of $\delta$-functions.  They
 are expressed in terms of the 3-point ``multiplicity''
\eqn\genmul{\eqalign{ N_3(P)=\sum_{m,n=0}^{\infty} \delta(P-e_0 +2mb
-2n/b).  }} It satisfies the two relations \homrel, while \recrr\ is
replaced by the difference identities
\eqn\genmula{\eqalign{
&N_{P_1- b,P_2,P_3} -N_{P_1+ b,P_2,P_3}
 = \sum_{n=0}^ {\infty}\delta(\sum_{i=1}^3 P_i -(2n+1)/b)\cr
&N_{P_1+ 1/b,P_2,P_3}-N_{P_1-1/b,P_2,P_3} = 
\sum_{m=0}^{\infty}\delta(\sum_{i=1}^3 P_i +(2m+1) b)\,.
}}
 We define the quantity $\langle P \ra$ as in \chanfor, but with the
 new 3-point multiplicity \genmula.  Using the properties of \genmul\
 one reproduces the functional relation
\eqn\contgen{\eqalign{
\langle P \rangle _{P_1 ; P_2,P_3,P_4}&+\langle P \rangle _{P_1+2b;
P_2,P_3,P_4}-
\langle P \rangle _{P_1+b; P_2 -\,b,P_3,P_4)}-\langle P \rangle _{P_1+b; P_2 +\,b,P_3,P_4)}\cr
&=2b \sum_{m,n=0}^{\infty}(n+1) \delta(\sum_i P_i-2 e_0 +2mb -2n/b)\, . 
}}
The identity is equivalent to
\contgena, when projected to a fixed sum of momenta, since the
irreducible part of the 4-point function satisfies the homogeneous
equation. A local correlator  with $\langle P \ra$ replaced by $\langle |P| \ra$ is also obtained.

\medskip
 
\rb Now let us turn to the diagonal theory defined by the action
\deltadiag.  We shall look for solutions for the 4-point function
assuming a ``diagonal'' ($m=n$) charge conservation condition \neutgg.
This leads to a single sum of $\delta$ functions representing the
3-point multiplicity
\eqn\thpmtdi{\eqalign{
 &N_{P_1,P_2,P_3}\equiv 
 N_3(P ) =
 \sum_{k=0 }^{\infty}\delta(P -(2k+1)e_0)  .
 }}
 The 4-point multiplicity is accordingly 
\eqn\fourd{
N_{P_1,P_2,P_3,P_4}\equiv \int_{-\infty}^\infty dP\, 
N_{P_1, P_2, -P}N_{P, P_3, P_4}=
 \sum_{m=0}^{\infty}\, (m+1)\,
 \delta(\sum_{i=1}^4 P_i -2(m+1) e_0)\, .
}
 Instead of \contgen\ one obtains
\eqn\newcont{\eqalign{&
\sum_{\sigma=\pm 1}    \< P\>_{P_1+\s  e_0; P_2,P_3,P_4} -
\sum_{\sigma=\pm} \< P\>_{P_1; P_2+\s e_0,P_3,P_4} 
=-2e_0\ 
 N_3(P_1+P_2+P_3+P_4)\, .
}}
The the r.h.s. of \fopbP, now with the multiplicities defined in
\thpmtdi, \fourd, provides a solution of \rrplupdia\foot{The
arbitrariness in the diagonal case is fixed comparing the first term,
$m=0$, with \fopbP, and furthermore, with the solutions with one
degenerate field, to be discussed in the next section.},
\eqn\chang{\eqalign{
&{\hat  \CG}_4^{(\e)}(P_1|P_2,  P_3,P_4)
=
{1\over 2
}\,
   \(Q\, N_{P_1,P_2,P_3,P_4}- \e\,\< P\>_{P_1; P_2,P_3,P_4}\)\cr
&={1\over 2  
}\sum_{m=0}^{\infty}\, \[Q (m+1)
+\e\,\sum_{s=2}^4 \sum_{k=0}^{m}\,
(P_1+P_s-(2k +1) e_0)\] \delta(\sum_{i=1}^4 P_i -2(m+1) e_0) \cr
&= 
\sum_{m=0}^{\infty}\, (m+1)
\big(\sum_i \a_i -Q-b^{\e} +\e\,m\, \frac{e_0}{2}\big)\delta(\sum_{i=1}^4
P_i -2(m+1) e_0) \,.
}}
Note that in contrast with \chang, in the non-diagonal theory the
solution cannot be expressed entirely in terms of the 3-point
``multiplicities'' and the inverse 2-point correlator due to the more
complicated form of the 1pi part.

From \chang\  one extrapolates the symmetric  correlator
\eqn\changmod{\eqalign{
&{\hat  \CG}
(P_1, P_2, P_3,P_4; m)
= \frac{Q}{2}(m+1) - \hf \sum_{s=2}^4
\sum^m_{_{r=-m \atop  {{\rm  mod} \ 2}}}\,
|P_1+P_s- (m+1) e_0 -r e_0|\,.
}}
 In this case there is no underlying  local matter theory to compare with, rather we preserve the fusion rules \thpmtdi.  A  formula of this type is reproduced in the microscopic approach in
\KPb, with the delta-functions replaced by periodic delta's.
 
\medskip

\rb 
The ``multiplicities'' introduced in this section are considered for
real momenta only, but they can be expressed in terms of meromorphic
functions defined in the whole complex plane.  Thus the 3-point
multiplicity \genmula\ is given by the discontinuity on the real axis
of a meromorphic function,
\eqn\genmuld{\eqalign{
N_3(P)={1\over 4\pi i}\, 
\Big( f(\frac{e_0-P}{ 2}+i0)-  f(\frac{e_0-P}{2}-i0)\Big),\cr
}}
namely the logarithmic derivative of the double $\G$-function
\eqn\derG{
f(z)
\equiv -\partial_z \log \G_b(z+b)
=-\int_0^\infty\, dt\( {e^{-zt}\over (1-e^{ bt})(1-e^{- t/b})}
-{z-{e_0\over 2}\over t} e^{-t}+{1\over t^2}\)
.
}
The diagonal  
multiplicity 
\thpmtdi\  is expressed as the discontinuity of 
$\psi(z)=\partial_z \log \Gamma(z).$ 
   

\newsec{Degenerate fields in the  diagonal theory}


\noindent 
The most interesting correlation functions, especially from
the point of view of comparing with the microscopic theory, are those
involving four degenerate fields.  In this section we solve the
difference equations for the spectrum of momenta corresponding to the
degenerate matter fields (order operators) in the diagonal theory.
This spectrum is given by the diagonal $e_0 \IZ$ of the grid \Latt,
with the point $P=0$ excluded.  We will assume that there is no
tachyon with $P=0$, $\!${\it i.e.$,\!$} the 
 tachyon
correlation functions vanish if one of the momenta is zero.

\rb To begin with, we will find the solutions of the diagonal ring
relations with one degenerate field $P_2\!=\!  (m+1)e_0, $ $\a_2=b
-\frac{m e_0}{2}$, and three generic.  As
 initial condition we take
\eqn\bcond{
   \hat \CG_4^{(+)}(P_1, e_0,P_3,P_4)=N_{m_1,m_3,m_4}
(\sum_{s\ne 2}\a_s -Q)=(\sum_{s\ne 2}\a_s -Q)\,,
}
where the generic solution with $N_{P_1,P_2,P_3}=1$ for the 3-point
correlator is inserted.  We solve \rrplupdia\ recursively, under the
assumption that at the point $P_2=0$ the correlator vanishes.  This is
achieved automatically if the fields are interpreted as linear
combinations of vertex operators antisymmetric under a composition of
matter and Liouville reflections:
\eqn\compos{\eqalign{
&\CV^{(\e, A)}_{P} =
\CV^{\e}_{P} -  \L^{{\e
P\over b}}\LM^{-{P\over b}}
\,\CV^{\e}_{-P} =- \L^{{\e P\over b}}\LM^{-{P\over b}}
\,{\CV}^{(\e, A)}_{-P}\,.
}}
Since the degenerate field is assumed anti-symmetric, the contact
terms cancel, as the generic solution \ttp{}\ satisfies the reflection
identity \drefl.  We get
\eqn\onedeg{\eqalign{
\hG_4^{(+)}(\a_1,\a_2,\a_3,\a_4) &=(m+1)(\a-Q-b + {e_0\over 2} m)\cr
&= \frac{Q}{2}(m+1) -  \sum_{s\ne2}
\sum^m_{_{r=-m \atop  {{\rm  mod} \ 2}}}\,
\hf (\e_s\,P_s -    r\,e_0)\,.
}}
This expression has the form \summar, with trivial multiplicities
$N_{P,P_s,P_r}=1 $ for $s,r\ne 2$ and a non-trivial multiplicity
$N_{P_t, P_2=(m+1)e_0,P}$, representing a continuation of the $sl(2)$
decomposition rule \sltwo\ to non-integer isospins; the shifts by $r
e_0$ in \onedeg\ correspond to the weight diagram of the irrep of
dimension $m+1$.  The solution \onedeg\ also justifies the choice of
the linear combination in \chang.  The meaning of the nonnegative
integer $m$ in the two types of solutions is different, but in both
cases $m+1$ counts the number of intermediate contributions in each
channel.  Projecting \chang\ to a fixed charge $m$ and inserting the
value $P_2=(m+1) e_0$ reproduces \onedeg.

\medskip

\rb Now let us consider correlators in which all tachyons correspond
to degenerate 
 fields, $P_i=\e_i m_i e_0$, $m_i \in
\IN$, $\!${\it i.e.$,\!$} $\a_i={Q\over 2}-{m_i e_0\over 2}$.  These
tachyons satisfy fusion rules given by the $sl(2)$ decomposition
multiplicity \sltwo, which is also expressed by an integral, in
general
\eqn\Nint{
      N_{m_1,..., m_p}=\frac{1}{\pi}
       \int _0^{2\pi} d\theta    \sin^2 \theta\ \prod_{i=1}^p
\frac{\sin(m_i\theta)}{\sin\theta}\,,
  }
in terms of the characters $\chi_{m_i}(\theta)=
\frac{\sin(m_i\theta)}{\sin\theta} $.
These multiplicities preserve the homogeneous identity, implied by
\homreldi, with respect to any pair of variables:
\eqn\homid{
N_{m_1+ 1,m_2,\dots, m_p}+N_{m_1-1,\dots, m_p}=
N_{m_1,m_2+1,\dots, m_p}+N_{m_1,m_2- 1,\dots, m_p}\,.
}
They are symmetric under permutations and extend to arbitrary integer
values of the weights $m_i$ by the (shifted) Weyl reflection property
$$
  N_{-m_1, m_2,..., m_p}=-N_{m_1, m_2,..., m_p},
$$
so that they vanish if some $m_i=0=P_i$. 
 
We start with a simple example in which $P_2= 2 e_0$, illustrating the
recursive determination of the 4-point correlators.  As an ``initial''
condition we take again the correlator in the first equality in
\bcond, but instead of the generic solution $N_{m_1,m_3,m_4}=1$ we
take the $sl(2)$ 3-point multiplicity \pardiag.  Up to the contact
terms, which we will neglect at this stage, eqn.  \rrplupdia\ gives
for $\hat \CG_4^{(+)}(P_1|2e_0, P_3,P_4)=\hat G_4(\a_1, b- \hf e_0,
\a_3, \a_4)$

\eqn\rectw{\eqalign{
&\hat G_4(\a_1,  b-  \hf  e_0, \a_3, \a_4)
=\sum_{\s=\pm 1} N_{m_1+\s, m_3, m_4}(\a -Q-b
+e_0\frac{1-\s}{ 2})+\dots \cr
&=N_{m_1, 2, m_3, m_4}(\a-Q-b )
+(e_0 N_{m_1-1,m_3, m_4}+\dots )\,.
}}
To obtain the first term in the second line we have used the
homogeneous relation \homid\ for the 4-point multiplicity, applied for
$m_2=1$ and using that
 $N_{m_1,  m_3, m_4}=N_{m_1,1,  m_3, m_4}$. 
 The result
should be symmetric with respect to $m_1,m_3,m_4$, so instead of the
incomplete second term in the last line of \rectw\ we should have a
symmetric expression, which vanishes if some $m_s=1$, recovering
\bcond.  A solution to these conditions is given by
\eqn\frecd{\eqalign{
& \hat G_4(\a_1,  b-  \hf  e_0, \a_3, \a_4)=\ N_{m_1,2, m_3,m_4}\(\a-Q-b
+{e_0\over 2}(N_{m_1, 2, m_3,m_4}-1)\)
.\cr
}}
The normalization of the added term is fixed to $+\frac{e_0}{2}$,
since generically $ N_{m_1,2, m_3,m_4}=2$ and this is in agreement
with our previous solution \onedeg\ taken for $m+1=m_2=2$.  In the
next step of the recursion we take $P_2=3e_0$ and use the result in
\frecd.  Once again we recover the first term $N_{m_1,3,
m_3,m_4}(\a-Q-b)$ uniquely, while we get an expression for the second
term which is not symmetric, and generically should be equal to
$2e_0=e_0(N_{m_1,3, m_3,m_4}-1)$, if compared with \onedeg.  The end result is
a formula in which $N_{m_1,2, m_3,m_4}$ in \frecd\ is replaced by
$N_{m_1,m_2,m_3,m_4}$.  This formula can be cast in the form
\eqna\fpconj
$$\eqalignno{
&\hat
G_4^{(+)}(m_1,m_2,m_3,m_4)= & \fpconj {} \cr
&
{1\over 2
}\,
\(Q\, N_{m_1,m_2,m_3,m_4}\,  -
\sum_{m=1}(N_{m_1,m_2,m} \,  (m e_0)\,
N_{m,m_3,m_4}  +   {\rm permutations})\)\cr
&=
  N_{m_1,m_2,m_3,m_4}\(\sum_i\a_i -Q-b
+ {e_0\over 2} (N_{m_1,m_2,m_3,m_4}-1)\)\,, \quad
\a_i=\frac{Q}{2}- m_i\frac{e_0}{2}\,.
}$$
To connect the two expressions in the second and the third lines we
have used the relation
\eqn\linrel{
\sum_{m=0} (N_{m_1,m_2,m} \, m \, N_{m,m_3,m_4}
  +   {\rm permutations})
= N_{m_1,m_2,m_3,m_4}
\Big(\sum_{i=1}^4 \, m_i -N_{m_1,m_2,m_3,m_4}\Big)\,.
}
This identity has a purely group theoretical formulation being
expressed in terms of the $sl(2)$ tensor product decomposition
multiplicities \sltwo\ and the dimensions $m_i$ of the irreps.  It is
derived using the definition \sltwo.  We stress that by construction
the diagonal degenerate fields satisfy closed fusion algebra, in
contrast with the standard $c<1$ matter quasi-rational theory.

\medskip

\noindent

\rb  We 
obtained recursively the solution \fpconj{}\ from the difference
equations \rrplupdia \ without referring to the exact form of the
contact terms.  Instead, we strongly used the expected symmetries of
the solution and the requirement that 
whenever
$N_{m_1m_2m_3m_4}=m_2$, the solution coincides with \onedeg, derived
for one degenerate and three generic momenta.  In the particular case
$P_2=(m+1) e_0=P_2=P_4=-P_1$ in which $\sum_iP_i=2(m+1) e_0$ all the
three formulae \fpconj{}, \onedeg\ and \chang\ coincide.

We shall now show that \fpconj{}\ satisfies the difference relations
\rrplupdia, but with a contact term proportional to the difference of
two $sl(2)$ multiplicities,
\eqn\contt{
[N]_{m_1,m_2+m_3,m_4}:=
N_{m_1, m_2+m_3, m_4}- N_{m_1,|m_2-m_3|,m_4} \, .
}
The quantity $[N]$ can take values $0,\pm 1$.  The second term
reflects the interpretation \compos\ of the fields.  Indeed, if we
represent the fields as in \compos\ and assume that the 3-point
functions of the initial fields are given by \sltwo, we have finally
to retain two of the four resulting contact terms -- namely the ones
with positive labels, as they appear in \contt.

To prove the above statement  we
need some identities for the $sl(2)$ multiplicities. In particular
we shall exploit
\eqn\tiden{
N_{m_1+ 1,m_2,m_3}-N_{m_1- 1,m_2,m_3}
=N_{m_1, |m_2-m_3|,1}-N_{m_1, m_2+m_3,1}\,.
}
This identity -- the r.h.s. of which represents the ``deviation'' from
the simpler relation in \recr{}, is derived using the general integral
representation \Nint; two of the initially four terms in the r.h.s.
survive, as in \tiden, when the equality is restricted to positive
indices, $\!${\it i.e.$,\!$} when the multiplicities of the l.h.s. are
given by \sltwo, as we assume throughout this section.  Applying
\tiden\ to both sides of the following equality
\eqn\fiden{\eqalign{
&\sum_{m=0} N_{m_1,m_3,m}(N_{m + 1,m_2,m_4}-N_{m- 1,m_2,m_4})=\cr
&-\sum_{m=0} (N_{m+ 1,m_1,m_3}-
N_{m- 1,m_1,m_3}) N_{m,m_2,m_4}\,
}}
we obtain
\eqn\fide{
\sum_m N_{m_1,m_3,m}(N_{m + 1,m_2,m_4}-N_{m- 1,m_2,m_4})=
- [N]_{m_2+m_4,m_1,m_3}=
[N]_{m_1+m_3,m_2,m_4}\,.
}
If $m_1\ge m_s\,, s=2,3,4$, the linear combination in  \contt, \fide\
is symmetric
with respect to the three variables $m_2,m_3,m_4$. Indeed in this
case
\eqn\newsymta{\eqalign{
[N]_{m_1+m_3,m_2,m_4}& =\frac{1}{\pi}
       \int _0^{2\pi} d\theta  \,
\frac{\cos m_1 \theta}{\sin\theta} \prod_{s=2,3,4}
\sin m_s\theta\, 
\cr
&=-
[N]_{m_1,m_2+m_4,m_3}=-
[N]_{m_1,m_2+m_3,m_4}=
-[N]_{m_1,m_4+m_3,m_2}\,.
}}
Thus choosing the largest of the labels $m_i$, say, $m_1$, as the one
corresponding to the negative chirality $\e_1=-1$, we arrive at the
symmetry relation \newsymta\ of the type of \symta.

We shall now check that \fpconj{}\ satisfies the ring relation with
the contact term given by the linear combination \contt.  Indeed if we
compute the shifts of the function \fpconj{}\ -- interpreted as $\hat
\CG^{(+)}(P_1|P_2,P_3,P_4)$ -- we get, using \homid, \fide, \newsymta,
\eqn\degrel{\eqalign{&
\sum_{\s=\pm 1}\hat  G_4^{(+)}(m_1+ \s, m_2,m_3,m_4)-
\sum_{\s=\pm 1}\hat  G_4^{(+)}(m_1, m_2+\s,m_3,m_4)\cr
&\qquad =
\frac{e_0}{ 2}\,
\([N]_{m_2+m_3,m_1,m_4}+
[N]_{m_2+m_4,m_1,m_3}\)
= e_0
[N]_{m_2+m_3,m_1,m_4}\,,\cr
& \cr
&\sum_{\s= \pm 1}\hat  G_4^{(+)}(m_1, m_2+\s, m_3,m_4)-
\sum_{\s=\pm 1}\hat  G_4^{(+)}(m_1,m_2, m_3+\s,m_4);
\cr
&\qquad =
\frac{e_0}{ 2}\,
\(
[N]_{m_1, m_3+m_4,m_2}
- [N]_{{m_1,m_2+m_4,m_3}}\)
=0.
}}
  The r.h.s of the second relation in \degrel\ vanishes due to
  \newsymta,
    so that it takes the form of the diagonal version of the homogeneous
  relation \rrminhom.  The two terms in the r.h.s of the
  first relation are identical and sum up to one term (which can now
  take the values $0,1$).  We stress that these identities hold in the
  region of validity of \newsymta, $\!${\it i.e.$,\!$} when the field
  of negative chirality is chosen to coincide with the largest of the
  integers $m_i$.  Otherwise the formula \fpconj{}\ is symmetric with
  respect to the four labels.  Eq.  \degrel\ is an analog of the
 formula \fourpDK\ in the sense that,
  similarly to \fourpDK, it reproduces solutions of the ring relations
  with $\sum_i\e_i= 2$ in certain regions of momenta ($\!${\it
  i.e.$,\!$} it does not distinguish the negative chirality sign
  unless we specify which $m_i$ is bigger.) What simplifies here the correlator and the
  shift equations is that the various local  regions are determined by the individial momenta and furthermore
  the intermediate momenta all have an identical sign. 

\medskip 
\rb The first line in \fpconj{}\ extends to negative $m_s$,
so that $ G_4^{(+)}(m_1,m_2,m_3, -m_4)=-(\frac{\LM}{\L})^{e_0\over
b}\,G_4^{(+)}(m_1,m_2,m_3, m_4)$.  The values $-m_4=-1, m_4=1$
correspond to $T^+_{1/b}$, $T_{b}^+$ respectively.  Restoring the prefactor
$-b/e_0$ in \diagresc\ we can write
\eqn\derdiag{\eqalign{
(-b\p_{\L} )\, &\L^{{1\over b}(Q-\sum_{s=1}^3\a_s)}\,
N_{m_1,m_2,m_3} =-\frac{e_0}{b}\,G_4^{(+)}(m_1,m_2,m_3,1)\cr
&=G_4^{(+)}(m_1,m_2,m_3, 1)+
\frac{1}{b^2}\,(\frac{\L}{\LM})^{e_0\over b}\,
G_4^{(+)}(m_1,m_2,m_3, -1)\,.
}}
We can interpret \derdiag\ as a ``boundary'' condition obtained from
the first two of the four terms in the diagonal action \deltadiag; the
differentiation with respect to $\L$ gives the linear combination in
the second line in \derdiag\ 
(taken with a prefactor $1/b$ due to the rescaling in \renormal). 
 In that sense the action defining our
correlators is given by the two positive chirality terms of the
diagonal perturbation \deltadiag.
 
\medskip

\rb The 4-point correlator with $\sum_i\e_i=-2$ is constructed in a
similar way, parametrizing the momenta as $P_i=-\e_i m_i e_0$ (so that
they are physical for $e_0<0$), \eqna\fpconjd
$$\eqalignno{
&\hat 
G_4^{(-)}(m_1,m_2,m_3,m_4):= & \fpconjd {} \cr
& 
{1\over 2}\,
\( Q\, N_{m_1,m_2,m_3,m_4}  - \sum_{m=0}(N_{m_1,m_2,m} \,  (-m e_0)\,
N_{m,m_3,m_4}
  +   {\rm permutations})\)\cr
&=
N_{m_1,m_2,m_3,m_4}\(\sum_i\a_i-Q-\frac{1}{b}
- {e_0\over 2}
(N_{m_1,m_2,m_3,m_4}-1)\)\,, \quad \a_i=\frac{Q}{2} +m_i\frac{e_0}{2}\,.
}$$
Then 
\eqn\derdiagd{\eqalign{
(-{1\over b}\p_{\tilde \L}) &{\tilde \L}^{ b(Q-\sum_{s=1}^3 \a_s)}\,
N_{m_1,m_2,m_3}= e_0 b\, G_4^{(-)}(m_1,m_2,m_3,1)\cr
&=G_4^{(-)}(m_1,m_2,m_3, 1)+b^2 \,(\frac{\tilde \L}{\tilde \LM})^{-e_0\, b}\,
G_4^{(-)}(m_1,m_2,m_3, -1)
}}
 can be interpreted as a boundary value related to the two negative
 chirality terms in \deltadiag.  
 \medskip

The solution 
\fpconj{}\  with $b e_0>0$ (or \fpconjd{}\  with $b e_0<0$)
reproduces the 4-point
correlation function
 of the microscopic  model \KPb.  Duality interchanges the  normalised with \partfa\
 correlators; effectively both \fpconj{}\ and \fpconjd{}\  get multiplied by $Q$ times the standard powers of $\L,\LM$. 
 On the other hand the transformations \dualmL\ lead to correlators of
the same type, in which the positive integers $m_i=2j_i+1$ parametrize
the diagonal degenerate Liouville points.  ($\!${\it i.e.$,\!$}
$Q-2\a= (2j+1) Q$, so that now the Liouville scaling dimension takes a
``Sugawara'' form $\triangle_L=-j(j+1)Q^2$).  These correlators are
solutions of a ring relation computed with the dual diagonal action
\sinL.

 \newsec{   Degenerate fields in the conventional theory}

\noindent
As we have discussed, in order to extend all ring relations 
to the whole lattice $\CL$,  one needs to know all
possible additional contact terms.  On the other hand when only
one of the tachyons in the correlator is degenerate, solving some of
the ring equations already determines the unique solution.  \medskip

\rb {\it One degenerate, three generic fields}

\medskip
 It will be convenient to shift the notation compared with
\degen, so that the matter degenerate momenta $P=e_0-mb+n/b$ are
parametrized by nonnegative integers $ m, n\in \IZ_{\ge 0}$.  We take
$\CW_{P_2}^{(\e)}$ as the degenerate tachyon, while the momenta of the
remaining three operators are assumed generic.  According to the
analysis in Appendix A.3.  there are no additional unknown contact
terms in this case.  We shall solve recursively the equations,
assuming that the tachyons at the border lines $n=-1$ and $m=-1$
have vanishing correlators.

Let us start with the ``thermal'' cases $n=0$, or $m=0$.  As before we
take as initial conditions 
\eqn\initco{
\hG_4^{(\e)}(\a_1,b^{\e},\a_3,\a_4)=\sum_{s\ne 2}\a_s-Q= \a-Q -b^\e\,.
}
 Solving 
recursively \rrminus\ we obtain
\eqn\ntexath{\eqalign{
& \hG_4^{(+)}(\a_1,\a_2=b+\frac{mb}{2}, \a_3,\a_4)
 = \sum^m_{{r=-m \atop {\rm mod}\ 2}}\,
\hG_4(\a_1-r\frac{b}{2},\a_2=b, \a_3,\a_4)
+(m+1)\frac{mb}{2}\cr
}}
which can be also rewritten as  
\eqn\exa{\eqalign{
\hat G_4^{(+)}(\a_1,\a_2=b +\frac{m b}{
2},\a_3,\a_4)&=
(m+1)(\sum_{s\ne 2}\alpha_s-Q +\frac{m b}{2})
=(m+1)(\alpha-Q -b)\cr
&=
(m+1)(\frac{Q}{2}+\frac{mb}{2})- \sum_{s\ne 2}
\sum_{k=0}^m (\frac{Q}{2}-\a_s -\frac{mb}{2} +kb)\,.
}}
Similarly \rrplus\  gives 
\eqn\exad{\eqalign{
\hat G_4^{(-)}(\a_1,\a_2=\frac{1}{b} +\frac{n}{2 b},\a_3,\a_4)&=
(n+1)(\sum_{s\ne 2}\alpha_s-Q +\frac{n}{2b})=(n+1)(\alpha-Q -\frac{1}{b})\cr
&=
(n+1)(\frac{Q}{2}+\frac{n}{2b})- \sum_{s\ne 2}
\sum_{k=0}^m (\frac{Q}{2}-\a_s -\frac{n}{2b} +kb)\,.
}}
The $m+1$ or $n+1$ terms in each of the three channels of the
above expansions correspond to the weight diagram of the $sl(2)$
irreps of dimension $m+1$, or $n+1$.

In deriving these formulae we have used only one of the
ring relations. In the  other channels one has to take into account
the additional
contact terms. 
If the degenerate field is represented by an
integrated tachyon,  then the 
accidental contact term due to \opeonetde, taken for $n=0$, precisely compensates the generic
one. 
 Indeed \exa, \exad\ satisfy
homogeneous relations, e.g. \foot{On the other hand 
 the homogeneous equation for the correlator of type $^{++++}$ is solved by
 $G^{++++}(\a_1,\a_2=b+\frac{mb}{2},\a_3,\a_4) 
 =(m+1)(\sum_{\ne 2}\a_s-Q) 
 $.}
$$\sum_{\pm} \hat G^{(+)}_4(\a_1\pm \frac{b}{2},\a_2,\a_3,\a_4)=
\sum_{\pm} \hat G^{(+)}_4(\a_1,\a_2,\a_3\pm \frac{b}{2},\a_4)\,.
$$

We can compare these solutions with the ones in \twobcond, extending
the latter to the values $P_2=e_0-mb$ or $P_2=e_0 +n/b$ for $\e=\pm 1$
respectively.  For these special values \twobcond\ coincides with
\exa,  or  \exad, and this justifies the choice in the 1pi-term in
\twobcond, obtained by a different argument.  \medskip
 
Now let us consider an arbitrary degenerate momentum $P_2=e_0-mb+n/b$.
 Solving \contgena\  recursively with $m$
we get instead of \ntexath\
\eqn\ntexa{\eqalign{
& \hG_4^{(+)}(\a_1,\a_2=b+\frac{mb}{2}-\frac{n}{2b}, \a_3,\a_4)
 =\cr
&\sum^m_{{r=-m \atop {\rm mod}\ 2}}\,
\hG_4^{(+)}(\a_1-r\frac{b}{2},\a_2=b-\frac{n}{2b}, \a_3,\a_4)
+(n+1)(m+1)\frac{mb}{2} .
}}
To proceed further we need to identify $\hG_4^{(+)}$ with a  negative
chirality correlator \exad\ with the same value of the degenerate
momentum $P_2$.  We choose
\eqn\whythis{\eqalign{
&\hG_4^{(+)}(\a_1,b-\frac{n}{2b}, \a_3,\a_4)\equiv
\hG_4^{(-)}(\a_1,\frac{1}{b}+\frac{n}{2b}, \a_3,\a_4)
\cr
&\hG_4^{(-)}(\a_1,\frac{1}{b}-\frac{m b}{2}, \a_3,\a_4)\equiv
\hG_4^{(+)}(\a_1, b+\frac{m}{2}, \a_3,\a_4)
\cr
}}
so that in particular  the initial condition \initco\ for $n=0=m$ is preserved.
 Inserting \whythis\ in \ntexa\ and using the first equalities in \exad, \exa,
we obtain
\eqn\ntexab{\eqalign{
&\hG_4^{(\e)}(\a_1,\a_2=b^{\e}+\e(\frac{mb}{2}-\frac{n}{2b}), \a_3,\a_4)=
(n+1)(m+1)\big(\a -Q-b^{\e} + \frac{n}{b}\frac{1+\e}{2}+m b \frac{1-\e}{2} 
\big)\cr
&= (n+1)(m+1)\big(\frac{Q}{2}+\frac{mb}{2}  +\frac{n}{2b}\big)
- \sum_{s\ne 2}
\sum^m_{{r=-m \atop {\rm mod}\ 2}}\,
\sum^n_{{t=-n \atop  {{\rm  mod} \ 2}}}
 (\frac{Q}{2}-\a_s     + \frac{r b}{2}+\frac{t}{2b})
 \,.
}}
The identification \whythis\ 
 is
suggested by the comparison with \twobcond\ - the latter coincides
with \ntexab\ if $P_2=e_0+n/b-mb$.  

The second line of \ntexab\  illustrates  the general form \summar.  The first fusion multiplicity 
corresponds to the shift of $P_s$ with the weight diagram of the degenerate field,  $\!${\it i.e. $\!$}
$P=P_s+P_2-e_0 +2k b-2l/b\,, k=0,\dots m\,, l=0,\dots n$, while the  multiplicity depending on three generic momenta corresponds to the trivial solution \ttp{}. As in section 7 we shall choose a solution
of the symmetry requirement preserving these fusion rules since once again they 
correspond to the fusion rules of the underlying local matter correlator. We obtain
 a  symmetric in the three generic momenta $P_s$
formula  
\eqn\exach{\eqalign{
&\hat \CG_4(P_1,P_2=e_0+n/b-mb,
P_3,P_4)  = \cr
& \qquad \hf\((n+1)(m+1)(Q+mb   +\frac{n}{b})
  -  \sum_{s\ne 2}
\sum^m_{{r=-m \atop {\rm mod}\ 2}}\,
\sum^n_{{t=-n \atop {\rm mod}\ 2}}\,
|P_s + rb +\frac{t}{b}|\)
\,.
}}

To check this result let us analyse directly the equation \recu\ for the local correlators similarly as we did in section 7. We  rewrite \recu\   as  
\eqn\recudeg{\eqalign{
&\hat G(P_1,P_2-b=e_0-m b,P_3,P_4)= \sum_{\pm}\hat G(P_1\pm b ,P_2,P_3,P_4) 
-\hat G(P_1,P_2+b,P_3,P_4)\cr
&+b +\sum_{s=3,4} 
\theta(-P_{2s}+b) (P_{2s}-b) \,,  \ \  \qquad P_{2s}=P_s-(m-1)b >0\,, s=3,4\,. 
}}
We shall illustrate  this identity  for $m=1$ in which case the last correlator corresponding to the border momentum $P_2=e_0+b$ drops.   We start with the local counterpart of   \initco\ as an initial condition  
  $$
\hat \CG(P_1,P_2=e_0,P_3,P_4) = \frac{Q}{2} - \hf \sum_{s\ne 2}  |P_s|\,.
$$
Then 
\eqn\recuc{\eqalign{
&\hat \CG(P_1,e_0-b,P_3,P_4) =\sum_{\pm} \hat \CG(P_1\pm b,e_0,P_3,P_4)  +b -\sum_{s=3,4}
\theta(-P_{2s}+b) |P_{2s}-b| \cr
&= Q+b  -\hf |P_1-b|-\hf |P_1+b|   - \hf \sum_{s=3,4}\( (P_s+b)  + |P_s-b|  (\theta(P_s-b)+\theta(-P_s+b))\)\cr
&=Q+b  -\hf \sum_{s\ne 2}( |P_s +b| +|P_s-b|) \,, \ \qquad  \ {\rm for} \  P_3, P_4 >0\, }}
thus   confirming formula \exach\   for this particular example.

The fixed chirality formulae \exa, \exad\ were presented in \KPlet{}.
The 
physical correlator \exach\ reproduces the expression
found by a different method in \BZ, in which the locality of the underlying 
correlators
is  automatically taken into account.

Similarly one  solves the equations in the case when
one of the tachyons is Liouville degenerate.

\medskip
\rb {\it Four  degenerate fields -- a conjecture}
\medskip

When all four fields are labelled by degenerate matter representations the 
3-point 
 function $\hat G_3=1$
is to be replaced by the fusion multiplicity in \sltwo, \sltwopr.  Accordingly
the initial value  \initco\  gets    multiplied by this multiplicity.
The equations themselves get more complicated
due to many additional contact terms and the possible cancellations
between them. We 
conjecture that the effect will be, like in the diagonal case, an
expression in which the 3-point $sl(2)$ fusion multiplicities
\sltwopr\ determine the expansion range, while the factors $n+1$ and
$m+1$ in \ntexab\ are replaced by the 4-point sl(2) multiplicities in \fourpmul,
symmetric under the change of sign of any of the momenta.  For $\a_i =
\frac{Q}{2} -\e (\frac{n_i}{2b}-\frac{m_i b}{2})\,, i=1,2,3,4$ and
$P_{m,n}=n/b-mb$, $ n_i,m_i,n,m \in \IN$,  this leads to
\eqn\conjec{\eqalign{
&{\hat G}_4^{(\e)}(\a_1,\a_2,\a_3,\a_4)
= \hf 
[ N_{P_1,P_2,P_3, P_4} \(b N_{m_1,m_2,m_3,m_4}
+\frac{1}{b}\,N_{n_1,n_2,n_3,n_4}\)\cr
& \ \  \qquad \  - \e\, \sum_{m,n=1} \(N_{P_1,P_2, P_{m,n}}
 \,  (\frac{n}{b}-m b)\,
N_{P_{m,n},P_3,P_4} +   {\rm permutations} \{2,3,4\}\)]\cr
&= N_{P_1,P_2,P_3, P_4}
\(\a-Q- b^{\e} +\frac{1-\e}{ 2} b (N_{m_1,m_2,m_3,m_4} -1)
+\frac{1+\e}{ 2b}  \,
(N_{n_1,n_2,n_3,n_4}-1)\)\,.
}}
 In the last equality we  used \linrel. 
The conjectured local 
correlator  
is given by a formula as in the first line of \conjec\ with intermediate momenta
$\e(\frac{n}{b}-m b)$ replaced by $|\frac{n}{b}-m b|$.

\newsec{Summary and discussion}

\noindent
 In this paper we reported the results of our study of 2d quantum
 gravity, or non-critical bosonic string theory, with generic
 non-rational values of the matter central charge \ccharge.  

The main point of our investigation is the systematic study of the
effects of including matter interactions in the 2d string.
Conventionally one adds to the gaussian action the two matter
screening charges, which together with the Liouville ones serve as
interaction terms. Motivated by the comparison with a discrete, microscopic approach, to
be discussed in a subsequent paper \KPb, we introduced and studied
also another deformation of the Liouville theory, defined by the
interaction action \deltadiag.  While in  the first, ``conventional''  theory, the 
 $c < 1$ (matter) and $c > 25$   (Liouville) 
parts 
factorize before moduli integration,  there is no such factorization in the second theory, which
we called ``diagonal''.  

To construct the tachyon correlators we have adopted and extended the ground ring  approach 
introduced long ago \refs{\Witten
\KlebPol\KMS
\kachru-\bershkut}.  In this  approach  the matter-Liouville 
factorization of the integrand of the 4-point tachyon correlators (in the conventional theory) is not directly exploited, and so the precise 
realization of any of these $c<1$ and $c>25$ correlators is  not  apriori required.
In particular no assumption on the 
 existence of a fully consistent non-rational matter theory is made.  Indeed
such a theory has not been rigorously established in the conventional theory, and  does not exist 
 in the second,  diagonal theory.   For our purposes it was sufficient to derive  a 3-point generic matter OPE
constant,    formula \conj{},    a  $c<1$ analog of the Liouville DOZZ formula,  
\Lc\  which extends the Dotsenko-Fateev  Coulomb gas constant. 
The ground ring method  is based on the derivation of  functional equations for the  tachyon  correlators, 
using the module action (operator product)
of the fundamental ground ring elements on a (0,0)-form tachyon $W_\a^\e$
in the presence of integrated tachyons.  
The OPE coefficients of  the ground ring action
are determined by well defined  free field correlators, computed 
either by using  the matter-Liouville Coulomb gas
representation, or exploiting the factorization into known $c<1$ and $c>25$ Coulomb gas correlators. 

The explicit 3-point OPE
coefficients in \mpd, \pmd\ 
confirm the ground  ring structure  
conjectured in \SeibergS. The functional relations for the 3-point functions are closely related
to a standard identity for the tensor product decomposition
multiplicities of sl(2) finite dimensional irreps, which are
reproduced as a particular case.\foot{ 
We stress that these  
multiplicities 
would not be allowed if the formal "matter $\times$ Liouville  factorisation"
was taken too literary.}
Besides those in the non-rational case one has more 3-point solutions
and some were used as a building block in the construction of the
4-point solutions we have described. The diagonal theory 
admits an action of the ground ring generated by the new deformations
of the product of ring generators $a_-a_+$.  The result \compra\ is an
effective projection of the ring action to a diagonal $sl(2)$ type
identities.

What complicates the case of $n$-point functions, $n>3$,  are the additional
contact terms in the functional relations  due to the fact that the
fourth, {\it etc.} field,  given by an integrated tachyon $T_\a^\e$, serves as a new ``screening
charge''.  Thus, besides the two operator terms in \mpd, \pmd\ which
correspond to perturbations by the screening charges in the
interaction actions \deltaS, \deltaSd, there are other channels in the
OPE of a  ring generator and a tachyon  $W_\a^\e$.  These OPE terms account for the  effect of the $Q_{\rm BRST}$-exact
terms,  skipped in the r.h.s. of \mpd, \pmd. 
We have computed two  series of 
4-point   OPE coefficients, \opeonet, \opeonetde, sufficient for the class of tachyon correlators
we consider. 
The diagonal model is more restrictive on
the content of the operator products and in particular leaves less
room for contact terms.

We have found basically two types of 4-point solutions of the
functional equations \wardid.  Apart from a particular example,
both involve an integer number of some of the screening charges.  We
have presented in more detail the solutions  with  matter screening
charges, however, because of the symmetry of the ring identities, in
the conventional non-diagonal theory some of these solutions  have Liouville analogs as
well. 

The first class of solutions, \twobcond,  appears for  generic values
of the four tachyon momenta, such that
their sum is restricted by a matter charge conservation, 
thus generalizing the tachyon correlators for gaussian matter of \DiK.
 The
arbitrariness in the solutions of the homogeneous equations, or,
effectively, in the determination of the 
1pi part
in \summar, is partially fixed by comparison with 4-point functions in which one
of the tachyons is a screening charge.  
The choice of these ``boundary'' conditions corresponds to
the type of interaction action, which otherwise enters the definition
of the correlators rather formally. To fix the remaining arbitrariness  we have 
required that different classes of solutions are 
related to each other,
whenever their partial wave expansions, as in \summar,  are comparable.

The second class of solutions found in sections 8 and 9 represents 4-point functions in which
one  field \ntexab, 
 or all 
fields (formulae \fpconj{}, \fpconjd{}\ in the diagonal theory) correspond to a degenerate Virasoro representation.  For the correlators with four degenerate fields in
  the conventional theory  we only give a conjecture. The problem is
complicated by the 
unknown additional
contact terms at degenerate
values.  

The equations we have derived and studied apply by definition to the correlators satisfying the chirality
rule.  Besides these  fixed chirality, and hence partially symmetric,  4-point functions  we have 
described   also correlators  
 symmetric with respect to the four (or the three generic) 
 matter 
charges. We interpreted this symmetry as tachyon "locality". Until this point locality of the underlying $c<1$ and $c>25$ correlators  is only partially exploted 
in the computation of the OPE coefficients. 
In the simplest example in section 6 the set of fixed chirality solutions serves as a local basis
for the local correlators. Then the original equations are rewritten equivalently as equations for these  
symmetrised correlators.
To fix in general 
the arbitrariness  in the solution of the symmetry requirement we have exploited the fact that all our solutions 
 admit the channel decomposition form \summar.  Our universal choice was to preserve the fusion 
 multiplicities in the symmetric counterpart of this expression
 - the formal rule is to replace the inverse propagator $\e P$ with $|P|$.  This choice indirectly takes into account the locality of the  underlying $c<1$ 
theory, since 
these fusion rules correspond to  the ones manifested by the local correlators of that theory. 
Furthermore
the local correlators 
 \fourpDK,  \chmodges, \exach\  are  invariant  under Liouville reflections.

The symmetric correlators  
(as well as their analogs, symmetrised  with respect to the Liouville labels)
do not satisfy globally 
the original ring relations, rather   satisfy shift equations
with modified and momenta dependent 
 inhomogeneous terms. We have   proposed an alternative recursive derivation of these equations for the local correlators 
starting with
the simplest case of sect. 6.1.
 It  also yields a full set of local representatives of the 
symmetric tachyon correlators, extending the set of fixed chirality correlators
obtained as solutions of the initial equations.   

 Our treatment  of the local correlators remains  however rather  ``phenomenological" 
and the direct derivation of  their equations  is still an open problem which requires an extension 
of the Coulomb gas based technique we had mostly exploited.  
Conceptually this is important since it is natural to interpret  the local correlators as the true "physical"  ones,  while the sets of partially symmetric, fixed chirality 
correlators, though basic in our construction, should be considered rather as auxiliary objects.
This is confirmed  by the matrix model approach \KPb\  (formula \changmod) and also by the comparison with  the recent paper \BZ,  in which the 
underlying matter and Liouville theories are explicitly exploited in the computation of 
the  
 4-point
function with one matter degenerate and three generic fields: the local correlator 
\exach\ coincides with the expression   in \BZ\ 
computed by this   more constructive method.
A notable exceptional case,
avoiding these problems and confirmed by the discrete model in \KPb, are the
4-point functions \fpconj{}\ of four degenerate fields in the diagonal theory.  

Our analysis has been restricted so far to the bulk quantities.
However, as it is well known from the studies in the rational matter
and the generic Liouville BCFT, the bulk 3-point correlators, $\!${\it
i.e.$,\!$} the (properly normalized) OPE coefficients,  give 
information about the boundaries, since the matrices diagonalizing
them are closely related to the disc 1-point functions, as briefly
discussed in Appendix B. In Appendix A.5.  we have also computed some
chiral OPE coefficients in the presence of matter charges, including
the four OPE coefficients of the boundary ground ring, which has a
similar to \mpd, \pmd\ two-term structure.  The functional relations
for the boundary tachyon correlators, generalizing the trivial matter
case \bershkut, \KostovCY, will be discussed elsewhere, see also the
paper \BM\ which appeared meanwhile, which deals with this problem
too, but in the minimal string theory.

\bigskip
\noindent
{\bf  Acknowledgments}
\smallskip

\noindent
We thank Al.  Zamolodchikov for numerous valuable discussions.  We
also thank S. Alexandrov, A. Belavin, Vl.  Dotsenko, V. Fateev, V.
Schomerus, M. Stanishkov and J.-B. Zuber for the interest in this work
and for useful comments.  Special thanks to Paolo Furlan for the
careful reading of the manuscript, the many useful remarks on it, as
well as for the explicit computation included in formulae \aminusda,
\aplusda.   I.K.K. thanks the Institute for Advanced Study, Princeton,
and the Rutgers University for their kind hospitality during part of
this work.  V.B.P. acknowledges the hospitality of Service de Physique
Th\'eorique, CEA-Saclay and the University of Northumbria, Newcastle.
This research is supported in part by the European Community
through
RTN  EUCLID, contract  HPRN-CT-2002-00325, 
MCRTN ForcesUniverse, contract MRTN-CT-2004-005104, 
MCRTN ENRAGE, contract MRTN-CT-2004-005616, 
MCRTN ENIGMA, contract   MRTN-CT-2004-005652, 
and by the Bulgarian National Council for Scientific Research, grant
F-1205/02.

\appendix{A}{Coulomb gas computations}

\def\gb{{\bf b}}
\def\gc{{\bf c}}
\def\bgc{\bar{\bf c}}

\noindent
In this appendix we shall compute some matrix elements of the type
\eqn\npointbcc{\eqalign{
&\langle \a{'}|\gc_{-1}\gc_0\, \int_{C_n} dz_n
\,V_{\a_n}^{\e_n}(z_n) \cdots
\int_{C_2} dz_2\, V_{\a_2}^{\e_2}(z_2)\,a_-(z)\,
\gc V_{\a_1}^{\e_1}(z_1)|0 \ra_{\rm free}=\cr
&\int \cdots \int \(({\a_1-e_1\over b}-1) {1\over z-z_1}-
\sum_{i=2}^n\,
{\a_i-e_i\over b} {1\over z_i-z}\) \, \times\cr
&\langle \a{'}| V_{\a_n}^{\e_n}(z_n) \cdots \,V_{({b\over
2},-{b\over 2})}(z)
\, V_{\a_1}^{\e_1}(z_1)|0\ra_{\rm free}=\cr
& \int \cdots \int\, 
\({1\over b^2}\,\langle... \ra_M\p_{z}\langle... \ra_L -
{1\over b^2}\,\langle...\ra_L\p_{z}\langle... \ra_M
-{1\over z -z_1} \langle... \ra_M\,\langle... \ra_L \)
}}
and their volume integral counterparts,
which determine the OPE coefficients of the ring generator $a_-$ with
the tachyon fields.  Everywhere here $V_{\a}^{\e}\,, $ or
$V_{(e,\a)}\, $ denote {\it unnormalized} products of vertex
operators, with no relation necessarily of the type in \mashell\ on
the pair of matter and Liouville charges $(e,\alpha)$.  There is a
similar formula for the other generator.

\medskip
\noindent
Conventions:
$$
\langle \phi(x_1)\phi(x_2) \ra = -{1\over 2}\log x^2_{12}=
\langle \chi(x_1)\chi(x_2) \ra
$$
$$
\phi(z)=\phi^{(+)}(z)+\phi^{(-)}(z):= {i\over
\sqrt{2}} \big( - a_0 \log z+
\sum_{n> 0} {a_n\over n} z^{-n}\big) +{i\over \sqrt{2}}
\big(-iq -\sum_{n> 0} {a_{-n}\over n} z^{n}\big)
$$
$$
[a_n,a_{-m}]=n\delta_{n,m}\,,\qquad  [a_0,q]=-i
$$
$$
V_{(e_1,\a_1)}^{(+)}(z_1)\,V_{(e_2,\a_2)}^{(-)}(z_2)=
z_{12}^{2(e_1e_2-\a_1\a_2)}\,
V_{(e_2,\a_2)}^{(-)}(z_2)\, V_{(e_1,\a_1)}^{(+)}(z_1)\,,  \qquad |z_1|>|z_2|
$$
\medskip
\noindent
Let us also recall some ghost field correlation functions.  The
correlators of the ghost $\gb,\gc$ fields decouple as the full
correlators factorize.  The 2-point function is computed in the vacua
$\langle 0|\gc_{-1} \gc_0\gc_1$ and $|0\ra$, normalizing $\langle
0|\gc_{-1}\, \gc_0 \,\gc_1|0\ra=1$, 
\eqn\gtwopoint{\eqalign{ &\langle \gb(z_1)\
\gc(z_2)\ra =\langle 0|\gc_{-1}\gc_0\, \gc_1
\sum_{k=-1}\,\gb_k\, z_1^{-k-2}\,
\sum_{m=-1}\gc_{-m}z_2^{m+1}\, |0\ra=\cr
&\langle 0|\gc_{-1}\gc_0 \gc_1|0\ra
{1\over z_1}\sum_{p=0}
\left({z_2\over z_1}\right)^p=
{1\over z_{12}}\,, \quad |z_1|>|z_2|\,.
}}
The 3-point ghost  $\gc$ function is
\eqn\gthree{ %
\langle 0|\gc(z_1)\,\gc(z_2)\, \gc(z_3)  |0 \ra =\langle 0| \,
\prod_{i=1}^3\sum_{k_i=-1}^1\gc_{k_i} z_i^{-k_i+1}\, |0\ra=
z_{12}\, z_{13}\, z_{23}\,.
}
The 3-point function with the insertion of one field $:\gb\gc:$ reads
\eqn\gcomm{\eqalign{ &\langle 0|\gc_{-1} \, \gc(z_2)\,
:\gb\gc(z):\,\gc(z_1)|0\ra=\cr &\langle 0|\gc_{-1}\, \sum_{s=0}\gc_s\,
z_2^{-s+1}\, \sum_m\, z^{-m-1}\Big(\sum_{k=2}
\gb_{-k}\gc_{m+k}-\sum_{k=-1} \, \gc_{m-k} \gb_k \Big)\,
\sum_{p=-1}\gc_{-p}z_1^{p+1} |0\ra\cr &= \langle 0|\gc_{-1} \gc(z_2)\,
\gc(z_1)|0\ra\ ({1\over z_2-z}- {1\over z-z_1}) \,, \quad
|z_2|>|z|>|z_1|\,, }} while
\eqn\gcommb{
\langle 0|\, \gc_{-1}\, \gc_0\, :\gb\gc(z):\,\gc(z_1)|0\ra=-
{1\over z-z_1}\,, \quad |z|>|z_1|\,.
}
The last formula is used in \npointbcc\ producing the shifts
by $-1$. In particular it leads to the last term in the matter-Liouville factorized expression
in the last line, where one is using the representation of the ring operator in terms of derivatives,
\eqn\derivrep{
a_-(z)=:e^{b i \chi} ({1\over b^2} 
\buildrel{\leftrightarrow}\over\partial_z+ \gb\gc(z) ) \, e^{-b \phi}:\,,
}
meaning action of the derivative to the right minus action to the left.

\subsec{ 3-point volume integral matrix elements}

\noindent
We start with some bulk correlators, most of which have been already
computed \refs{\KMS,\kachru,\bershkut}.  In these examples we shall
use the first representation in \npointbcc, while 
in the next subsection we will exploit the
matter-Liouville factorized expression in the second line.

The factor $\gc_{-1}\,
\bgc_{-1}\, \gc_0 \bgc_0$ is denoted
$(\gc\bgc)_{-1}\, (\gc\bgc)_0$. Consider first the matrix element
\eqn\noint{\eqalign{
&\langle \a' |(\gc\bgc)_{-1}\, (\gc\bgc)_0\,
\, a_-(x_0) (\gc\bgc V_{\a}^{\e})(x_1) \ra =
\langle \a'|(e+{b\over 2},
\a-{b\over 2})\ra\,
({\a-e-b\over b})^2 \, (x_{01}^2)^{b(\a+e-{1\over b})}\cr
& =  \cases{ 0 & if  $\e=1$,\cr
({2\a-Q\over b})^2=-{\g({1\over b}(Q-2\a+b)\over \g({1\over b}(Q-2\a)}
& if  $\e=-1$}
}}
and $\a{'}=Q-\a +{b\over 2}$,
recovering
\eqn\actaa{a_{-}     W_\a^{-} =-
   W_{\a- {b\over 2} }^{-}\,, \qquad
a_{+}    W_\a^{+} = -
    W_{\a-  {1\over 2b } }^{+} \,,
}
\eqn\actaab{
  a_{-}  W_\a^{+} = a_{+} W_\a^{-} =0\,.
  }
Our next example is
\eqn\oneint{\eqalign{
&\int {d^2x_2\over \pi}\,\langle \a' |(\gc\bgc)_{-1}\, (\gc\bgc)_0\,
a_-(x_0) (\gc\bar{\gc} V_{\a}^{\e})(x_1)\,
V_{\a_2}^{\e_2}(x_2) \ra_{\rm free} =
\langle \a' |(e+e_2+{b\over 2},
\a+\a_2-{b\over 2})\ra\,\cr
&\int {d^2x_2\over \pi}\,\left|{\a-e-b\over b} {1\over z_{01}}+ {\a_2-e_2\over b}
{1\over z_{02}}\right|^2
\, (x_{01}^2)^{b(\a+e)}\,(x_{02}^2)^{b(\a_2+e_2)}
\,(x_{21}^2)^{2e e_2-2\a\a_2}
\cr
}}
In the three of the four possible cases this integral vanishes for generic momenta, either
due to factors $\g(1)$, or, because of sign compensation of the
various terms. 
 It survives only for $\e=\e_2=1$ producing the
constant
$$
{\g(b(2\a_2-b))\, \g(b(2b-2\a-2\a_2)+1)\over \g(b(b-2\a)+1)} = 
{\g(b(Q -2(\a+\a_2-{b\over 2}))\over
\g(b(Q-2\a))\,\g(b(Q-2\a_2))} 
$$
which precisely provides the leg factor normalization of the three
tachyons, thus recovering the first formula in \aminus.\foot{In particular the non-generic value $\a+\a_2- \frac{b}{2} =\frac{Q }{2} $
corresponds to a tachyon of no definite chirality, for which the numerator and equally the compensating leg factor,  become singular. }  In agreement
with the BRST invariance, both in \noint\ and the integrated \oneint\
only the combination satisfying the  mass-shell condition survives,
while all the other terms, possible in the analogous pure matter or
Liouville 3-point matrix elements, now cancel out automatically, due
to the effect of the raising prefactor in the ring generator.  In
particular choosing $\a_2=b$ or $\a_2=0$ one recovers the generic two
term action in the first line in \mpd.

Finally for $\e_2=\e_3=1$ and $\a_2+\a_3=b$ there is a double integral
matrix element
\eqn\twoint{\eqalign{
&\int\, {d^2 x_2\over \pi}\,\int\, {d^2 x_3\over \pi}\,
\langle \a' |(\gc\bgc)_{-1}\, (\gc\bgc)_0\, a_-(x_0) 
(c\bc V_{\a}^{-})(x_1)\,
V_{\a_2}^{+}(x_2)\, V_{b-\a_2}^{+}(x_3) \ra_{\rm free} =\cr
&= - 
 {\g({1\over b} (Q-2\a-b))\over \g({1\over b}(Q-2\a))\,
\g(b(Q-2\a_2))\, \g(b(Q-2(b-\a_2)))}\,.
}}
This constant reproduces again the relevant leg factors and thus we obtain
for the normalized fields the first of the relations \di.

\subsec{The general 3-point  constant}

\noindent
In general accounting for all possible matter and Liouville screening
charges one computes the 3-point function of the ring generator with
two tachyons using the representation in the last line in \npointbcc.
The result for the 2d integral (for the unnormalized tachyons) is
proportional to the product of the 3-point $c<1$  Coulomb gas OPE
constant computed in \DF\ and its 
$c>25$  counterpart given by an
analytic continuation of the same formula,
\eqn\baconst{\eqalign{
&\langle 0|
(\gc\bgc V_{(e_0-e_3,Q-\a_3)}^{\e_3})(\infty)\,
(\gc\bgc)_0\,a_-\,(\gc\bgc V_{(e_2,\a_2)}^{\e_2})\,|0 \rangle=
c(\a_2,\a_3)\times \cr & \GM({b\over 2}, e_2,e_0-(e_2+{b\over 2} -k_1 b+k_2/b))
\hat{C}^{\rm Liou}(-{b\over 2}, \a_2,Q- (\a_2-{b\over 2} +s_1 b+s_2/b))\,, \cr
}}
where
\eqn\baconstb{\eqalign{
&c(\a_2,\a_3)={1\over b^4}\,
\((\a_3-\a_2+\frac{b}{2})(\a_3+\a_2-Q-\frac{b}{2}) +(\a_3-\a_2-\frac{b}{2})
(\a_3+\a_2-Q+\frac{b}{2})\)^2\cr
&={1\over b^4}\,
\((e_3-e_2+\frac{b}{2})(e_3+e_2-e_0-\frac{b}{2}) +
(e_3-e_2-\frac{b}{2})(e_3+e_2-e_0+\frac{b}{2})\)^2
}}
Here $e_3=e_2+\frac{b}{2} -k_1b+\frac{k_2}{b}\,, \a_3=\a_2-\frac{b}{2}
+s_1b+\frac{s_2}{b}$ and
the four integers $s_1,s_2,k_1,k_2$ - the number of screening charges
of type $T^+_b\,, T^-_{1/b}\,, T^+_0\,,T^-_0\,, $ are restricted by \mashell\
 depending on the combination of chiralities,
$\!${\it i.e.$,\!$}
\eqn\mscon{
(\e_3-\e_2)e_2+b{1+\e_3\over 2}
+b^{\e_3}-b^{\e_2}=(s_1+\e_3 k_1)b +(s_2-\e_3 k_2)\frac{1}{b}\,.
}
The OPE coefficient for the normalized tachyons is given, by the
r.h.s., of \baconst\ times the ratio of leg factors, $\!${\it
i.e.$,\!$}
\eqn\opeconst{
a_- W_{\a_2}^{\e_2} =\sum_{\a_3\,, \e_3} c(\a_2,\a_3) 
\GM({b\over 2}, e_2, e_0-e_3)
\,\hat \GL(-{b\over 2}, \a_2, Q-\a_3)\,
{\g(b^{\e_2}(Q-2\a_2))\over \g(b^{\e_3}(Q-2\a_3))}\,
W_{\a_3}^{\e_3}
}
\noindent
The coefficient in the r.h.s.  is examined either using directly the expressions of the two
$c<1$ and $c>25$ Coulomb gas  constants, or by exploiting the compact formula  \conj{}\ for the matter
constant and the relation of $\hat \GL$ to 
$\GL$ in \Lc, regularizing $\a_1=-e_1=-b/2+\epsilon$.
We get that the overall constant goes to zero like $\epsilon^2$, unless one of the four factors
$$
(e_{123}-e_0)^2\, (e_{23}^1- e_0)^2\,  (e_{12}^3 )^2\, (e_{13}^2)^2
$$
vanishes as well.

 The values $e_3=e_2 \pm b/2$ are
equivalent to $k_2=0\,, k_1=0,1$.  When the chirality is preserved,
$\e_2=\e_3$, we obtain taking into account the condition \mscon\ two
solutions for each sign \eqn\genr{\eqalign{ k_2=0=s_2\,,& \
k_1=1-s_1=0,1\,, \quad {\rm for}\ \e_2=1=\e_3\,,\cr k_2=0=s_2\,,&\
k_1=s_1=0,1\,, \quad {\rm for} \ \e_2=-1=\e_3\,, }} altogether leading
to the generic OPE relations \fcons, \mpd.

For $e_3=e_2 \pm b/2$ but $\e_2=-\e_3=1$ the values of $e_2$ become
restricted by \mscon\
\eqn\lra{\eqalign{
e_0-2e_2&=P_2=Q-2\a_2=s_1b +\frac{s_2}{b} -\frac{b}{2} \pm \frac{b}{2}\,,\cr
&(e_3,\a_3)=(e_2\pm \frac{b}{2}, Q-\a_2\mp \e_2 \frac{b}{2})
}}
The resulting $(e_3,\a_3)$ correspond to the Liouville reflected
counterparts of the two terms in \mpd; they have to be added whenever
the momenta take the special discrete values in \lra.  These values
include the Liouville degenerate points  (with the plus sign in
\degenL{}).  Similar formula arises for $\e_2=-1=-\e_3$.

One gets a nonzero expression also for $ e_3=e_0-e_2\pm b/2$.  The
combination of chiralities $\e_2=-\e_3$ reproduces the matter
reflected points occurring for
\eqn\mra{\eqalign{
e_0-2e_2&=P=
-k_1b+\frac{k_2}{b}  +\frac{b}{2} \mp \frac{b}{2}\,, \cr
& (e_3, \a_3)=(e_0-e_2\pm \frac{b}{2}, \a_2 \mp \e_2 \frac{b}{2})
}}
with
$s_2=0\,, s_1=0,1\,,$ for $ \e_2=1$, and $s_2=0,s_1=1,0$ for $\e_2=-1$
respectively.  These values include the matter degenerate momenta
(with the positive sign in \degen{}).  Finally for $ e_3=e_0-e_2\pm
b/2$ and $\e_2=\e_3$ there are two series of solutions, corresponding
to both matter and Liouville reflections
\eqn\lmr{\eqalign{
e_0- 2e_2 &= \frac{b}{2}\mp \frac{b}{2}-k_1b+\frac{k_2}{b}\,,\cr
&(e_3,\a_3)=(e_0-e_2\pm \frac{b}{2},Q-\a_2 \pm \e_2 \frac{b}{2}),
}}
with $s_1+k_1=1\,, s_2=k_2$ for $\e_2=1$ and $k_2=0=s_2$, $s_1=k_1$
for $\e_2=-1$.

In all cases the constant in \baconstb\ becomes
$c(\a_2,\a_3)=\frac{(Q-2\a_2)^2}{b^2}$ and the final result simplifies
to powers of $\L,\LM$\ and $b$,  e.g., in 
the case \mra,  $e_0-2e_2=Q-2\a_2=k_1b -k_2/b$  we have 
\eqn\mraap{\eqalign{
&C_{-\frac{b}{2}, \a}^{(+-)}{}^{\a-\frac{b}{2}}=- {\LM^{k_1-\frac{k_2}{b^2}}\over b^2}=
 {(Q- 2\a)^2\over b^2} {\g(b(Q-2\a))\over \g(\frac{1}{b}(Q-2\a+b))}\,
 \GM(\frac{b}{2}, e_2, e_2-\frac{b}{2}) 
}}
using that  $\hat \GL(-\frac{b}{2}, \a_2,Q-\a_2+\frac{b}{2})=1$, while $ \GM(\frac{b}{2}, e_2, e_2-\frac{b}{2})$ is determined by matter reflection as in \reflm\ from \ex{}.  
This  constant is finite for positive integers $k_1,k_2 \ne 0$,
which are the values of the degenerate representations in \degen. Similar formulae hold for the other cases in \lra, \mra\ and the constituent matter or Liouville Coulomb gas constants 
$\GM\,, \, \hat \GL$ are finite for the degenerate
values of the momenta.
On the other hand in the last case \lmr,  which involves values on the boundary 
of the degenerate regions appear, there may appear  singularities in the  constants or the leg factors.
 In all these considerations we have assumed that $b^2$ is
generic, non-rational.  \medskip

\rb In contrast to the above result in the diagonal case the 3-point
function computed with $k_1, k_2, s_1, s_2$ tachyons of type $T^-_b,
T^+_{1/b}, T^+_b, T^-_{1/b}$ respectively, is more severely
restricted.  We have $e_3=e_2-\frac{e_0}{2}+(k_1+k_2) e_0\,,
\a_3=\a_2 -\frac{Q}{2}+(s_1+k_1)b + \frac{s_2+k_2}{b}$ and the
 condition \mscon\ is replaced by
\eqn\mascondd{
(\e_3-\e_2) e_2+\frac{Q}{2}-\e_3 \frac{e_0}{2}
+b^{\e_3}-b^{\e_2}=(s_1+k_1 +\e_3(k_1+k_2)) b +
(s_2+k_2-\e_3(k_1+k_2))\frac{1}{b}\,.
}
Thus in the generic case $\e_2=\e_3$ we have $s_1=1\,,
s_2=k_1=0=k_2$, or $k_2=1\,, s_2=k_1=0=s_1$ for $\e_2=1$ and $s_2=1\,,
\ s_1=k_2=0=k_1$, or $k_1=1\,, s_1=k_2=0=s_2$, - for $\e_2=-1$.  These
solutions all involve one of the interaction terms
in  \deltadiag.  The action of $a_-a_+$ can be understood as the
composition of the free field formulae \actaa, \actaab, \aminus\ for
the two generators
\eqn\compr{\eqalign{
&a_+\,a_- T_{\a_1}^+ \, W_{\a}^+=a_+\, W_{\a+\a_1-\frac{b}{2}}^+
=-W_{\a+\a_1-\frac{Q}{2}}^+\ \  \Rightarrow\cr
&a_+(-\,a_- T_{b}^+) \, W_{\a}^+=
W_{\a-\frac{e_0}{2}}^+\cr
&a_+\,(-a_- T_{\frac{1}{b}}^+) \,
W_{\a}^+=W_{\a+\frac{e_0}{2}}^+\cr
}}
and
\eqn\comprd{\eqalign{
&a_+\,T_{\a_1}^-\, a_-  \, W_{\a}^-=-a_+\, T_{\a_1}^-\,
W_{\a-\frac{b}{2}}^-
=-W_{\a+\a_1-\frac{Q}{2}}^- \ \Rightarrow \cr
&(-a_+\,T_{\frac{1}{b}}^-)\, a_-  \, W_{\a}^-=
W_{\a+\frac{e_0}{2}}^-\cr
&(-a_+\,T_{b}^-)\, a_-  \, W_{\a}^-
=W_{\a-\frac{e_0}{2}}^-  \,.
}}
If the chirality is inversed there are more solutions.  However,
restricting to 
diagonal momenta
the only
soltions are $P=\pm e_0, 0$. Since the tachyon of $P=0$ has no definite chirality,
all these solutions
 effectively fit   the generic formula \compra.

\subsec{Mass-shell  restrictions on the contact terms}

\noindent
The contact terms in the difference equations for the 4-point tachyon correlators
are determined by the OPE coefficients computed by
the  4-point Coulomb gas functions
\eqn\bafconst{\eqalign{
C_{-\frac{b}{2}\, \a_2\,\a_3}^{\ (\e_2\e_3\e_4)}{}^{\a_4}&=
{1\over \gamma((Q-2\a_4)b^{\e_4}) \gamma((2\a_4-Q)b^{\e_4}) }\,
\langle 0|W_{Q-\a_4}^{\e_4}(\infty)
\, (\gc\bgc)_0\,a_-\,W_{\a_2}^{\e_2}\,T_{\a_3}^{\e_3}
|0 \rangle\,, \cr
{}& {}\cr
&\a_4=\a_2+\a_3 -\frac{b}{2}+s_1 b+\frac{s_2}{b}\,, \ \ 
e_4=e_2+e_3 +\frac{b}{2}-k_1 b+\frac{k_2}{b}\,.
}}
 As in the computation of the 3-point OPE coefficients the denominator  
 comes from the leg factors in the 
 trivial 2-point matrix element, cf. \twopP,
$$
\langle 0|W_{Q-\a_4}^{\e_4} (\infty)(\gc\bar \gc)_{0} W_{\a_4}^{\e_4} |0\rangle = 
\langle 0| \gc(z)\p_z \bar \gc(\bar z)\p_{\bar z} W_{Q-\a_4}^{\e_4} (z,\bar z)  W_{\a_4}^{\e_4} (z', \bar z')|0\rangle \,.
$$
 If $\a_2$ coincides with one of the four values of the screening charges \bafconst\ reduces to a 3-point function. 
 The mass-shell 
condition \mashell\   implies
\eqn\mscond{
(\e_4-\e_2)e_2+(\e_4-\e_3)e_3+{1+\e_4\over 2}\,b
+b^{\e_4}-b^{\e_2}-b^{\e_3}=(s_1+\e_4 k_1)b +(s_2-\e_4 k_2)\frac{1}{b}\,.
}
For generic momenta $P_2,P_3\not \in \CL$ and $P_2+P_3\not \in \CL$
the only solution of the  mass-shell condition occurs for
$\e_2=\e_3=\e_4=1$, with $k_2=s_2=k, \, k_1+s_1=0$, whence
$k_1=0=s_1$.   The case $k=0$,  
  corresponds to \aminus\ 
while $k\ge 1$  to \aminusda, \aplusda.

 The relation \mscond\  is generalized to a product $W_{\a_2}^{\e_2}T_{\a_3}^{\e_3}\dots 
T_{\a_{N-1}}^{\e_{N-1}}$  in the N-point analog of the Coulomb gas correlator \bafconst, 
which contains  $N-3\ge 1$  integrated tachyons. 
If the momenta are generic, i.e, $P_{i_1}+... + P_{i_s}\not
\in \CL$ for any partial sum, 
the mass shell condition implies $\e_i=\e_N\,, i=2,3,\dots, N-1$. Hence for $\e_N=1$
one gets $s_1+ k_1+N-4=0\,,  k_2=s_2$, while for $\e_N=-1$, the constraint is
  $k_1=s_1,\,  k_2+s_2+N-3=0$. In both cases there are no solutions 
for $N-3>1$. 
As a consequence, there are no new
contact terms coming from two or more integrated tachyons in the equations for the  n-point functions, $n\ge 5$ for those  generic  values of the momenta.
If the given  chiralities $\e_i\,, i=2,\dots, N-1$ are all identical, 
 imposing the
condition $\sum_{i=2}^{N-1} P_i \ne \CL$ forces $\e_N$ to be of the same sign
which again implies that there are no solutions of the mass shell condition
for $N>4$. Similarly  for $\e_2=-1$ and $\e_i=1\,, i=3,\dots, N-1$ imposing 
$P_2\ne \CL\,, \, \sum_{i=3}^{N-1} P_i \ne \CL$   excludes any contact terms. 
These properties are  taken into account in writing the  equations \oldring.

Apart from the above series of contact terms for $\e_2=\e_3=\e_4$,
which occurs for generic momenta, there are various possibilities
taking place for particular values of $\a_2$ or $\a_3$, determined by
the choice of the signs $\e_i$ in \mscond.  E.g.,
\eqn\mpp{\eqalign{
&\e_2=-1=-\e_3=-\e_4\,, \   \   e_0-2e_2=P_2=(k_2-s_2)/b -(s_1+k_1)b \,,\cr
&\e_2=-1=-\e_3=\e_4\,, \   \   e_0-2e_3=P_3=(k_2+s_2+1)/b -(k_1-s_1)b \,.
}}
In particular these conditions admit solutions for $P_2$ or $P_3$
which correspond to the degenerate matter values when the interaction
involves the two matter charges.  Another example of \mscond\ is given
by $\e_2=\e_3=1=-\e_4$ which may occur for generic values of $P_2$ and
$P_3$ but their sum restricted by $P_2+P_3=(s_1-k_1) b+(s_2+k_2+1)/b$.

For the N-point generalization of \bafconst\ with  $p=N-3 >1 $  integrated tachyons, there are further possibilities. 
E.g., the second condition
in \mpp\ is replaced by $\sum_{i =3}^{p+2}\
P_i=(k_2+s_2+p)/b -(k_1-s_1) b$.
etc.   The simplest example with $p=2$ and no screening charges
was computed in  \twoint.   

In the diagonal theory described by the action \deltadiag\ the
 mass-shell condition for the 4-point function \bafconst\ is again more
restrictive.  We have $e_4=e_2+e_3-\frac{e_0}{2}+(k_1 +k_2) e_0 \,,
\a_4=\a_2+\a_3 -\frac{Q}{2}+(s_1+k_1)b + (s_2+k_2)/b$ and in the
generic case $\e_2=\e_3=\e_4=\pm 1$ the condition admits the unique
solution $k_1=0=k_2=s_1=s_2$.  It reproduces the OPE in the first
lines in \compr\ and \comprd.  The analog of the second
example in \mpp\ is
$$
e_0-2e_3=P_3= (s_2+2k_2 +k_1)/b-(k_2-s_1)b\,.
$$
Restricting to diagonal values $P_3=ke_0$, we obtain
$s_1+k_1+k_2+s_2=0$, $\!${\it i.e.$,\!$} 
 the
only possible value is $P_3=0$, which is beyond the degenerate matter
range.

The above conditions on the momenta are kinematical.  As in the
analysis of the general 3-point function \baconst, further
restrictions appear from the fusion rules of the degenerate fields
dictated by the 3-point constants in the decompositions of the 4-point
matter and Liouville functions.  In the next section we consider some examples.

\subsec{ 4-point OPE coefficients}

 We derive  here the  4-point OPE coefficients in  \aminusda\ and 
 \opeonetde. 
 
 The  vertex part of the function \bafconst\
 for 
 \eqn\genpf{
 e_4=e_2 +e_3+\frac{b}{2}+\frac{n}{b}\,, \ \ \a_4=\a_2+\a_3 -\frac{b}{2}+\frac{n}{b}\,, \ \e_i=1\,, i=2,3,4
 }
is  realized by
 a  $2n$-multiple integral coming from the power   of screening charges  $(T_0^-\, T_{1/ b}^-)^n$. We shall use instead    the matter-Lioville factorization formula as in the last line in \npointbcc, 
with an  alternative realization of each of the two types of correlators.
  Since one of the fields is the simplest degenerate field,  these $c<1$ and $c>25$ correlators are 
standard, given by sums of products of hypergeometric functions. In particular in our  example only one of the OPE channels survives. E.g.,   the matter matrix element
 is given for the values $e_i$ in \genpf\ by
the product of blocks with intermediate charge $e_2+\frac{b}{2}$,
\eqn\hyperm{\eqalign{
&\la e_0-e_4| V_{e_3}(x_3) \,  V_{\frac{b}{2}}(x_0)\, V_{e_2}(x_2) |0\ra_M=(x_{32}^2)^{\triangle_M(e_4)-\triangle_M(e_2)-\triangle_M(e_3)-\triangle_M(b/2)}\cr
&\times\,  {}_2F_1(-n, (2e_2+2e_3+2b)b +n-1; (2e_2 +b)b; {z_{02}\over z_{32}})\, \ ({\rm same, with }\ z \to \bar z )\cr
& \times \GM(\frac{b}{2}, e_2, e_0-e_2-\frac{b}{2})\, \GM(e_2+\frac{b}{2}, e_3, e_0-e_2-e_3-\frac{b}{2}-\frac{n}{b})\  f_M(z)f_M(\bar z)\,, \cr
&{} \cr
& f_M(z)=
 \, \big({z_{02}\over z_{32}}\big)^{b e_2}\, \big({z_{30}\over z_{32}}\big)^{b e_3}\,.
}}
The expression for the Liouville correlator $\la Q-\a_4| V_{\a_3}(x_3) \,  V_{-\frac{b}{2}}(x_0)\, V_{\a_2}(x_2) |0\ra_L $ is analogous, with $be_i\,,\,
\,b^2\, $ replaced by $b\alpha_i \,, \, 
 -b^2$, etc., while the constants $\GM$ are replaced by the $c>25$ Coulomb gas constants $\hat \GL$.
The first of the OPE constants in both cases is trivial,  $\GM(\frac{b}{2}, e_2, e_0-e_2-\frac{b}{2})=1=
\hat \GL(-\frac{b}{2}, \a_2, Q-\a_2+\frac{b}{2})$. The
hypergeometric function is the same, using that $e_i=\a_i-b$,
and it  reduces to a finite series of $n+1$ terms.  We then apply the derivatives with respect to $z_0$
and $\bar z_0$  term by term.
Using that the difference of the powers
of $z_{0i}\,, i=2,3$ from the matter and the Liouville 
functions  is  a constant $b(\a_i-e_i) =b^2$ one gets simply an overall factor 
\eqn\combder{\eqalign{
&\int \frac{d^2\, x_3}{\pi}\, 
(\sum_i {1\over z_{0i}} -  {1\over z_{02}}
)
\, 
f_M(z)\, f_L(z) \big({}_2F_1(-n, (2\a_2+2\a_3-2b)b +n-1; (2\a_2 -b)b; {z_{02}\over z_{32}})\big)^2
\cr 
&\qquad  \times\, ({\rm same, with }\ z \to \bar z )\cr
&=(x_{02}^2)^{u} \, \int \frac{d^2\, x_3}{\pi}\, 
\big({}_2F_1(-n, -w +n-1; u; {z_{02}\over z_{32}})\big)^2\,
\big({}_2F_1(-n, -w +n-1; u; {\bar z_{02}\over \bar z_{32}})\big)^2
 (x_{30}^2)^{v-1} \, (x_{32}^2)^{w}  \cr 
 &{}\cr
 &=\(\sum_k {(-n)_k \, (-w+n-1)_k\over k! (-w)_k }\, {}_3F_2(-n,-w+n-1, u+k; -w+k, u;1)
 \)^2\
 {
 \g(v) \g(w+1)\over \g(-u +1)}\cr
 &= \(n! \, 
 \,{(v)_n\over (u)_n\, (-w)_n}\, {w-n+1\over w-2n+1}\)^2
\,{
\g(v) \g(w+1)\over \g(-u +1)}=  :C(\a_2,\a_3; n) \cr
}}
where
$$ u:=(2\a_2 -b)b\,, \, v: =(2\a_3 -b)b\,, w: =-u-v =(b-2\a_4)b +2n \,.
$$ 
Altogether one obtains
for the values in \genpf\  the OPE coefficient in the $n$-th term in \opeonet\ 
\eqn\conami{\eqalign{
(\tilde  \L\, \tilde \LM)^n &=C_{-\frac{b}{2}\, \a_2\,\a_3}^{\ (+++)}{}^{\a_4}=
C(\a_2,\a_3; n)\ {\g(b(Q-2\a_2))\, \g(b(Q-2\a_3))\over  
 \g(b(Q-2\a_4))}\cr
& \times\, \GM(e_2+\frac{b}{2}, e_3, e_0-e_4)\, 
\hat \GL(\a_2-\frac{b}{2}, \a_3, Q-\a_4)\,. 
 }}
This result has been derived for generic values of the momenta. If however  $\a_2$ takes a degenerate value,
 $\a_2=b \pm \frac{m b}{2} -\frac{p}{2}\,, \ m,p\in \IZ_{\ge 0}, $
one of the Coulomb gas constants in \conami\ vanishes, 
 the other becomes singular.  The vanishing of the constant reflects a null vector factorization so it is natural
 to resolve this ambiguity for such fields  by restricting the validity of \conami\ to the values of $n$ avoiding the singularity, i.e., $n\le p$. On the other hand we can still keep \conami\ for any $n$ for fields with degenerate
 labels but not obeying the factorization conditions.

This  derivation generalizes to 
other cases discussed in the previous section. In particular let us consider  the case described by the second line
in \mpp\ 
$$
\a_2=-e_2+\frac{1}{b}\,, \ \a_3=e_3+b= \frac{(k_1-s_1 +1) b}{2}-\frac{k_2+s_2}{2b}\,, \ 
 \a_4=\a_2+\a_3-\frac{b}{2}+s_1 b+\frac{s_2}{b}=-e_4 +\frac{1}{b}\,.
$$
Examining the product $\GM(e_2+\eta \frac{b}{2}, e_3, e_0-e_4)\, 
 \GL(\a_2+\eta' \frac{b}{2}, e_3, Q-\a_4)$ one observes that for generic values of $\alpha_2$
 it becomes singular only if $ \eta=\eta'=1$,  $k_1=0\,, s_1\ge 1$, or $ \eta=\eta'=-1$, $s_1=0\,, k_1 \ge 1$; alternatively for these values the corresponding products $\GM \hat \GL$ are finite (as well as each of the two constants itself).  This implies that
 in each of these two cases only one of the 
four possible products of matter and Liouville  blocks survives in \bafconst. Furthermore
the hypergeometric functions corresponding to the matter and Liouville 
local correlators are identical again and the differences of overall  powers of $z_{0i}$ are
$b^2$ 
as before. E.g., in the case $s_1=0$,
the chiral factor $ f_M(z) \,{}_2F_1$ in \hyperm\ is replaced by
$$
z^{b(e_0-\e_2)}(1-z)^{be_3}{}_2F_1(-s_2, b(2e_3-2e_2)+s_2+1 ;1+b(e_0-2e_2);z)
$$
$$
=z^{b\a_2-b^2}(1-z)^{b\a_3-b^2}{}_2F_1(-s_2, (2\a_2+2\a_3-2b)b+s_2-1; (2\a_2-b)b;z)
$$
Then all the steps in the derivation of \combder\ are repeated with $n $ replaced by $s_2$ or $k_2$ respectively. We summarize
these results by relations analogous to \conami\ 
 for $s_1=0\,, k_1 \ge 1$,  and $k_1=0\,, s_1 \ge 1$, respectively:
\eqn\conamide{\eqalign{
 &\LM^{k_1}\, \tilde \LM^{k_2}\, \tilde \L^{s_2} =C_{-\frac{b}{2}\a_2\, \a_3}^{\ (-+-)}{}^{\a_4}=
C(\a_2,\a_3; s_2)\ {\g(\frac{1}{b}(Q-2\a_2))\, \g(b(Q-2\a_3))\over   \pi\,  \g(\frac{1}{b}(Q-2\a_4))}\cr
& \times\,  \GM(e_2-\frac{b}{2}, e_3, e_0-e_4)\,\GM(\frac{b}{2}, e_2, e_0-e_2+\frac{b}{2})\,
\hat \GL(\a_2-\frac{b}{2}, e_3, Q-\a_4)\,, \cr
&{}\cr
&\tilde \LM^{k_2}\, \L^{s_1}\, \tilde \L^{s_2}\,  =C_{-\frac{b}{2}\a_2\, \a_3}^{\ (-+-)}{}^{\a_4}=
C(Q-\a_2,\a_3; k_2)\ {\g(\frac{1}{b}(Q-2\a_2))\, \g(b(Q-2\a_3))\over   \pi\,  \g(\frac{1}{b}(Q-2\a_4))}\cr
& \times\, \GM(e_2+\frac{b}{2}, e_3, e_0-e_4)\, \hat \GL(-\frac{b}{2}, \a_2, Q-\a_2-\frac{b}{2})\,
\hat \GL(\a_2+\frac{b}{2}, e_3, Q-\a_4)\,.
 }}
Denoting $s_2+k_2=n\,,  s_2-k_2=2s-n$ and $ k_1 =m+1$, or $s_1=m+1$, respectively we arrive at \opeonetde. The case in the first line of \mpp\ can be analysed similarly.

On the other hand the alternative  multiple integral representation of  any of these
4-point correlators is not of the type in \DF.  Comparing with \conamide\ one effectively computes these nonstandard integrals.

 One can compute also the 4-point correlator with $W^+_{Q-\a_4}$ replaced by the tachyon $W^-_{\a_4}$
 (or by $W^-_{Q-\a_4}$)  so that the labels $\a_i$ (or the labels $e_i$)  satisfy a Liouville (matter) reflected version of the respective charge conservation condition. 
 This does not change the hypergeometric functions and it remains to use \reflL\ or \reflm\ respectively.
 The result  is 
 \eqn\reflagain{
 b^2\L^{2\a_4-Q\over b} C_{-\frac{b}{2}\, \a_2\,\a_3}^{\ (++-)\, Q-\a_4}=
  C_{-\frac{b}{2}\, \a_2\,\a_3}^{\ (+++)\,\a_4}= b^2 \LM^{2e_4-e_0\over b} C_{-\frac{b}{2}\, \a_2\,\a_3}^{\ (++-)\,\a_4}= (\frac{\L}{\LM})^{2\a_4-Q\over b}  C_{-\frac{b}{2}\, \a_2\,\a_3}^{\ (+++)\, Q-\a_4}
  }
 Thus the
  reflection properties of the underlying 4-point  Liouville  and matter correlators ensure the validity
  of  \drefl\  on the level of these  particular string 4-point correlators.

\subsec{ Some chiral OPE coefficients}

\noindent
Now we consider a few chiral matrix elements,
some of which have been computed in
\bershkut, \KostovCY.
The chiral analog of the simplest matrix element \noint\ reads
for $|z_0|> |z_1|$
\eqn\nointch{\eqalign{
&\langle \a' |\gc_{-1}\gc_0\, a_-(z_0) (\gc V_{\a}^{\e})(z_1) \ra
=  \cases{ 0 & if  $\e=1$,\cr
{2\a-Q\over b}=- {\G({1\over b}(Q-2\a+b)\over \G({1\over b}(Q-2\a)}
& if  $\e=-1$}
}} 
We recognize in the r.h.s. the leg factor normalization exploited in
the boundary theory, which is obtained replacing in \tachyons\ $\g(x)
\to \G(x)$.

The fields are radially ordered as above, accordingly the bounds on
the integrals are given by the arguments of the neighbouring fields,
the utmost left one being at $+\infty$, the utmost right one - in
$-\infty$.  E.g., let us look at the chiral analog of \oneint\ for
$\e=1=\e_2$.  We choose $|z_0|>|z_1|$ and send these two arguments to
$1$ and $0$ respectively.  The coordinate $z_2$ is `floating' and we
can collect the result for the three possible insertions of the
integral by writing the linear combination with coefficients
indicating the contours of integration
\eqn\oneintch{\eqalign{
&\sum_{(ij)} c_{i,j}\int_{C_{ij}} \, dz_2
\langle \a' |\gc_{-1}\gc_0\, a_-(z_0) (\gc V_{\a}^{+})(z_1)\,
V_{\a_2}^{+}(z_2) \ra =
{\pi\, \G(b(Q-2\a-2\a_2 +b))\over \Gamma(b(Q-2\a))\G(b(Q-2\a_2))}\cr
&\Big(- c_{\infty, 1}{ \sin\pi b(Q-2\a-2\a_2+b)\over \sin \pi b(Q-2\a)\,
\sin\pi b(Q-2\a_2))}+ c_{1,0} {1\over \sin \pi b(Q-2\a_2)}+
c_{0, -\infty} {1\over \sin \pi b(Q-2\a)}\Big)\,
}} 
The overall constant reproduces the chiral leg factor normalization of
the three fields.  Taking $\a_2=b$ or $\a_2=0$ the r.h.s. can be
identified 
with a (linear combination of) Liouville or matter
matrix elements of three chiral vertex operators (CVO).
The
intermediate states are described by a proper choice of the
coefficients in \oneintch, 
as it has been done in the boundary
Liouville case \FZZb; the boundary fields are linear combinations of
CVO\foot{See \PZb\ for the precise meaning of this statement in the
rational case.}.  Each of the two constants determine the
corresponding OPE of CVO, $\!${\it i.e.$,\!$} has the meaning of a
particular Liouville or matter fusing matrix elements in a certain
gauge.  The parametrization of the Liouville or matter intermediate
states (boundary conditions) is taken as $ z_L(\s)= \L^{1/2}\,\cos \pi b
(Q-2\s)$ or $z_M(a)= \LM^{1/2}\,\cos \pi b (e_0-2a)$, with
\eqn\parind{\eqalign{
&c_{\infty,1}=-z_L(\s_3)\,, \ c^{\pm}_{1,0}=-z_L(\s_3\mp \frac{b}{2})\,, \ 
c_{0,-\infty} = -z_L(\s_1)\cr
&c_{\infty,1}^{'}=-z_M(a_3)\,, \ c^{'\,\pm}_{1,0}
=-z_M(a_3\pm \frac{b}{2})\,, \  c_{0,-\infty}^{'} = -z_M(a_1)\,.
}}
Up to the overall leg factor normalization this gives for
the constants in the r.h.s.
\eqn\bopeL{\eqalign{
C^L_{\s_3\mp {b\over 2}\,,
\a+{b \over 2}}\left[\matrix{-{b\over 2}&\a \cr
\s_3& \s_1} \right]=-\L^{\hf}\,{2\sin \pi b(\a -\frac{b}{2} \pm
(\s_3+\s_1-Q))\,\sin \pi b(\a
-\frac{b}{2}\pm (\s_3-\s_1)) \over  \sin \pi b(Q-2\a)}\,,
}}
\eqn\bopeM{\eqalign{
C^M_{a_3\pm {b\over 2}\,,
e-{b \over 2}}\left[\matrix{{b\over 2}&e \cr
a_3& a_1} \right]
=-\LM^{\hf}
{
2\sin \pi b(e +\frac{b}{2}\pm
(a_3+a_1-e_0))\,
\sin \pi b(e+\frac{b}{2}\pm (a_3-a_1)) \over  \sin \pi b(e_0-2e)}
}}
and the case without screening charges \nointch\ corresponds to (a
product) of trivial constants
$$
C^M_{a_2\,,
e+{b \over 2}}\left[\matrix{{b\over 2}&e_2 \cr
a_3& a_1} \right]=1=C^L_{\s_2,\a-{b\over 2} }\left[\matrix{-{b\over 2}&\a \cr
\s_3& \s_1} \right]\,.
$$
The first of these expressions \bopeL\  has been derived in \KostovCY\
combining the formulae in \bershkut; it differs by an overall constant from the boundary Liouville constant
computed in \FZZb. The analytic continuation of the latter is similarly related to the matter constant \bopeM; vice versa \bopeM\ is  obtained from \bopeL\  via the analytic continuation formula \dualmL.  Finally let us look at the
chiral analog of the double integral matrix element \twoint\ with the
various possible positions of the two inserted vertex operators.  
E.g. for $\a_2=b$ we obtain for
$|z_0|>|z_1|$
\eqn\bconstb{\eqalign{
\int_{C_2}d z_2\,\int_{C_3}d z_3
\la \a' |&\gc_{-1}\gc_0\, a_-(z_0) (\gc V_{\a}^{-})(z_1)\,
V_{b}^{+}(z_2)\, V_{0}^{+}(z_3) \ra =
{\pi^2\,\G(\frac{1}{b}(Q-2\a-b))\,
\over \G(\frac{1}{b}(Q-2\a)) \,
\G(bQ)\,\G(be_0)}\times
\cr
&{\sin\pi b(Q-2\a)\over \sin \,\pi  2b\a}\,
C^L_{\s_2\,,
\a+{b \over 2}}\left[\matrix{-{b\over 2}&\a \cr
\s_3& \s_1} \right]\, 
C^M_{a_2\,,
e-{b \over 2}}\left[\matrix{{b\over 2}& e \cr
a_3& a_1} \right]\,,
}}
where the ratio of leg factors is extracted in the r.h.s. of the first line.

These formulae will be applied to the boundary tachyon operators at
generic values of momenta.  
Whenever the relation is applied to the left,
$\!${\it i.e.$,\!$} with the opposite order of the fields, there appears an  overall
minus sign. The four 
constants, \bopeL, \bopeM\ for
plus chirality, 
and  $-1\, $ 
in \nointch, and the constant in the second line in \bconstb\ for minus chirality,
correspond to the four
OPE coefficients in the bulk identities \mpd.
One can write down a formula analogous to \fcons, 
expressing any of the four OPE coefficients 
 in terms of  a  product of the corresponding 
Liouville and matter boundary OPE constants.

The formula \oneintch\ for arbitrary $\a_2$, but keeping one of the
terms, provides the simplest contact term in the boundary counterpart
of the recurrence relation \rrminus.  It remains to compute the boundary  
4-point functions determining the analogs of \opeonet, \opeonetde\
and hence of
the contact terms in \contgena.  
The details will be
presented elsewhere.

\appendix{B}{ More 3-point solutions and boundary CFT interpretation}

\noindent
We have encountered several examples of solutions of the ring
relations for the 3-point functions 
described by the
various ``fusion'' multiplicities $N_{P_1,P_2,P_n}$.  They generate
$n$-point multiplicities which can be cast formally into the general
form
\eqn\homrels{\eqalign{ &N_{P_1,P_2,\cdots ,
P_n}=\int d\mu(a) \prod_i \chi_{P_i}(a)\,,\cr &\qquad
\chi_{_{P+b^{\e}}}(a) +\chi_{_{P-b^{\e}}}(a)
\ (\ =  2\cosh (b^{\e}\p_P)\, \chi_{P}(a) )=
f_{b^{\e}}(a)\, \chi_{_{P}}(a)\,, \qquad \e=\pm 1\,.
}}
The relation in \homrels\ is a sufficient condition ensuring the
validity of the homogeneous relations \homrel; analogous identity
holds in the diagonal case.

This formula is specialized by certain range of the variable $a$, dual
to the spectrum of momenta $P$, and by some choice of the measures in
the two spaces, the ``characters'' $\chi_{_{P}}(a)$ and the function
$f_b(a)$.  E.g., the simplest example \detp\ corresponds to
$\chi_{_{P_i}}(a)=e^{ia (\e_ie_0-P_i)}$ with $\sum_{i=1}^p \e_i =p-2$,
and $f_b(a)=2\cos b a$.
Another explicit
solution of the homogeneous relations \homrel\
is given by a formula dual to \sltwopr,
\eqn\degbound{\eqalign{
&N_{P_1,P_2,P_3} =  \sum_{m=0}\sum_{n=0} \big(4\sin \pi m e_0 b\,
\sin \pi
n \frac{e_0}{b}\big)^2 \, \chi_{_{P_1}}(m,n)
\chi_{_{P_2}}(m,n)\, \chi_{_{P_3}}(m,n) \,, \cr
&\chi_{_{P(m,n)}}={\sin \pi m Pb \over
\sin \pi m e_0 b} {\sin \pi n P/b\over \sin \pi n e_0/b}
=\chi_{_{-P}}(m,n)\,,\quad 
f_b(a_{m,n})=2\cos (b(\frac{n}{b}\pm mb))\,.
}}
\vskip 0.11 cm
The ``characters'' satisfying the relation in \homrels\    coincide up
to a normalization with the tachyon disk  1-point functions
   $\ \langle \CV_{P}^{\e}
\ra_a
\sim {\L^{\e}
}^{P\over 2b}
\,\chi_P(a)$,  with  a boundary  label $a$,
\eqn\tpbound{
\L\, \langle \CV_{P-\e b}^{(\e)} \ra_a+
\langle \CV_{P +\e b}^{(\e)} \ra_a
=- \langle a_- \CV_{P}^{(\e)}\ra_a
=\sqrt{\L
}\, f_b(a)\,
  \langle \CV_{P}^{(\e)} \ra_a\,
  }
In the first equality we have used \mpd\
  (with $\LM$ set to 1).  The second is a version of the standard bulk
  - boundary equations, yet to be established in this context.  It
  implies \SeibergS\ that the eigenvalue of the operator $a_-$ is
  identified up to a power of $\L$ with the function $f_b(a)$ in
  \homrels\ for any solution $N_{P_1,P_2,P_3}$ of the ring relations
  \homrel; in \SeibergS\ this reasoning has been used in the rational
  case, assuming the validity of the OPE relations \mpd,\pmd.

The solution \degbound\ provides in the boundary CFT interpretation an
example of boundaries parametrized by the degenerate matter (or
Liouville, as in \ZZPseudo{}) representations; the two transformations
\dualmL\ preserve the formula for the characters inverting the sign
$(m,n) \to (m,-n)$ or $(m,n) \to (-m,n)$ in agreement with the $c<1$
versus $c>25$ parametrizations
\degen\ and \degenL. In this
case taking $P=e_0$ and $\e= \pm 1$ 
the first (or the
second) term in \tpbound\ disappears, respectively, so that we have
for $a=a_{m,n}$ that
$\L\, \langle \CV_{e_0-b}^{(+)} \ra_a =
-\langle a_- \CV_{e_0}^{(+)} \ra_a
=\p_{\L} \langle a_- \ra_a
$. This determines the 1-point function  $\langle a_- \ra_a$.

Finally let us mention another symmetric under the change of sign
$P\to - P$ solution of
\homrel,  represented as in \homrels,
\eqn\densityf{\eqalign{
&N_{P_1,P_2,P_3}\sim  4\,\int_{0}^{\infty}\, dt\(
{\prod_{i=1}^3\cosh  P_i t
\over \sinh {bt }\sinh {t\over b}} - {1\over  t^2}\) \cr
&=- {\p \over \p_{\a_1}} \log \(S_b(\a_{123}-Q)S_b(Q-\a_{23}^1)
S_b(\a_{12}^3) S_b(\a_{13}^2)\)
}}
The formula applies to 
complex values  of the momenta $P_i$.  This 3-point function
is similar to the density $\rho(P_1)$ which appears in the disk
partition function \ZZtp, \Tes, with the two boundary parameters
replaced by the two momenta $P_2,P_3$.  It is interpreted as the
derivative of the $\log$ of a particular fusion matrix element.  The
'diagonalizing' matrix here is a disc bulk 1-point function $\cosh
(Q-2\a)t$, analogous to the solution in \FZZb.

 \listrefs
 \bye